\documentclass[12pt,a4paper]{article}

\usepackage{latexsym}
\usepackage{amsmath}
\usepackage{amsfonts}
\usepackage{amssymb}
\usepackage{amsthm}
\usepackage{amscd}
\usepackage{mathrsfs}

\newcommand{\be}{\begin{equation}}
\newcommand{\ee}{\end{equation}}
\newcommand{\bea}{\begin{eqnarray}}
\newcommand{\eea}{\end{eqnarray}}

\def\no{\noindent}                      %
\def\cP{{\cal P}_y}                     %
\def\cG{{\cal G}}                       %
\def\G{{\cal G}}                        %
\def\gl{GL(2,\bZ)}                      %
\def\cA{{\cal A}}                       %
\def\cT{{\cal T}}                       %
\def\cN{{\cal N}}                       %
\def\cH{{\cal H}}                       %
\def\cI{{\cal I}}                       %
\def\ri{{\mathrm{i}}}                   %
\def\cR{{\cal R}}                       %
\def\cJ{{\cal J}}                       %
\def\bR{{\mathbb R}}                    %
\def\bC{{\mathbb C}}                    %
\def\bN{{\mathbb N}}                    %
\def\bZ{{\mathbb Z}}                    %
\def\bT{{\mathbb T}}                    %
\def\1{{\mbox{\boldmath $1$}}}          %
\def\tr{\mathrm{tr\,}}                  %
\def\cL{{\cal L}}                       %
\def\e{\epsilon}                        %
\def\lm{\lambda}                        %
\def\bt{\beta}                          %
\def\dt{\delta}                         %
\def\jp{\frac{1}{2}}                    %
\def\om{\omega}                         %
\def\omfs{ \omega_{\mathrm{FS}}}        %
\def\al{\alpha}                         %
\def\mmo{\mu^{-1}(\mu_0)}               %
\def\hb{\hat\beta}                      %
\def\ha{\hat\alpha}                     %
\def\val{\vert  y \vert}                %
\def\cp{\bC P(n-1)}                     %
\def\cE{{\cal E}}                       %
\def\cD{{\cal D}}                       %
\def\Hb{H^\mathrm{loc}}                 %
\def\tN{\tilde N}                       %
\def\fR{{\mathfrak{R}}}                 %
\def\fS{{\mathfrak{S}}}                 %
\def\stw{\stackrel{\wedge}{,}}          %
\def\cM{{\cal M}}                       %
\def\cC{{\cal C}}                       %
\def\cK{{L}}                            %
\def\IIIb{\mathrm{III}_\mathrm{b}}      %
\def\Aut{\mathrm{Aut}}                  %
\def\flat{\mathrm{flat}}                %
\def\res{\mathrm{res}}                  %
\begin{document}

\vspace*{0.5cm}
\begin{center}
{\Large \bf  Self-duality of the compactified Ruijsenaars-Schneider
system from quasi-Hamiltonian reduction}
\end{center}

\vspace{0.2cm}

\begin{center}
L. Feh\'er${}^{a}$ and C. Klim\v c\'\i k${}^b$  \\

\bigskip

${}^a$Department of Theoretical Physics, WIGNER RCP,  RMKI\\
H-1525 Budapest, P.O.B.~49,  Hungary, and\\
Department of Theoretical Physics, University of Szeged\\
Tisza Lajos krt 84-86, H-6720 Szeged, Hungary\\
e-mail: lfeher@rmki.kfki.hu

\bigskip

${}^b$Institut de math\'ematiques de Luminy
 \\ 163, Avenue de Luminy \\ F-13288 Marseille, France\\
 e-mail: klimcik@univmed.fr

\bigskip

\end{center}

\vspace{0.2cm}

\begin{abstract}
The Delzant theorem of symplectic topology is used to
derive the completely integrable compactified
Ruijsenaars-Schneider $\IIIb$ system   from a quasi-Hamiltonian
reduction of the internally fused double $SU(n)\times SU(n)$.
In particular, the reduced spectral functions depending respectively
on the first and second   $SU(n)$ factor of the double engender two
toric moment maps on the $\IIIb$  phase space $\cp$ that play the
roles of  action-variables and particle-positions.   A suitable
central extension of the  $SL(2,\bZ)$ mapping class group  of the torus with  one
boundary component  is shown to act on the quasi-Hamiltonian double by
automorphisms and, upon reduction,  the standard generator $S$
of the mapping class group
is proved to descend  to the
Ruijsenaars self-duality symplectomorphism  that exchanges the toric moment
maps.  We give also two new presentations of this duality map: one as
the composition of two Delzant symplectomorphisms  and the other as
the composition of three Dehn twist
symplectomorphisms realized by Goldman twist flows.
 Through the well-known relation between quasi-Hamiltonian manifolds and moduli
spaces, our results  rigorously establish the validity of the
interpretation [going back to Gorsky and Nekrasov] of the $\IIIb$ system in terms
of flat $SU(n)$ connections on the
one-holed torus.
\end{abstract}

\newpage

\tableofcontents

\newpage
\section{Introduction}
\setcounter{equation}{0}

A remarkable feature of the non-relativistic \cite{Cal,Sut,Mos} and
relativistic \cite{RS}
integrable many-body systems
of Calogero type
is their duality relation discovered by Ruijsenaars \cite{SR-CMP}.
The phase space of a classical  integrable many-body system is always
equipped with  two Abelian algebras
of distinguished observables:
the particle-positions tied to the physical interpretation and the
action-variables
tied to the Liouville integrability.
The `Ruijsenaars duality' between systems (i) and (ii) requires
the existence of a symplectomorphism between the pertinent phase spaces
that converts the particle-positions of system (i) into the
action-variables
of system (ii), and vice versa.
One talks of self-duality if the leading Hamiltonians of systems (i)
and (ii), which underlie
the many-body interpretation, have the same form possibly with
different coupling constants.
In addition to being fascinating in itself, the duality proved
useful for studying the dynamics, and it also appears  at the
quantum mechanical level where its manifestation is the bispectral
property \cite{DG} of the many-body
Hamiltonian operators \cite{SR-Kup,SR-CRM}.

The duality relation has been established in \cite{SR-CMP,RIMS95} with
the help of a direct method
  for all non-elliptic Calogero type
systems associated with the $A_n$ root system.
The present paper is part of the series \cite{FKinJPA,FKinCMP,FA}
aimed at understanding
all Ruijsenaars dualities
by means of the reduction approach.
The basic tenet of this approach, which
originated from the pioneering papers \cite{OP,KKS},
is that the integrable many-body systems descend from certain
natural `free' systems that can be reduced using their large symmetries.
Regarding the duality,
it is envisioned \cite{JHEP} that the starting phase spaces to be
reduced actually carry two `free' systems
that turn into a dual pair in terms of two models of a single reduced
phase space.
The existence of the symplectomorphism between two models
(that arise in the simplest cases as two gauge slices) of a single
reduced phase space is entirely automatic.
In this way the reduction approach may yield  considerable
technical simplification over the direct
method, where the proof of the symplectic character of the duality map
is very laborious.
However, nothing guarantees that this approach must always work;
finding the
correct reduction procedure relies on inspiration.

To this date, the reduction approach has been successfully implemented
for  describing all but two
cases of the known Ruijsenaars dualities.
The remaining  two cases are
the self-dualities of the hyperbolic  and of the  compactified
trigonometric Ruijsenaars-Schneider systems.
The physical interpretation of the latter system (also called $
\mathrm{III}_\mathrm{b}$ system in
\cite{RIMS95}, with
  `b' for `bounded')
is based on its local description valid before compactification.
Since it is needed subsequently, next we briefly summarize this local
description.

The definition of the $\mathrm{III}_\mathrm{b}$ system begins with the
local
Hamiltonian\footnote{The index $k$ in the next product $\prod_{k\neq
j}^n$ runs over
$\{1,2,...,n\}\setminus \{j\}$, and similar notation is used
throughout.}
\be
\Hb_{y}(\delta, \Theta) \equiv \sum_{j=1}^n \cos p_j \prod_{k\neq j}^n
\left[1 - \frac{\sin^2 y}{
\sin^2 (x_j-x_k)} \right]^{\frac{1}{2}},
\label{I.1}\ee
where  $y$ is a real non-vanishing parameter, the $\delta_j= e^{\ri 2
x_j}$ $(j=1,...,n)$
are interpreted as the positions
of $n$ `particles' moving on the circle, and the canonically conjugate
momenta $p_j$ encode the compact
variables  $\Theta_j = e^{-\ri p_j}$.
Here,  the
center of mass condition $\prod_{j=1}^n \delta_j = \prod_{j=1}^n
\Theta_j =1$ is also adopted.
Denoting the standard maximal torus of $SU(n)$ as $S\bT_n$, the local
phase space is
\be
M_y^{\mathrm{loc}}\equiv  \{ (\delta, \Theta)\,\vert\,
\delta= (\delta_1,...,\delta_n)\in  \cD_y,\, \Theta =
(\Theta_1,...,\Theta_n)\in  S\bT_n\},
\label{I.2}\ee
where the domain $\cD_y \subset S\bT_n$ (a so-called Weyl alcove with
thick walls \cite{vDV}) is chosen in such a way
to guarantee that $H_{y}^{\mathrm{loc}}$ takes real values.
The non-emptiness of $\cD_y$ is ensured by the restriction
$ \vert y \vert < \frac{\pi}{n}$.
The symplectic form on $M_{y}^{\mathrm{loc}}$ reads
\be
\Omega^{\mathrm{loc}}\equiv \frac{1}{2} \tr\!\left( \delta^{-1} d
\delta \wedge \Theta^{-1} d\Theta\right)
= \sum_{j=1}^n  dx_j \wedge dp_j  .
\label{I.3}\ee
The Hamiltonian $H_{y}^{\mathrm{loc}}$ can be recast as
the real part of the trace of the unitary Lax matrix
$L_{y}^{\mathrm{loc}}$:
\be
L_{y}^{\mathrm{loc}}(\delta,\Theta)_{jl}\equiv
\frac{e^{\ri y} - e^{-\ri y}}{e^{\ri y}\delta_j \delta_l^{-1} - e^{-
\ri y} }
W_j(\delta,y) W_l(\delta,-y) \Theta_l
\label{I.4}\ee
with the positive functions
\be
W_j(\delta,y):= \prod_{k\neq j}^n  \left[ \frac{ e^{\ri y} \delta_j  -
e^{-\ri y} \delta_k }
{\delta_j - \delta_k}  \right]^{\frac{1}{2}}.
\label{I.5}\ee
The flows generated by the spectral invariants of $L_{y}^{\mathrm{loc}}
$ commute, but are not
complete on $M_{y}^{\mathrm{loc}}$.
Ruijsenaars \cite{RIMS95} has shown that one can realize
$(M_{y}^{\mathrm{loc}}, \Omega^{\mathrm{loc}})$ as a dense open
submanifold
of the complex projective space $\bC P(n-1)$ equipped with a multiple
of the  Fubini-Study symplectic form,
  and thereby the commuting local flows generated by
$L_{y}^{\mathrm{loc}}$
  extend to complete Hamiltonian flows on the compact phase space $\bC
P(n-1)$.
The self-duality of the resulting
  compactified $\mathrm{III}_\mathrm{b}$ system was also proved in
\cite{RIMS95}.

 Besides its appearance in soliton theory,
 the importance of the $\IIIb$ system resides mainly in its
interpretation in terms of an appropriate  symplectic
  reduction of the space of $SU(n)$ connections on the torus with one
boundary component (i.e.~the one-holed torus).
  In fact, the local  version of the  $\IIIb$ system had been derived
by means of such a  symplectic reduction by Gorsky and his collaborators
\cite{GN,JHEP} who moreover  conjectured  that the
Ruijsenaars duality originates from the geometrically natural
action of the $SL(2,\bZ)$ mapping class group of   the torus  on the
reduced  phase space. However, important global issues such as the
compactification of the local  phase space and the  problem of the
completeness of the Liouville flows were not addressed in their
approach, and
they have not proved that the Ruijsenaars
self-duality symplectomorphism  of \cite{RIMS95} indeed originates  from the action of
the standard mapping class generator $S\in SL(2,\bZ)$.

The principal achievement of the present paper  is  a complete, global
reduction treatment of the compactified $\mathrm{III}_\mathrm{b}$
system including a simple proof of its self-duality.
The self-duality map will automatically arise as the composition of two
Delzant symplectomorphisms, which will pave the way to also prove its conjectured relation \cite{JHEP}
to the  mapping class group.
 To obtain these results,
we do  not  proceed by developing further the infinite-dimensional
approach of
\cite{GN}, but shall rather work in a suitable
{\it finite-dimensional} framework  based on
a non-trivial generalization of the Marsden-Weinstein symplectic
reduction, called
quasi-Hamiltonian reduction  \cite{AMM}.

  The quasi-Hamiltonian reduction was invented  \cite{AMM} as a
finite-dimensional alternative for describing the
  symplectic structures on various moduli spaces of flat connections
on Riemann surfaces whose investigation was initiated by
  Atiyah and   Bott  in the infinite-dimensional reduction context
  (see e.g. the book \cite{Khes} and references therein).  From this angle,
  it is not surprising that quasi-Hamiltonian
methods can be applied
  for finite-dimensional reduction treatment of  integrable systems
\footnote{It was remarked by Oblomkov
   \cite{Ob}
  that the Fock-Rosly treatment \cite{FR} of the complexified
trigonometric Ruijsenaars-Schneider system
  could be replaced by quasi-Hamiltonian reduction based on $GL(n,\bC)$.
  This is very close in spirit to our framework, but the compact case
that we consider
  is very different technically. }.
Nevertheless, we find it remarkable how naturally the
quasi-Hamiltonian
geometry together with the  Delzant theorem of symplectic topology
\cite{De}
lead to an understanding of the global structure of the compactified
Ruijsenaars-Schneider system.
In fact, the Delzant theorem will be applied to establish the existence
of two suitably equivariant symplectomorphisms, $f_\alpha$ and $f_\beta$, that both map
$\cp$ onto the quasi-Hamiltonian reduced phase space.
By utilizing their main features,  we also will be  able to construct these
Delzant symplectomorphisms explicitly,
and then shall recover the Ruijsenaars self-duality symplectomorphism
as the composition
$\fS = f_\alpha^{-1} \circ f_\beta: \cp \to \cp$.
A re-phrasing of this formula  will allow us to interpret the self-duality of the compactified
Ruijsenaars-Schneider system as a direct consequence
of the `mapping class democracy' between the $SU(n)$ factors of the
quasi-Hamiltonian
double.
In fact, we shall prove the presentation
$\fS = f_\beta^{-1} \circ S_P \circ f_\beta$ where $S_P$ stands for the natural action of $S\in SL(2,\bZ)$
on the quasi-Hamiltonian reduced phase space.
Inspired by results of Goldman \cite{Gold}, $S_P$ itself will be decomposed into a product
of three  Dehn twist sympletomorphisms realized as special cases of certain Hamiltonian flows.

The paper is essentially self-contained and
its organization is as follows.
      In Section 2, we  first recall the concept of
     quasi-Hamiltonian dynamics
     and the method of quasi-Hamiltonian
     reduction. Then  we describe the internally fused
     quasi-Hamiltonian double of the group $SU(n)$ and
     define two torus actions on it that will descend to the  reduced
phase space of our interest.
     In Section 3,
     we perform the reduction, we prove that the reduced phase space
is a
     Hamiltonian toric manifold
     (in two alternative but equivalent ways)
     and we
     find its topology and symplectic structure by  identifying
     the Delzant polytope  corresponding to the moment map of the
torus action.
     In Section 4,  we construct the Delzant symplectomorphisms
     $f_\alpha$ and $f_\beta$ explicitly.
     The local Lax matrix (\ref{I.4}) and its global extension will
arise naturally as building blocks
     of these maps.
     In Section 5,
     we recover the compactified Ruijsenaars-Schneider system
     and its self-duality from our reduction. In Section 6, we
demonstrate that the action of the  Ruijsenaars
     self-duality symplectomorphism on the $\IIIb$ phase space is  the standard action of
 the mapping class generator $S\in SL(2,\bZ)$.
Theorems 7, 8  of Section 5 and Theorem 9 of Section 6 are our main results representing the final outcome
of our analysis. Their implications  are further discussed in Section
7, together  with an outlook on open problems.

   \section{Preliminaries}
   \setcounter{equation}{0}

Quasi-Hamiltonian systems can be useful since they
can be reduced to honest Hamiltonian systems
by a generalization of the standard Marsden-Weinstein reduction
procedure,  and this can give an effective tool for studying the resulting reduced systems.
Below we first recall from \cite{AMM} the relevant notions, and then
describe those quasi-Hamiltonian dynamical systems that later will be shown to
yield the compactified Ruijsenaars-Schneider system upon reduction.

   \subsection{Quasi-Hamiltonian systems and their reductions}

Let $G$ be a compact Lie group with Lie algebra $\G$.
Fix an invariant scalar product
$\langle\cdot,\cdot\rangle$ on $\G$ and denote by $\vartheta$ and $\bar\vartheta$,
   respectively, the left- and right-invariant Maurer-Cartan forms on $G$.
   For a  $G$-manifold $M$ with action $\Psi: G\times M\to M$, we
   use $\Psi_g(m):= \Psi(g,m)$ and
   let $\zeta_M$ denote the vector field on $M$
   that corresponds to $\zeta\in \G$; we have $[\zeta_M, \eta_M] = -[\zeta, \eta]_M$
   for all $\zeta, \eta \in \G$.
   The adjoint action of $G$ on itself is given by $\mathrm{Ad}_g (x):= g x g^{-1}$,
   and $\mathrm{Ad}_g$ denotes also the induced action on $\G$.

   By definition \cite{AMM}, a quasi-Hamiltonian $G$-space $(M,G,\om,\mu)$ is a $G$-manifold $M$  equipped
   with an invariant $2$-form $\om\in\Lambda(M)^G$ and with an equivariant  map $\mu:M\to G$,
$\mu \circ \Psi_g = \mathrm{Ad}_g \circ \mu$,
 in such way that the following conditions hold.

   \noindent (a1) The differential of $\omega$ is given by
\be
   d\omega =-\frac{1}{12} \mu^*\langle\vartheta,[\vartheta,\vartheta]\rangle.
\ee
\noindent (a2) The infinitesimal action is related to $\mu$ and $\omega$  by
\be
\om(\zeta_M,\cdot) = \jp\mu^*\langle\vartheta+\bar\vartheta,\zeta\rangle, \quad \forall \zeta\in\G.
\ee
\noindent (a3) At each $x\in M$, the kernel of $\om_x$ is provided by
\be
   \mathrm{Ker}(\om_x)=\{\zeta_M(x)\,\vert\,
   \zeta\in \mathrm{Ker}(\mathrm{Ad}_{\mu(x)}+ \operatorname{Id}_\cG)\}.
\ee
The map $\mu$ is called the moment map.

   A quasi-Hamiltonian dynamical system  $(M,G,\om,\mu,h)$ is a quasi-Hamiltonian
   $G$-space with a distinguished
   $G$-invariant function $h\in C^\infty(M)^G$, the Hamiltonian.
   It follows from the axioms that there exists a unique
    vector field $v_h$ on $M$  determined by the following two requirements:
\be
\omega( v_h, \cdot)=dh, \qquad {\cal L}_{v_h}\mu=0.
 \label{hamact}\ee
The `quasi-Hamiltonian vector field' $v_h$ is $G$-invariant and it preserves $\omega$,
${\cal L}_{v_h}\om=0$.
Thus, $G$-invariant Hamiltonians on a quasi-Hamiltonian $G$-space define evolution flows
in much the same way as
arbitrary Hamiltonians do on symplectic manifolds.
One can also introduce an honest Poisson bracket on $C^\infty(M)^G$.
Naturally, if $f$ and $h$ are  $G$-invariant functions and
$v_f$ and $v_h$ the corresponding
quasi-Hamiltonian vector fields, then this Poisson bracket is given by
\be
\{f, h\}:=\omega(v_f,v_h).
\label{novy}\ee
Indeed, it is not difficult to check that the result $\{f,h\}$ is
again  an invariant function and all the usual properties (including the Jacobi identity)
are verified by this Poisson bracket.
It is worth emphasizing
that the general quasi-Hamiltonian manifold $M$ is not
symplectic and the quasi-Hamiltonian form $\omega$ does not induce a proper
Poisson algebra on the smooth functions on $M$ but just
on the $G$-invariant smooth functions.

The quasi-Hamiltonian reduction of a quasi-Hamiltonian dynamical system  $(M,G,\om,\mu,h)$ that
interests us
is determined by choosing an element $\mu_0\in G$. We say that
    $\mu_0$ is {\it strongly regular} if it satisfies the following two conditions:
\begin{enumerate}
 \item{
 The subset  $\mu^{-1}(\mu_0):=\{   x\in M\,\vert\,  \mu(x)=\mu_0\}$  is an embedded submanifold of $M$.}
\item{
   If  $G_0 \subset G$ is  the isotropy group of $\mu_0$ with respect to the adjoint action, then the
   quotient  $\mu^{-1}(\mu_0)/G_0$ is a manifold for which the canonical projection
$p:\mu^{-1}(\mu_0)\to \mu^{-1}(\mu_0)/G_0$ is a smooth submersion.}
\end{enumerate}
The  result of the reduction based on a strongly regular element $\mu_0$  is
 a standard Hamiltonian system,  $(P,\hat\om,\hat h)$.
The reduced phase space $P$ is the manifold
\be
P \equiv \mu^{-1}(\mu_0)/G_0,
\ee
which carries
the reduced symplectic form $\hat\om$ and reduced Hamiltonian $\hat h$ uniquely defined by
\be
p^*\hat\om=\iota^*\om, \quad p^*\hat h=\iota^*h,
\label{genred}\ee
where $\iota:\mu^{-1}(\mu_0)\to M$ is the tautological embedding.

We stress that $\hat \omega$ is a symplectic form in the usual sense, whilst $\omega$ is neither  closed
nor globally non-degenerate in general.
It follows from the above definitions that
the Hamiltonian vector field and the flow defined by
$\hat h$ on $P$ can be obtained by first restricting
the quasi-Hamiltonian
vector field $v_h$ and its flow to the `constraint surface' $\mu^{-1}(\mu_0)$ and then applying
the canonical projection $p$.
The Poisson brackets on $(P, \hat \om)$ are inherited from
the Poisson brackets (\ref{novy})  of the $G$-invariant functions as in
standard symplectic reduction.

\subsection{Evolution flows on the internally fused double of $SU(n)$}

Consider    a  quasi-Hamiltonian space $M$ and a set of $k$
distinguished $G$-invariant functions on it. In the sense of the
preceding subsection,
these data define a family of quasi-Hamiltonian dynamical systems. We
shall speak about a `commuting $k$-family' if the corresponding
quasi-Hamiltonian   vector fields all commute among each other.

In this paper, we shall deal with {\it two} commuting $(n-1)$-families
of  quasi-Hamiltonian dynamical systems, which both live on a {\it
single}
  quasi-Hamiltonian $G$-space.   The quasi-Hamiltonian $G$-space in
question is the so-called
\emph{internally fused double} of the group  $G:=SU(n)$  \cite{AMM},
which as a
manifold is provided by the direct product
\be
D:=  G \times  G =\{ (A,B)\,\vert\, A,B\in   G\}.
\label{2.1}\ee
The invariant scalar product on $\cG:= su(n)$ is given by
\be
\langle \eta, \zeta\rangle := -\frac{1}{2} \tr(\eta \zeta),
\qquad
\forall \eta, \zeta  \in \cG.
\label{scalarprod}\ee
The group $G$ acts  on $D$ by componentwise
conjugation\footnote{Later in some equations we apply  $g\in U(n)$ in the formula of the action,
which is harmless since only the factor group $SU(n)/\bZ_n \simeq U(n)/U(1)$ acts effectively.}
\be
\Psi_g: (A,B)\mapsto (  gA  g^{-1}, gB g^{-1}).
\label{2conj}\ee
The $2$-form $\omega$ of $M:=D$  reads
\be
2 \omega = \langle A^{-1} dA \stackrel{\wedge}{,} dB B^{-1}\rangle
+\langle  dA A^{-1} \stackrel{\wedge}{,} B^{-1} dB \rangle
- \langle (AB)^{-1} d (A B) \stackrel{\wedge}{,} (BA)^{-1} d (BA)
\rangle,
\label{2.2}\ee
  and the $G$-valued moment map $\mu$  is defined by
  \be
\mu(A,B)= AB A^{-1} B^{-1}.
\label{1.1}\ee

Consider a real class function $h\in C^\infty(G)^G$.
Define the derivative $\nabla h\in C^\infty(G,\G)^G$ by the equation
\be
{\left.\frac{d}{dt}\right\vert_{t=0}} h(e^{t \zeta}g) = \langle \zeta,
\nabla h(g)\rangle,
\qquad
\forall g\in G,\,\forall \zeta \in \G.
\label{grad}\ee
Associate to $h$ the following $G$-invariant functions on $D$,
\be
h_1(A,B):= h(A),
\qquad
h_2(A,B):= h(B).
\label{f1f2}\ee
The evolution flow of the quasi-Hamiltonian
system $(D, G, \omega, \mu, h_1)$ through
the initial value $(A_0, B_0)$ is furnished by
\be
(A(t), B(t)) = (A_0, B_0 e^{-t \nabla h(A_0)}),
\label{flow1}\ee
while the system $(D, G, \omega, \mu, h_2)$ has the flow
\be
(A(t), B(t)) = (A_0 e^{t \nabla h(B_0)}, B_0).
\label{flow2}\ee
Indeed, the evolution vector field given by the $t$-derivative
of the flow (\ref{flow1}) at
the point $(A(t),B(t))$ of the double equals $(0 \oplus -B(t)\nabla h(A_0))$
and one can easily verify
that it satisfies the defining relations (\ref{hamact}) of the
quasi-Hamiltonian vector field belonging to the function $h_1$.

In order to specify  the Hamiltonians of two commuting $(n-1)$-families
of  quasi-Hamiltonian dynamical systems on the double,
 we have to introduce
the so-called spectral functions on the group $G$.

As a preparation,
we define the \emph{alcove} $\cA$  by
\be
\cA:=\Bigl\{ (\xi_1,..., \xi_n)\in \bR^n\,\Big\vert\, \xi_j \geq 0,
\quad j=1,...,n, \quad \sum_{j=1}^n \xi_j=\pi\Bigr\},
\label{par}\ee
and the open alcove $\cA^0$ by
\be
\cA^0:=\Bigl\{ (\xi_1,..., \xi_n)\in \bR^n\,\Big\vert\, \xi_j > 0,
\quad j=1,...,n, \quad
\sum_{j=1}^n \xi_j=\pi\Bigr\}.
\label{par0}
\ee
We then consider the injective map $\dt$ from $\cA$ into the subgroup
$S\bT_n$ of  the diagonal elements
  of $SU(n)$  given by
  \be
  \dt_{11}(\xi):=e^{\frac{2\ri}{n}\sum_{j=1}^n j\xi_j},\quad  \dt_{kk}
(\xi):=e^{2\ri \sum_{j=1}^{k-1}\xi_j}\dt_{11}(\xi),
  \quad k=2,...,n.
  \label{par1}\ee
The image of $\delta$ is a fundamental domain for the action of the
Weyl group of
$SU(n)$ (i.e.~the permutation group) on $S\bT_n$, which is often
called a Weyl alcove.
For this reason, we may also  refer to $\cA$ as a Weyl alcove.
With the aid of the fundamental weights $\lambda_k$ of $su(n)$
represented by the diagonal matrices
$\lambda_k \equiv \sum_{j=1}^k E_{j,j} - \frac{k}{n} \1_n$,
the matrix $\delta(\xi)$ can be recast in the form
\be
\delta(\xi) = \exp\left(-2 \ri \sum_{k=1}^{n-1} \xi_k \lambda_k\right).
\label{fund}\ee
Here we denoted by  $E_{j,j}$ the $n\times n$-matrix featuring $1$ in
the intersection of the $j^{\mathrm{th}}$-row with
the $j^{\mathrm{th}}$-column and $0$ everywhere else.

Every element $A\in SU(n)$ can be written as
\be
A=g(A)^{-1}\dt(\xi)g(A),
\label{car}\ee
for some $g(A)\in SU(n)$ and {\it unique}  $\xi\in \cA$. Moreover,
whenever $\xi\in\cA^0$,
the element $A$  is {\it regular}
and $g(A)$ is then   determined   up to left-multiplication by an
element of $S\bT_n$.
By definition, the $j^{\,\mathrm{th}}$ component
of the alcove element $\xi$
entering the decomposition (\ref{car}) is
the value of  the \emph{spectral function} $\Xi_j$ on $A\in SU(n)$.
In other words, the conjugation invariant function $\Xi_j$
on $G=SU(n)$ is  characterized by
the equation
\be
\Xi_j( \delta(\xi)) = \xi_j,
\qquad \forall \xi \in \cA,
\quad j=1,...,n.
\label{bigXi}\ee
It is easily seen that the spectral function $\Xi_j$ is smooth on
$G_{\mathrm{reg}} \subset G$,
but it develops singularities at the non-regular points of $G$.
Note also that $\Xi_n = \pi - \sum_{j=1}^{n-1} \Xi_j$ according to (\ref{par}).

\emph{We are now in the position to define
the  $2(n-1)$ distinguished $G$-invariant Hamiltonians  $\al_j$, $\bt_j
$ on the double $D$
as follows:}
\be
\al_j(A,B):=\Xi_j(A),\quad \bt_j(A,B):=\Xi_j(B), \quad j=1,...,n-1.
\label{Ham}\ee
We call $\al_j$ and $\bt_j$   `spectral Hamiltonians' and our next
task is to show that they respectively define
{\it commuting} $(n-1)$-families of quasi-Hamiltonian dynamical
systems.
To be more precise, it must be noted that the domain of
the $\alpha_j$-Hamiltonians (resp.~$\beta_j$-Hamiltonians)
is the  dense open subset $ D_a\subset D$ (resp.~$D_b\subset D$)
  consisting of the couples $(A,B)\in D$ with
 $A\in G_{\mathrm{reg}}$ (resp.~$B\in G_{\mathrm{reg}}$), which is stable under the
 corresponding flows.
In order to describe the flows, we now prove the following lemma.

\medskip
\noindent
{\bf Lemma 1.} \emph{The derivative of the spectral function
$\Xi_j \in C^\infty(G_{\mathrm{reg}})$ ($j=1, ..., n-1$)
   reads
\be
\nabla \Xi_j (A) = g(A)^{-1} (  \ri (E_{j+1,j+1} - E_{j, j})) g(A),
\qquad
\forall A\in  G_{\mathrm{reg}}.
\label{A6}\ee}

\begin{proof}
The $G$-invariance of $\Xi_j$ implies the $G$-equivariance of $\nabla \Xi_j$,
and therefore it is enough
to calculate $\nabla \Xi_j$ at the points of the open Weyl alcove. But
at the points of the
Weyl alcove $\nabla \Xi_j$ must be a diagonal matrix, because of
invariance under $S\bT_n$
inherited from the $G$-equivariance on $G_{\mathrm{reg}}$. Then
$(\nabla \Xi_j)(\delta(\xi))$ is  readily calculated to be
$\ri (E_{j+1,j+1} - E_{j, j})$, which implies (\ref{A6})  on account
of (\ref{car}).
\end{proof}

Note that $\nabla \Xi_j(A)$ is well-defined by formula (\ref{A6}) since
  $g(A)$ is determined up to left-multiplication by the elements
of the maximal torus, and
its smoothness on $G_{\mathrm{reg}}$ follows directly from the
smoothness of $\Xi_j$.

By combining the formulae
(\ref{flow1}), (\ref{flow2}) and (\ref{A6}), we find that  the
following $2\pi$-periodic
  curve  in $D_a$ passing through $(A,B)$  is the integral curve of
the quasi-Hamiltonian vector field $v_{\al_j}$:
  \be
  \left(A,Bg(A)^{-1}{\rm diag}(1,1,...,1,e^{\ri t},e^{-\ri t},
1,...,1)g(A)\right), \quad t\in\bR,
  \label{cura}\ee
   and the  following   $2\pi$-periodic curve  in $D_b$ is an integral
curve of the vector field $v_{\bt_j}$:
  \be
  \left(Ag(B)^{-1}{\rm diag}(1,1,...,1,e^{-\ri t},e^{\ri t},
1,...,1)g(B),B\right), \quad
  t\in\bR.
  \label{curb}\ee
  In particular, the formulae (\ref{cura}) and (\ref{curb}) trivially
imply that the {\it $\alpha$-flows commute among themselves
  and so do the $\beta$-flows.}  In other words, the
infinitesimal actions of  the commuting quasi-Hamiltonian vector
fields $v_{\alpha_j}$   integrate to
a (smooth free) action of the torus
\be
\bT_{n-1}:= U(1)^{(n-1)}
\label{torus}\ee
on $ D_a\subset D$. The formula (\ref{cura})
   gives the action of the $j^{\,\mathrm{th}}$ $U(1)$ factor of $
\bT_{n-1}$,
  the phase $e^{\ri t}$ sits in the $j^{\,\mathrm{th}}$ place of the
diagonal and $g(A)$ is given by
  the decomposition (\ref{car}).
In spite of the ambiguity in the definition of $g(A)$, the curve
(\ref{cura}) is defined unambiguously.
To display the action map  $\Psi^a: \bT_{n-1} \times D_a \to D_a$
more explicitly,
we introduce
\be
\rho(\tau):= \exp\Bigl(\ri \sum_{j=1}^{n-1} t_j (E_{j,j} - E_{j+1, j
+1})\Bigr)
\quad\hbox{for all}\quad
\tau = (e^{\ri t_1}, ..., e^{\ri t_{n-1}}) \in \bT_{n-1}.
\label{rho}\ee
Then we have
\be
\Psi^a_\tau: ( A, B)\mapsto (A, B g(A)^{-1} \rho(\tau) g(A)).
\label{acta}\ee
Similarly, the commuting quasi-Hamiltonian vector fields $v_{\bt_j}$
generate
a $\bT_{n-1}$-action
  on the dense open subset $ D_b\subset D$, and
   the corresponding action map $\Psi^b: \bT_{n-1} \times D_b \to D_b$
reads
  \be
\Psi^b_\tau: ( A, B)\mapsto (A g(B)^{-1} \rho(\tau)^{-1} g(B), B).
\label{actb}\ee

We   observe also  that
\be \{\al_j,\al_l\}=0=\{\beta_j,\beta_l\},\ee
where the Poisson bracket of $G$-invariant functions was defined in
Eq.~(\ref{novy}). Indeed, we
have
\be   \{\al_j,\al_l\}\equiv \omega(v_{\al_j},v_{\al_l})\equiv \cL_{v_{\al_l}}
\al_j=0,
\label{Pocom}\ee
where the last equality holds since the
$\al_l$-generated flow acts only on the $B$-component of the
double (see Eq.~(\ref{flow1})) leaving therefore the $\al_j$-functions
invariant.
The Poisson-commutativity (\ref{Pocom}) of the spectral Hamiltonians
$\alpha_j$ (and that of the $\beta_j$) survives any quasi-Hamiltonian
reduction, and this fact will provide one of the  underpinnings
of  our  approach to the compactified Ruijsenaars-Schneider system.

\medskip\no
{\bf Remark 1.}
The spectral  Hamiltonians $\alpha_j$, $\beta_j$ can be viewed as the
respective generators of the  Poisson-commutative rings $\cC_a$ and
$\cC_b$  consisting of  smooth
invariant functions  defined with the help of Eq.~(\ref{f1f2}):
\be
\cC_a:= \{ h_1 \in C^\infty(D_a)^G\,\vert\, h\in C^\infty(G_{\mathrm{reg}})^G
\},
\quad \cC_b:= \{ h_2 \in C^\infty(D_b)^G\,\vert\, h\in C^\infty(G_{\mathrm{reg}})^G\}.
\label{2.11}\ee
The rings $\cC_a$ and $\cC_b$ can be of course generated also by
other  generators,  e.g.~by the invariants $H_{m}(A,B)\equiv\Re \tr(A^m)$,
$H_{-m}(A,B)\equiv \Im \tr(A^m)$
and, respectively,
by $F_{m}(A,B)\equiv \Re \tr(B^m)$,
$F_{-m}(A,B)\equiv \Im \tr(B^m)$ for $m\in \bN$.
Although the generators
$H_{\pm m}$, $F_{\pm m}$ have the apparent advantage
of being globally smooth on $G$, it  is more suited for our purpose to
use   the generators $\alpha_j$ and
$\beta_j$  since   their flows are $2\pi$-periodic (this circumstance
will be crucial for our arguments in
Subsections 3.3 and 3.4). It will be shown that
   after our quasi-Hamiltonian reduction  the matrix $A$ yields the Lax matrix of the
Ruijsenaars-Schneider system, the generators $H_{\pm m}$  become
the Ruijsenaars-Schneider Hamiltonians,
   the $\alpha_j$  become the action-variables and the
$\beta_j$ will parametrize the particle-positions.
We shall also establish a dual interpretation of
the reduction, where  $B$ yields the Lax matrix, the generators
$F_{\pm m}$  become the   Hamiltonians,
   the $\beta_j$  become the action-variables and  the
 $\alpha_j$ the parameters of the particle-positions.

\medskip\no
{\bf Remark 2.}
We note that from the viewpoint of the corresponding moduli spaces of
flat connections
the flows (\ref{flow1}) and (\ref{flow2}) are special
cases of the Goldman flows \cite{Gold}.
The fact that the spectral functions are not smooth at the non-regular
points of
$G$ will cause no problem, since we shall consider a quasi-Hamiltonian
reduction for which the constraint surface $\mu^{-1}(\mu_0)$ turns out
to be a submanifold of
$G_{\mathrm{reg}}\times G_{\mathrm{reg}} \subset D$.

\section{Reduction of the internally fused double of $SU(n)$}
\setcounter{equation}{0}

   As we already know, the starting point of the reduction is the choice of an  element $\mu_0\in G$,
   and the corresponding constraint surface $\mu^{-1}(\mu_0)$
 is  the space of those $(A,B)\in D$ that solve
 the moment map
 constraint\footnote{
 A similar constraint equation was studied previously in a different local context
 \cite{GN,JHEP}
and in complex holomorphic settings \cite{FR,Ob}.}
   \be
   AB A^{-1} B^{-1}=\mu_0.
   \label{con1}\ee
The simplest non-trivial possibility is to take $\mu_0$ from a conjugacy class of minimal but
non-zero dimension.
As seen from simple counting,  in this case
we may hope
to obtain a non-trivial reduced system
of dimension $2(n-1)$.
Obviously,  different choices from the same conjugacy class yield equivalent reduced systems.
We here choose $\mu_0$ {\it diagonal} of the form
    \be
 \mu_0={\rm diag}(e^{2\ri y},...,e^{2\ri y},e^{2(1-n)\ri y}),
 \quad y\in \bR.
 \label{1.2}
 \ee
 Anticipating its eventual identification with the parameter of the Hamiltonian (\ref{I.1}),
 in the next subsection we restrict $y$ to the range $0< \vert y \vert < \frac{\pi}{n}$ and then
 prove that $\mu_0$ (\ref{1.2}) leads to a smooth,
 compact  reduced phase  space $P= \mu^{-1}(\mu_0)/G_0$.

In the end, we shall
identify the reduced phase space with the complex projective space
$\bC P(n-1)$  and shall also
obtain a full characterization of the reduced spectral Hamiltonians
$\hat \alpha_j$ and $\hat \beta_j$
in terms of the standard parametrization of $\bC P(n-1)$.

\subsection{The reduced phase space is smooth and compact}

\noindent
{\bf Theorem  1.} \emph{Consider the diagonal matrix
$\mu_0 = {\rm diag}(e^{2\ri y},...,e^{2\ri y},e^{2(1-n)\ri y}) \in SU(n)$
with a real parameter $y$ verifying
\be
0<\vert y\vert <\frac{\pi}{n}.
\label{yn}\ee
Any such $\mu_0$ is a  strongly regular value of the moment map $\mu$ (\ref{1.1}),
and the corresponding  reduced phase space $P= \mu^{-1}(\mu_0)/G_0$ is a smooth,
compact manifold of dimension $2(n-1)$.}

\begin{proof}
We first remark that
$\mu^{-1}(\mu_0)$ is non-empty since
every element of  any connected, compact semi-simple Lie group can be written as a commutator
\cite{Goto}.

To continue, note that
the action (\ref{2conj}) of $G$ on the double naturally descends to an action of the factor group
$\bar G:= G/\bZ_n$, where $\bZ_n$ is the center of $G=SU(n)$.
Similarly, the action of the adjoint isotropy group $G_0 \subset G$ of $\mu_0$ on the constraint surface
$\mu^{-1}(\mu_0)$ descends to an action of the factor group
\be
\bar G_0:= G_0/ \bZ_n.
\ee
It is sufficient to prove that this latter action is free.
Indeed, the free action of $\bar G_0$ implies the embedded nature of $\mu^{-1}(\mu_0)$
by statement 3 of Proposition 4.1 of \cite{AMM} (which shows that
the locally free nature of the action of the isotropy group on the constraint surface
is equivalent to the regularity of the moment map value).
The fact that the compact Lie group $\bar G_0$  acts freely on the smooth compact manifold $\mu^{-1}(\mu_0)$
then ensures that
\be
\mu^{-1}(\mu_0)/G_0 = \mu^{-1}(\mu_0)/ \bar G_0
\ee
also becomes a smooth compact manifold.
As for its dimension, we  have  $\operatorname{dim}(\bar G_0) = (n-1)^2$, since
\be
G_0 = S( U(n-1) \times U(1)),
\ee
and therefore
\be
\operatorname{dim}\left(\mu^{-1}(\mu_0)/ \bar G_0 \right) =   (n -1)(n+1) - (n-1)^2 = 2 (n-1).
\label{1.25}
\ee

It remains to prove
that if $(A,B)\in \mu^{-1}(\mu_0)$ is fixed by some $g\in SU(n)$, then
$g$ belongs to the central subgroup $\bZ_n$.
For this, suppose that $(gA g^{-1}, gB g^{-1})=(A,B)$ holds for some $(A,B)\in \mu^{-1}(\mu_0)$ and
$g\in G$.
This implies that both $A$ and $B$  belong
to the centralizer subgroup
\be
G(g):= \{ \eta \in SU(n)\,\vert\, \eta g \eta^{-1} =g \} \subseteq SU(n),
\ee
and $\mu_0=AB A^{-1} B^{-1}$ belongs to the corresponding derivative subgroup $G(g)'$
that contains the group-commutators in $G(g)$.
Now observe that if $g$ is not central, then it is conjugate
to an element $g_0$ of the maximal torus of $SU(n)$ whose centralizer $G(g_0)$
is a block-diagonal subgroup
\be
G(g_0) = S (U(n_1) \times U(n_2) \times \cdots \times U(n_k)),
\label{Gg0}\ee
for some $k\geq 2$ and positive integers for which $n_1 + n_2 + \cdots + n_k =n$
($k=1$ occurs for $g\in \bZ_n$).
Accordingly, if $g$ is not central, then $\mu_0$ must be conjugate to an element of the
commutator subgroup $G(g_0)'$ of $G(g_0)$ (\ref{Gg0}).
It is readily seen that $G(g_0)'$ is  provided by
\be
G(g_0)' = SU(n_1) \times SU(n_2) \times \cdots \times SU(n_k),
\label{DerGg0}\ee
which leads to a contradiction.
Indeed, it follows from (\ref{yn}) that  in whatever way we partition the $n$ eigenvalues of $\mu_0$ into
 $k>1$ parts,
the product of the  eigenvalues in at least one  part (actually in each part)  will not be equal  to $1$.
Thus $\mu_0$ cannot be conjugate to an
element of $G(g_0)'$  (\ref{DerGg0}) for $k>1$.
\end{proof}

\medskip

We remark in passing that the above arguments show also the strong regularity of any
such moment map value
from  $SU(n)$ which is not conjugate to a  block-diagonal $SU(n)$ matrix whose blocks
themselves have determinant $1$.

\subsection{The images of  the Hamiltonians $\al_j$, $\bt_j$ restricted to $\mu^{-1}(\mu_0)$}

   Having established that the reduced phase space $\mu^{-1}(\mu_0)/  G_0$ is a compact  smooth manifold, the
   next step is to determine the reduced symplectic form $\hat\om$ on it.
   Remarkably, the shortest way
   to this goal leads through the study of the images of  the  spectral Hamiltonians $\al_j$,
   $\bt_j$ (\ref{Ham}) restricted to the constraint surface $\mu^{-1}(\mu_0)$.

 \medskip
 \noindent
 {\bf Theorem 2.} \emph{For $\mu_0=  {\rm diag}(e^{2\ri y},...,e^{2\ri y},e^{2(1-n)\ri y})$
 with $0< \vert y \vert < \frac{\pi}{n}$,
 the convex polytope
 \be \cP:=\Bigl\{(\xi_1,...,\xi_{n-1})\in \bR^{n-1}\,\Big\vert\,
  \xi_j \geq \vert y \vert, \,\,\, j=1,...,n-1, \,\,\,
  \sum_{j=1}^{n-1} \xi_j\leq \pi-\val\Bigr\}\label{poly}\ee
 is the common image of
 the vector-valued Hamiltonian functions
 $(\al_1,...,\al_{n-1})$ and $(\bt_1,...,\bt_{n-1})$ restricted to  the
 constraint surface $ \mu^{-1}(\mu_0)$.}

\begin{proof}
 The formulation of  Theorem 2 in terms of the convex polytope $\cP$ will play an important role
in Section 3.4.   However,  from the technical point of view, it is more convenient
to include into the analysis  also
the functions $\al_n(A,B):=\Xi_n(A)$, $\bt_n(A,B):=\Xi_n(B)$  (cf.~(\ref{bigXi})--(\ref{Ham})) and to prove
the following equivalent statement:

\no {\it The common image of
 the vector-valued   functions $(\al_1,...,\al_{n-1},\al_n)$ and $(\bt_1,...,\bt_{n-1},\bt_n)$
 restricted to  the constraint surface $ \mu^{-1}(\mu_0)$ is the set
  \be
  \cA_y:= \Bigl\{(\xi_1,...,\xi_n)\in \bR^n\,\Bigl\vert\, \xi_j \geq \vert y \vert, \,\,\, j=1,...,n,
  \,\,\, \sum_{j=1}^n \xi_j=\pi\Bigr\}.
  \label{paral}\ee}
The constraint $ABA^{-1}B^{-1}={\rm diag}(e^{2\ri y},...,e^{2\ri y},e^{2(1-n)\ri y})$
is invariant under the interchange of $A$ and $B$ accompanied with
 a simultaneous  change  of  the sign of the parameter $y$.  Since $\cA_y$ does not depend
 on the sign of $y$,  it is enough to show that the image of
 $(\bt_1,...,\bt_{n-1},\bt_n)$ restricted to  $ \mu^{-1}(\mu_0)$
is $\cA_y$.

  \noindent
  {\it \underline{Part 1}:}
  First we show that  if $\xi\in \cA_y$ then there exist $g(\xi)\in SU(n)$ and $A(\xi)\in SU(n)$ such that
 $A(\xi)$ and $B(\xi):=g(\xi)^{-1}\dt(\xi)g(\xi)$ solve the moment map constraint (\ref{con1}).
 (Recall
 that the map $\dt:\cA\to S\bT_n$ was defined in Eq.~(\ref{par1}) in connection with
 the decomposition (\ref{car}); below we use $\delta_j:= \delta_{jj}$.)

Consider an arbitrary $\xi=(\xi_1,...,\xi_n)\in\cA_ y$ and define
   \be
   \xi_{kn+j}:=\xi_j,  \quad k\in\bZ, \quad j=1,...,n.
   \label{328}\ee
   As an immediate consequence of Eq.~(\ref{par1}),
   note the validity of the following relation:
   \be
   \delta_j(\xi)\delta_l(\xi)^{-1}=\exp{\biggl(2\ri \sum_{k=l}^{j-1}\xi_k\biggl)},
   \qquad 1\leq  l<j \leq n.
\label{novy2}\ee
Moreover, using the convention (\ref{328}), we have
\be
\cot\val \geq\vert \cot( \sum_{k=l}^{j-1}\xi_k)\vert, \qquad
l=1,...,n; \quad j=l+1,...,l+n-1,
\label{330}\ee
because $\sum_{k=l}^{j-1}\xi_k$ in (\ref{330}) always lies in the
closed interval $[\vert y\vert , \pi - \vert y \vert\, ]$.
Thus for  $\xi\in\cA_ y$ and $l=1,...,n$  we obtain the reality and
\emph{non-negativity} of the
quantities $z_l(\dt(\xi), y)$ defined by
    \be
    z_l(\dt(\xi), y):=
  \frac{ e^{2\ri y}-1}{ e^{2n\ri y}-1}
\prod_{j\neq l}^{n}\frac{ \dt_j(\xi) -   e^{2\ri y} \dt_{l}(\xi)}
{\dt_j(\xi)-\dt_l(\xi) }=
\frac{(\sin{\val})^n}{\sin{(n\val)}}\prod_{j=l+1}^{l+n-1}\Big(\cot\val
-\frac{ y}{\val}
\cot( \sum_{k=l}^{j-1}\xi_k)\Big).
\label{5.8}\ee
Note that the second equality in (\ref{5.8}) follows from
(\ref{novy2}) and from the following trigonometric identity:
\be \cot{y} -\cot{\bt}\equiv \frac{\sin{(\beta-y)}}{\sin{y}\sin{\beta}}
\equiv 2\ri  \frac{e^{2\ri \beta}-e^{2\ri y}}{(e^{2\ri y}-1)(e^{2\ri
\beta}-1)}.\ee
Now consider an arbitrary  map $v:\cA_ y\to \bC^n$ such that
\be
\vert v_l(\xi)\vert^2:= z_l(\dt(\xi), y).
\label{mala}\ee
Let us show then that
\be \vert\vert v(\xi)\vert\vert^2:=\sum_{l=1}^n\vert v_l(\xi)\vert^2=1.
\label{norm}\ee
For this, we first check the equality of the following two polynomials in an auxiliary complex
variable $\lm$:
 \be
\prod_{j=1}^n (\dt_j(\xi) - \lambda)=\prod_{j=1}^n (\dt_j(\xi)\e^{2\ri y}- \lambda)
+ (e^{2\ri(1-n) y}-e^{2\ri y}) \sum_{k=1}^n \Bigl(\vert v_k(\xi) \vert^2 \dt_{k}(\xi)
\prod_{j\neq k}^n (\dt_j (\xi)e^{2\ri y} -\lambda)\Bigr).
\label{5.6a}\ee
Indeed, it is easy to verify (\ref{5.6a})  for  the $n$ (all distinct) values
$\lm_j=\dt_j(\xi)e^{2\ri y}$ , $ j=1,...,n$.
Consequently, (\ref{5.6a}) holds
true for any $\lm$, and we obtain (\ref{norm}) by evaluating (\ref{5.6a})
for $\lm=0$.

We note that
the polynomial identity (\ref{5.6a}) can be understood as the equality of the characteristic polynomials
of the diagonal matrix $\dt(\xi)$ and of the matrix $\mu_{v(\xi)} \dt(\xi)$,
 \be
 \det(\dt(\xi)-\lm\1_n)=\det(\mu_{v(\xi)}\dt(\xi)-\lm\1_n),
 \label{det}\ee
 where the matrix $\mu_{v(\xi)}$ reads
 \be
 \mu_{v(\xi)}:=e^{2\ri y}\1_n+ (e^{2\ri(1-n) y}-e^{2\ri y})  v(\xi)v(\xi)^\dagger.
 \label{3.35}\ee
 Because of the normalization property  (\ref{norm}), there certainly exists an $SU(n)$ matrix $g(\xi)$
 having the vector $v(\xi)$ as
 its last column, i.e., $v_j(\xi)=g(\xi)_{jn}$.
 It is then easily seen that the diagonal moment map value $\mu_0$ (\ref{1.2}) can be written as
 \be
 \mu_0=g(\xi)^{-1}\mu_{v(\xi)}g(\xi),
 \label{cosi}\ee
    and the determinant identity (\ref{det}) can be therefore rewritten as
     \be
     \det(g(\xi)^{-1}\dt(\xi)g(\xi)-\lm\1_n)=\det(\mu_0 g(\xi)^{-1}\dt(\xi)g(\xi)-\lm\1_n).
     \label{detb}\ee
This means that the matrix $B(\xi):=g(\xi)^{-1}\dt(\xi)g(\xi)$ has the same spectrum as the matrix
$\mu_0B(\xi)$, which implies
the existence of a  matrix $A(\xi)\in SU(n)$ such that
\be
A(\xi)B(\xi)A(\xi)^{-1}=\mu_0B(\xi).
\label{3.38}\ee
Part 1 of the proof of Theorem 2 is thus complete.

\noindent
{\it \underline{Part 2}:}
 It remains to show that if   $(A,B)\in D$ satisfies the moment map
constraint (\ref{con1}), then $B$ can be written as
\be
B=g^{-1}\dt(\xi)g
\label{Bform}\ee
with some $g\in SU(n)$ and some $\xi\in \cA_{y}$ (\ref{paral}).
Using that any $B\in SU(n)$ has the form (\ref{Bform}) with uniquely determined $\xi\in \cA$ (\ref{par}),
it will be convenient to    distinguish two cases: i) $\xi$ is in the open Weyl alcove  $\cA^0$ (\ref{par0});
ii) $\xi\notin\cA^0$.
 We  consider first i) and then ii). More precisely, we shall first prove
the  statement:

\noindent {\it i) If $(A,B) \in \mu^{-1}(\mu_0)$,  $B=g^{-1}\dt(\xi)g$ for
some $g\in SU(n)$ and $\xi\in \cA^0$, then
$\xi\in \cA_ y$.}

\noindent Then we prove the statement:

\noindent {\it ii) If $\xi\notin\cA^0$ then whatever is $g\in SU(n)$, the matrix $B=g^{-1}\dt(\xi)g$
cannot be the second component
of some solution $(A,B)$ of the constraint (\ref{con1}).}

\noindent  {\it Proof of statement i):}  Define $A_g$ and $\mu_g$ as
 \be
 A_g:=gAg^{-1},\quad \mu_g:=g\mu_0g^{-1}.
 \label{rs}\ee
The validity of   (\ref{con1}) implies
\be
A_g\dt(\xi)A_g^{-1}=\mu_g \dt(\xi).
\label{mcon}\ee
 Note that the matrix $\mu_g$ depends only on the last column of the matrix $g$.
 To see this we rewrite $\mu_0$ as
\be
\mu_0=e^{2\ri y}\1_n+ (e^{2\ri(1-n) y}-e^{2\ri y}) v_0v_0^\dagger,
\ee
where the  vector  $v_0\in\bC^n$  is defined by its components  $(v_0)_n:=1$,
$(v_0)_j:=0$, $j=1,...,n-1$.
 This means that $\mu_g$ can be written as
 \be
 \mu_g=e^{2\ri y}\1_n+ (e^{2\ri(1-n) y}-e^{2\ri y}) (gv_0)(gv_0)^\dagger=
 e^{2\ri y}\1_n+ (e^{2\ri(1-n) y}-e^{2\ri y}) vv^\dagger,
 \label{mg}\ee
 where  $v:= g v_0$ is the last column of the matrix $g$,  i.e.,  $v_j=g_{jn}$.

 We observe from (\ref{mcon}) that the spectrum of the matrix $\mu_g\dt(\xi)$ must be equal to
 the spectrum of $\dt(\xi)$, which
 entails the equality of the  characteristic polynomials
  \be
  \det(\dt(\xi)-\lm\1_n)=\det(\mu_{g}\dt(\xi)-\lm\1_n).
  \label{detc}\ee
 Both determinants in (\ref{detc}) can be easily evaluated so that (\ref{detc}) becomes
  \be
\prod_{j=1}^n (\dt_j(\xi) - \lambda)=\prod_{j=1}^n (\dt_j(\xi)\e^{2\ri y}- \lambda)
+ (e^{2\ri(1-n) y}-e^{2\ri y}) \sum_{k=1}^n \Bigl(\vert v_k \vert^2 \dt_{k}(\xi)
\prod_{j\neq k}^n (\dt_j (\xi)e^{2\ri  y} -\lambda)\Bigr).
\label{detd}\ee
Due to the assumption $\xi\in \cA^0$, we know that the elements of the diagonal matrix $\dt(\xi)$
have $n$ distinct values.
By evaluating the relation (\ref{detd}) for the
$n$ distinct values $\lm_l=\dt_{l}(\xi)e^{2\ri y}$, we
immediately find
\be
\vert v_l\vert^2=
 \frac{ e^{2\ri y}-1}{ e^{2n\ri y}-1}
\prod_{j\neq l}^n\frac{ \dt_j(\xi) -   e^{2\ri y} \dt_{l}(\xi)}{\dt_j(\xi)-\dt_l(\xi) }=
\frac{(\sin{\val})^n}{\sin{(n\val)}}\prod_{j=l+1}^{l+n-1}\Big(\cot\val -
\frac{ y}{\val}\cot ( \sum_{k=l}^{j-1}\xi_k)\Big).
\label{dete}\ee
Now we have to distinguish whether $ y>0$ or $ y<0$. We start with $ y>0$.
Then the first term in the last product in (\ref{dete}) is $\cot  y -\cot \xi_l$. If
 $\xi_l$ was strictly inferior to $\val$ for a certain $l$, we would have obviously
  \be
  \cot \val -\cot \xi_l<0.
  \ee
If we then had $\cot \val -\cot(\xi_l+\xi_{l+1})>0$,  this would imply
 $\cot\val -\cot (\xi_l+\xi_{l+1}+\xi_{l+2})>0$ etc,
 which would give $\vert v_l\vert^2<0$ by (\ref{dete}). In order to avoid such a contradiction,
  we see that the assumption $\xi_l<\val$
 leads to
 \be
 \cot\val -\cot(\xi_l+\xi_{l+1})\leq 0,
 \ee
 and hence to
 \be
 \xi_{l}+\xi_{l+1}\leq \val.
 \label{mh}\ee
 Because $\xi\in\cA^0$, we have  $\xi_l>0$ and $\xi_{l+1}>0$.
 This  fact together with (\ref{mh}) gives
  \be
  \xi_{l+1}<\val.\ee
  Thus, we have shown that if $\xi_l<\val$ for some $l$
  then $\xi_{l+1}<\val$, and hence $\xi_j<\val$ for all $j=1,...,n$.
  This is a contradiction since we know, respectively,  from
  (\ref{par}) and (\ref{1.2})
  that  $\sum_{j=1}^n\xi_j=\pi$ and $n\val<\pi$ hold.
  We conclude that $\xi_l\geq  \val$ for all $l=1,...,n$,
  whereby  statement i) is proved for $ y>0$.

  If $ y<0$, note that the {\it last} term in the last product in (\ref{dete}) is
  equal to $\cot \val -\cot \xi_{l-1}$. If
 $\xi_{l-1}$ was strictly inferior to $\val$, we would have obviously
  \be
  \cot \val -\cot \xi_{l-1}<0.
  \ee
   If moreover the next to last term,  $\cot \val -\cot(\xi_{l-1}+\xi_{l-2}) $, was strictly positive,
    this would give $\vert v_l\vert^2<0$ because
  all preceding terms would have to be strictly positive, too. Thus, the assumption $\xi_{l-1}<\val$
  leads to $\cot \val -\cot (\xi_{l-1}+\xi_{l-2})\leq 0$.
  This implies $\xi_{l-2}<\val$, and consequently  $\xi_{j}<\val$ for all $j$. But this creates
   the same contradiction as in the case $y>0$, whereby the proof of
statement i) is complete.

\medskip
\noindent  {\it Proof of  statement ii):}
To start, we note that the condition
 that $\xi\in \cA$ but $\xi\notin \cA^0$ (\ref{par0}) means that there exists
at least one index $l\in\{1,...,n\}$ for which $\xi_l=0$. We call such a configuration $\xi$
\emph{degenerate}, since it is characterized
by the fact that the phases $\dt_j(\xi)$  take only $r<n$ distinct values. We
find it more convenient to describe the degenerate configurations directly by their phases $\dt_j(\xi)$.
Since the map $\dt:\cA\to S\bT_n$ is injective, such a description is equivalent to the
previous  description in terms $\xi$ and from now on we  simply write $\dt_j$ instead of $\dt_j(\xi)$.

Fixing an arbitrary degenerate configuration, we partition  $n$ as a sum of $1\leq r<n$ positive integers,
\be
n=k_1+k_2+...+k_r,
\label{3.52}\ee
in such a way that
\be \dt_1=\dt_2=...=\dt_{k_1},
\,\,\,
\dt_{k_1+1}=\dt_{k_1+2}=...=\dt_{k_1+k_2},\,   \dots,\,  \dt_{\sum_{i=1}^{r-1}k_i+1}
=\dt_{\sum_{i=1}^{r-1}k_i+2}=...=\dt_n.
\label{5.9b}\ee
Plainly, at least one integer $k_s$ ($1\leq s\leq r$)  must be superior or equal to $2$.

Define the matrices $A_g$, $\mu_g$
and the vector $v$ in the same way
as in the proof of  statement i).
Then the assumed validity of   the relation $A_g\dt A_g^{-1}=\mu_g\dt$
entails  the  equality of the characteristic polynomials of the matrices $\dt$ and $\mu_g \dt$,
  which now yields
   \be \prod_{j=1}^r(\Delta_j-\lm)^{k_j}=\prod_{j=1}^r(\Delta_je^{2\ri y}-\lm)^{k_j}+
   (e^{2\ri(1-n)y}-e^{2\ri y})\sum_{m=1}^rZ_m\Delta_m(\Delta_me^{2\ri y}-\lm)^{k_m-1}
   \prod_{j\neq m}^r(\Delta_je^{2\ri y}-\lm)^{k_j}.
   \label{Detd}
   \ee
   Here we introduced $r$ distinct variables $\Delta_s$ ($s=1,..., r$),
 \be
 \Delta_1:=\dt_{k_1},
 \,\,\, \Delta_2:=\dt_{k_1+k_2},\, \ldots\,, \Delta_r:=\dt_{k_1+k_2+...+k_r}\equiv \dt_n,
 \label{5.9f}\ee
 and $r$ \emph{non-negative} real variables $Z_s$,
   \be
   Z_1:=\vert v_1\vert^2+\vert v_2\vert^2+...+\vert v_{k_1}\vert^2,\dots,
   Z_{j+1}:=\vert v_{k_1+...+k_{j}+1}\vert^2+\vert v_{k_1+...+k_{j}+2}\vert ^2+...+
   \vert v_{k_1+k_2+...+k_{j+1}}\vert ^2
   \label{5.10f}\ee
   for all $j=1,... , r-1$.
 Due to the degeneracy of $\dt$,
  the implications of (\ref{Detd})
  are  qualitatively different from the implications of its relative (\ref{detd}) obtained in case i).
  To see this,
  we now rewrite equation (\ref{Detd}) as a relation between two rational functions of $\lm$:
  \be
  Q(\Delta,y,\lm):=\prod_{j=1}^r\frac{(\Delta_j-\lm)^{k_j}}{(\Delta_je^{2\ri y}-\lm)^{k_j-1}}=
  \prod_{j=1}^r(\Delta_je^{2\ri y}-\lm)+(e^{2\ri(1-n)y}-e^{2\ri y})
  \sum_{m=1}^rZ_m\Delta_m \prod_{j\neq m}^r(\Delta_je^{2\ri y}-\lm).
  \label{deg1}\ee
  Eq.~(\ref{deg1}) says that the variables
   $\Delta_s$ (\ref{5.9f}) must be such that the rational function $Q(\Delta,y,\lm)$  is  a polynomial in $\lm$.
  This means that all putative poles of $Q(\Delta,y,\lm)$ must be cancelled by appropriate monomials
  present  in the numerator.
The necessary and sufficient condition for this to occur is easily seen to be the following:

\noindent {\it  For every index $m\in\{1,...,r\} $ such that $k_m>1$, there must exist an index
$s\in\{1,...,r \}$ such that $\Delta_s=\Delta_me^{2\ri y}$
  and $k_s\geq k_m-1$.}
 \hfill{(*)}

    From now on we consider only the \emph{admissible} degenerate $\delta$-configurations
 that, by definition, satisfy the condition (*).
 (We saw that other degenerate  configurations
 cannot occur in the spectrum of the matrix $B$ solving the constraint (\ref{con1}).)
 Taking any such configuration, with $\Delta$ in (\ref{5.9f}),
 we can find the uniquely determined quantities $Z_m=Z_m(\Delta,y)$ $(m=1,... , r)$ for which the relation
 (\ref{deg1}) is satisfied.
 For this,
 it is sufficient to use  $r$ different values of the parameter $\lm$ given by
  $\lm_m=\Delta_me^{2\ri y}$, whereby we obtain $Z_m(\Delta,y)$ from (\ref{deg1}).
  However, in distinction
 to the non-degenerate cases, here  three possibilities may occur.
 First, if $k_m=1$
 and there exists no such index $s$ for which $\Delta_s = \Delta_m e^{2\ri y}$,
 then we find
  \be
Z_m(\Delta, y)=\frac{ e^{2\ri y}-1}{ e^{2n\ri y}-1}
\prod_{j\neq m}^r\biggl(\frac{ \Delta_j -   e^{2\ri y} \Delta_{m}}{\Delta_j-\Delta_m}\biggr)^{k_j}
\neq 0.
\label{deg2}\ee
Second, if $k_m>1$ and $k_s=k_m-1$,   then
\be Z_m(\Delta, y)=(-1)^{k_m+1}e^{2(k_m-1)\ri y}\frac{ e^{2\ri y}-1}{ e^{2n\ri y}-1}
\prod_{j\neq m,s}^r\biggl(\frac{ \Delta_j -   e^{2\ri y} \Delta_{m}}{\Delta_j-\Delta_m }\biggr)^{k_j}
\neq 0.
\label{deg3}\ee
Here and below, it should not cause any confusion that we suppressed the dependence
of $s$ on $m$ as given by the condition (*).
Third, in the rest of the cases,
for which either $k_m=1$ and there exists an index $s$ with $\Delta_s = \Delta_m e^{2\ri y}$
or  $k_m>1$ and $k_s>k_m-1$, we obtain
  \be
  Z_m(\Delta,y)=0.
  \label{deg4}\ee

  Let $S(\delta)$ denote the set of the integers $m$ that occur in
  Eqs.~(\ref{deg2}) and (\ref{deg3}). This set cannot be empty, since otherwise all components
  of the vector $v= g v_0$ of unit norm were zero (cf.~(\ref{5.10f})).
  We are going to finish the proof of statement ii) by showing
  that  Eqs.~(\ref{deg2}) and (\ref{deg3})  imply  that at least  one of the a priori non-negative
  quantities $Z_m(\Delta,y)$
   is necessarily strictly negative, whatever is the admissible
   degenerate $\delta$-configuration that we consider.
  To this end,  we introduce a real positive parameter $\e$ and  associate to every admissible degenerate
  $\delta$-configuration
  a continuous
 $\e$-family of  configurations $\dt_\e$ in the open Weyl alcove $\delta(\cA^0) \subset S\bT_n$:
\be
\dt_{\e,p}:= \Delta_{1}e^{\ri p\e},\,\,\, p=1,..., k_1;\qquad
\dt_{\e,k_1+...+k_{j-1}+p}:= \Delta_{j}e^{\ri p\e}, \,\,\, p=1,...,k_{j}, \,\,\, j=2,...,r,
\label{5.11a}\ee
where we use the partition (\ref{3.52}).
It is evident that for sufficiently small values of $\e>0$ the
 configurations  $\dt_\e$  are   {\it non-degenerate}, i.e.,
  they sit in the $\dt$-image of $\cA^0$ (defined in (\ref{par0}) and (\ref{par1})).
   Consider now  for $l=1,...,n$ the   quantities  $z_l(\dt_\e,y)$,
   \be z_l(\dt_\e,y):=
 \frac{ e^{2\ri y}-1}{ e^{2n\ri y}-1}
\prod_{j\neq l}^n\frac{ \dt_{\e,j} -   e^{2\ri y} \dt_{\e,l}} {\dt_{\e,j}-\dt_{\e,l} },
\ee
which appeared also in the  formula  (\ref{5.8}).
  From the fact that the configuration is admissible (*)  it follows easily
  that for some $l$'s the quantities  $z_l(\dt_\e,y)$ vanish.
 More precisely, we first observe that
 \be z_l(\dt_\e,y)=0 \qquad {\rm if} \qquad l\notin\{k_1,k_1+k_2,...,k_1+...+k_r\}
 \label{gr1}\ee
 and  also
 \be z_{k_1+k_2+...+k_m}(\dt_\e,y)=0\quad   \quad \hbox{$\forall\, m$\, for
 which\, $\exists\, s$\,
 such that\, $\Delta_s=e^{2\ri y}\Delta_m$\,  and\, $k_s\geq k_m$}.
 \label{gr2}\ee
Moreover, it turns out that  the {\it limits }  $\lim_{\e\to 0}z_l(\dt_\e,y) $
exist and do not vanish for all
 other $l$. That is they do not vanish  for all $l=k_1+k_2+...+k_m$ for which
 one of the following alternatives occurs:
 a) $k_m=1$ and  there is no $s$ such that $\Delta_s=e^{2\ri y}\Delta_m$; b)  $k_m>1$,
 $\Delta_s=e^{2\ri y}\Delta_m$ and $k_s=k_m-1$.
 Those non-vanishing limits read
 \be
 \lim_{\e\to 0} z_{k_1+...+k_m}(\dt_\e,y)=Z_m(\Delta,y),
  \label{lims}\ee
where $Z_m(\Delta,y)$ is given
by Eqs.~(\ref{deg2}) or (\ref{deg3})  for the cases a) and b), respectively.
In other words, the $m$-values occurring in (\ref{lims})  form the set $S(\delta)$
 defined after (\ref{deg4}).

  Now turning to the crux of the argument,  we note that for small $\e$ the configuration
   $\dt_\e$  does not belong to the $\dt$-image of $\cA_y$ (\ref{paral}).
   Indeed, whenever $k_j>1$ (recall that such $1\leq j\leq r$ exists), we observe that
   in the $\xi$-parametrization (\ref{par1}) of
  the configuration $\dt_\e$ we have $\xi_{k_j-1}=\e/2<\val$.  Following the proof
  of statement i),
  the quantity  $z_l(\dt_\e, y)$ must be therefore  strictly negative
  at least for one (in principle $\e$-dependent) value of  $l$.
  As an obvious consequence, there must also exist a fixed  index $l$ and
  a decreasing series $\e_p\to 0$  such that $z_l(\dt_{\e_p},y)$ is strictly negative for
  all positive integers $p$.
  From (\ref{gr1}) and (\ref{gr2}), we conclude the existence of an integer
   $m^*\in S(\delta)$ such
  that the above mentioned $\e$-independent $l$ is given by $l=k_1+k_2+...+k_{m^*}$.
 We know from (\ref{lims}) that the limit
  \be
  \lim_{p\to \infty}z_{k_1+...+k_{m^*}}(\dt_{\e_p},y)=Z_{m^*}(\Delta,y)
  \ee does not
  vanish,
   which implies that  $Z_{m^*}(\Delta,y)$ must be strictly negative. This is a contradiction
  with the non-negativity
  of the variables $Z_m$ (\ref{5.10f}).
\end{proof}

   \subsection{The reduced phase space is a Hamiltonian toric manifold}

   By definition,  a Hamiltonian toric manifold\footnote{The general theory of these
   compact completely integrable systems is reviewed, for example, in \cite{Cannas}.}
   is a compact, connected symplectic manifold of
   dimension $2d$  equipped with an effective, Hamiltonian action of a torus of dimension $d$.
   We already know that the reduced phase space $P=\mu^{-1}(\mu_0)/G_0$ is a compact
   symplectic manifold. The following Lemmas 2 and 3 show that $P$ is  a
    Hamiltonian toric manifold.
    Lemma 2 refers to the $\beta$-generated torus action (\ref{actb}), but
    of course an analogous result holds also for the $\alpha$-generated action (\ref{acta});
    and eventually this will explain the Ruijsenaars self-duality.

   \medskip\no
   {\bf Lemma 2.} \emph{The $\bt$-generated $\bT_{n-1}$-action  on the
   open submanifold $D_b$
   of the internally fused double $D$, given by (\ref{actb}),
   descends to the reduced phase space $P=\mu^{-1}(\mu_0)/G_0$,
    where it becomes Hamiltonian and effective.}

   \begin{proof}
   It follows from Theorem 2 that the  constraint surface  $\mu^{-1}(\mu_0)\subset D$
    lies completely
    in the open submanifold $D_b\subset D$ (recall  that
    $D_b$ is the set of pairs $(A,B)\in D$ for which $B$ is regular).
    Thus,
   the statement that the torus action $\Psi^b$  (\ref{actb}) descends to a Hamiltonian
   torus action on the reduced phase space follows immediately from the general
   theory of quasi-Hamiltonian reduction \cite{AMM}, which we briefly summarized around equation
   (\ref{genred}).
   In fact, the reduced torus action
    $\hat \Psi^b: \bT_{n-1} \times P \to P$ can be  defined by means of the equality
    \be
    \hat \Psi^b_\tau \circ p = p \circ \left(\Psi^b_\tau\vert_{\mu^{-1}(\mu_0)}\right) ,
    \qquad
    \forall \tau \in \bT_{n-1},
    \label{hatpsib}\ee
    where $p: \mu^{-1}(\mu_0) \to P$ is the canonical projection.
    The corresponding infinitesimal torus action on $P$ is generated by the vector fields
     $\hat v_{\hat\bt_j}$ ($j=1,..., n-1)$ that are
     the projections of
     the vector fields $ v_{\bt_j}$ (\ref{curb}) restricted to $\mu^{-1}(\mu_0)$.
  These projected vector fields are Hamiltonian,
 \be
 \hat\om( \hat v_{\hat \bt_j}, \cdot)=d\hat\bt_j,
 \ee
 where  $\hat \omega$ is the
 reduced symplectic form on $P$ and
 the reduced Hamiltonians $\hat \beta_j\in C^\infty(P)$ are characterized  by
 $\hat \beta_j \circ p = \beta_j \circ \iota$ using the embedding
 $\iota: \mu^{-1}(\mu_0) \to D$.
 In other words,
 \be
 \hat\beta \equiv (\hat \beta_1,..., \hat \beta_{n-1}): P \to \bR^{n-1}
 \label{hatbeta}\ee
 is the moment map for the $\bT_{n-1}$-action $\hat \Psi^b$  on $P$.

 Suppose now that the $\bT_{n-1}$-action $\hat \Psi^b$ on $P$ is \emph{not} effective.
 We observe from (\ref{actb}) that this is equivalent to the
 existence of a \emph{non-unit} element
 $\rho \in S\bT_n$ such that  for all
 $(A,B)\in \mu^{-1}(\mu_0)$ there exists an element $h(A,B)\in G_0$ satisfying
 \be
 (Ag(B)^{-1}\rho g(B),B)=(h(A,B)Ah(A,B)^{-1},h(A,B)Bh(A,B)^{-1}).
 \label{eff}\ee
 This means that $\hat \Psi^b_\tau(p(A,B)) = p(A,B)$ for the
 element $\tau \in \bT_{n-1}$ for which
 $\rho= \rho(\tau)$ according to (\ref{rho}).
 Note from (\ref{eff}) that $h(A,B)$ must commute with $B$.
 Because $(A,B)\in \mu^{-1}(\mu_0)$,
 $B=g(B)^{-1}\dt(\xi)g(B)$ is regular by Theorem 2 and therefore there exists some
 $(A,B)$-dependent
 $\zeta=\mathrm{diag}(\zeta_1,..., \zeta_n)\in S\bT_n$ such that
 \be
  h(A,B)=g(B)^{-1}\zeta g(B).
  \label{3.71}\ee
 The fact that $h(A,B)\in G_0$  then says that
 \be
 g(B)^{-1}\zeta g(B) \mu_0=\mu_0g(B)^{-1} \zeta g(B),
 \label{rov1}\ee
 or, equivalently,
 \be
 \zeta g(B)\mu_0 g(B)^{-1}= g(B) \mu_0g(B)^{-1}\zeta  .
 \label{rov2}\ee
 We know from the proof of Theorem 2 that the last column of the matrix $g(B)$ is given by a
 vector $v(\xi)$ verifying
 \be
 \vert v_l(\xi)\vert^2
=\frac{(\sin{\val})^n}{\sin{(n\val)}}\prod_{j=l+1}^{l+n-1}\Big(\cot\val -
\frac{ y}{\val}\cot( \sum_{k=l}^{j-1}\xi_k)\Big),
\qquad \forall\, l=1,..., n,
\label{5.8b}
\ee
where $\xi\in\cA_y$ (\ref{paral}).
If $\xi$ belongs to the interior $\cA^0_y$ of $\cA_y$,
\be
\cA^0_y:= \Bigl\{(\xi_1,...,\xi_n)\in \bR^n\,\Big\vert\, \xi_j > \vert y \vert, \quad j=1,...,n,
\quad \sum_{j=1}^n \xi_j=\pi\Bigr\},
\label{paralo}
\ee
then all components of the
vector $v(\xi)$ are non-vanishing.
In this case we compare the last columns of the matrices on  the two sides of Eq.~(\ref{rov2}).
By using the formula of $\mu_0$, this leads to the relation
\be
 \zeta v(\xi)   =   v(\xi)  \zeta_n,  \ee
from which we conclude that $\zeta=\zeta_n\1_n$. It then follows
from (\ref{3.71}) that $h(A,B)=\zeta_n \1_n$, and thereby
(\ref{eff}) implies
 that $\rho=\1_n$. This contradicts our assumption that $\rho$ is a  non-unit element of
$S\bT_n$. Therefore the $\bT_{n-1}$-action $\hat \Psi^b$ on $P$ is effective.
\end{proof}

The
statement of the following lemma can be obtained as an  immediate consequence
of Theorem 7.2 of \cite{AMM} (the proof of which itself is based on
results of \cite{MW}). We give here  a direct proof  since we shall
need some details of it subsequently.

\medskip\no
{\bf Lemma 3.} \emph{The reduced phase space  $P=\mu^{-1}(\mu_0)/G_0$ is connected.}

\begin{proof}
It is enough to prove that any two points of $P$ can be connected
by a continuous path.
We fix a point $x\in P$ and define $P_x:=\{z\in P\,\vert\, \hat\bt(z)=\hat\bt(x)\}$,
where $\hat\bt=(\hat\bt_1,...,\hat\bt_n)$
is the $\bR^n$-valued function on $P$ that descends  from the
$\bR^n$-valued\footnote{The fact that we here consider $\hat \bt$
 and $\bt$ as $\bR^n$-valued functions,
but elsewhere view them
as $\bR^{n-1}$-valued functions should not lead to any confusion;  we have
$\hat\bt_n \equiv \pi - \sum_{k=1}^{n-1} \hat \beta_k$  and similarly for $\bt$.}
invariant function $\bt$ on $\mmo$.
We  pick an arbitrary $z\in P_x$, and next show that $x$ can be connected to $z$.
To begin,
denote some  representatives of $x$ and $z$ in $\mmo$
by $(A,B)$ and $(A',B')$, respectively.  Referring to Section 2.2, we have
$g(B) B g(B)^{-1} =\delta(\hb(x)) =g(B') B' g(B')^{-1}$
for some $SU(n)$ matrices $g(B)$ and $g(B')$.
We see from the proof of Theorem 2 that
 $g(B)$ and $g(B')$ can be chosen to have the same  last  column (in fact, one may take
  $g(B)_{jn}=g(B')_{jn}=z_j(\delta(\hb(x)), y)^{\frac{1}{2}}$  defined in (\ref{5.8})).
 Then it follows that  $g(B) \mu_0 g(B)^{-1}= g(B') \mu_0 g(B')^{-1}$, which in turn implies that
 $h:=g(B)^{-1}g(B')$ is in $ G_0$.
 This tells  us that the representative of $z$ in $\mmo$ can be replaced by
  $(A'',B) = (h A' h^{-1}, h B' h^{-1})$.
  Then, it  must be true that  $A''=AM$ where $M=g(B)^{-1} \zeta g(B)$ for some $\zeta\in S\bT_n$.
  This holds because
the moment map constraint for $(A,B)$ and $(A'', B)$ implies that  $ABA^{-1}=A''BA''^{-1}$, and
  $B$ is regular  by Theorem 2.
Next, by using the $\bT_{n-1}$-action (\ref{actb}), we can rewrite
the equality $(A'',B) = (AM,B)$ as
   $(A'',B) = \Psi_\eta^b(A,B)$, where $\eta \in \bT_{n-1}$ is defined by the relation
  $\zeta^{-1} = \rho(\eta)$ with (\ref{rho}).
  Finally, we choose a continuous curve $ [0,1] \ni s \mapsto \tau(s) \in \bT_{n-1}$ for which
  $\tau(0)$ is the identity and $\tau(1)= \eta$, whereby we  obtain
  the continuous path
  $\hat \Psi^b_{\tau(s)}(x)$ in $P$  that connects $x$ to $z$.
  Notice that $\hat \Psi^b_{\tau(s)}(x)\in P_x$ for all $s$, and thus
  we have also shown that the $\hat \Psi^b$ action (\ref{hatpsib}) of
  $\bT_{n-1}$ is transitive on $P_x$.

We now take two arbitrary points $x_0,x_1 \in P$ for which $ \hat \bt(x_0)\neq  \hat\bt(x_1)$,
and prove the existence of a  continuous path $[0,1]\ni s \mapsto x(s)\in P$ for which
$x(0)=x_0$ and $x(1)=x_1$.
The following argument relies on the first part of the proof of Theorem 2.
We begin by choosing a continuous path $\xi(s) \in \cA_y$  in such a way
that $\xi(0)= \hat \bt(x_0)$ and $\xi(1)=\hat\bt(x_1)$.
Next we define the vector function $v(s)$ by putting $v_l(s):= z_l(\delta(\xi(s)), y)^{\frac{1}{2}}$
using (\ref{5.8}).
Since $v(s)$ is continuous in $s$, we can find (actually could give explicitly)
an $SU(n)$-valued continuous function $g(s)$ that solves $\mu_0 = g(s)^{-1}\mu_{v(s)} g(s)$,
where $\mu_{v(s)}$ is obtained by replacing $v(\xi)$ in (\ref{3.35}) by $v(s)$
(see also (\ref{cosi})).
We continue by defining $B(s):= g(s)^{-1} \delta(\xi(s)) g(s)$, and then  note the
existence of a continuous function $A(s) \in SU(n)$ for which $A(s) B(s) A(s)^{-1}
=\mu_0 B(s)$. Such function exists since $B(s)$ is similar to $ \mu_0 B(s)$,
as can be seen from the discussion around equations (\ref{det})-(\ref{3.38}),
and the eigenvectors of $B(s)$ and that of $ \mu_0 B(s)$ can be chosen as continuous
functions of $s$.
Now the projection  of the curve $(A(s), B(s))\in\mu^{-1}(\mu_0)$
yields a continuous curve $\tilde x(s):= p(A(s), B(s)) \in P$ for which
$\hat \bt(\tilde x(0)) = \hat\bt(x_0)$ and $\hat \bt(\tilde x(1)) = \hat\bt(x_1)$.
By the previous part of the proof, it is obviously
possible to find a continuous curve $\tau(s) \in \bT_{n-1}$ such that
$x(s) := \hat \Psi^b_{\tau(s)}(\tilde x(s))$ gives the path connecting $x_0$ with $x_1$.
\end{proof}

\medskip\no
{\bf Remark 3.}  The main message of the present subsection is that
the reduced phase space $(P, \hat \omega)$ is naturally equipped with  \emph{two} effective
Hamiltonian actions of the torus $\bT_{n-1}$.
The first is the action $\hat \Psi^b$ (\ref{hatpsib}), which we call the
$\hat \beta$-generated action since its moment map is given by $\hat \beta$ (\ref{hatbeta}).
The second is the $\hat \alpha$-generated action, which can be defined by the formula
 \be
    \hat \Psi^a_\tau \circ p = p \circ \left(\Psi^a_\tau\vert_{\mu^{-1}(\mu_0)}\right),
    \qquad
    \forall \tau \in \bT_{n-1},
    \label{hatpsia}\ee
where $\Psi^a$ is given by (\ref{acta}). The corresponding moment map is
\be
 \hat\alpha \equiv (\hat \alpha_1,..., \hat \alpha_{n-1}): P \to \bR^{n-1},
 \label{hata}\ee
where the functions $\hat \alpha_j$ descend from the spectral
Hamiltonians $\alpha_j$ introduced in (\ref{Ham}).
The data $(P, \hat\omega, \hat \alpha)$ and $(P, \hat \omega, \hat \beta)$
both encode Hamiltonian
toric manifolds, or in other words we have two completely integrable systems
on the reduced phase space $(P, \hat \omega)$.

\medskip\no
{\bf Remark 4.} Let us introduce the following open submanifolds of $P$:
\be
P_0^a:= \hat \alpha^{-1}( \cP^0) \quad\hbox{and}\quad
P_0^b:= \hat \beta^{-1}( \cP^0).
\label{3.79}\ee
Observe from the proofs of Lemmas 2 and 3 that the $\hat \Psi^b$ action is \emph{free}
and \emph{transitive} on $P_x$ for all
$x\in P_0^b$ (the transitivity holds for all $x\in P$).
Thus $P_0^b$ is  a principal $\bT_{n-1}$-bundle over the base $\cP^0$.
This bundle is topologically trivial  since its base
is contractible. Consider now $x_0,x_1\in P$ such that $x_0\notin P^b_0$ and $x_1\in P^b_0$.
Then $x_0$ can be connected to $x_1$ by
  a curve $x(s)$ as in the proof of Lemma 3 in such a way
that $\hb( x(s))\in \cP^0$ for all $0< s \leq 1$ (since $\cP$ is a convex polytope). This in turn
implies that $P_0^b$ is a \emph{dense} open submanifold of $P$.
By the reasoning used
below Eq.~(3.27), $P_0^a$ is a also a dense open submanifold of $P$ and, equipped with
the $\hat \Psi^a$ action, it is
a principal $\bT_{n-1}$-bundle over $\cP^0$.

\subsection{The global structure of the reduced systems $(P, \hat \omega, \hat \alpha)$
and $(P, \hat \omega, \hat\beta)$}

We below identify the reduced systems by utilizing a celebrated result of Delzant \cite{De} that
characterizes Hamiltonian toric manifolds in terms of the image of the moment map.

\medskip
  \no
  {\bf Delzant's first theorem (Th.~2.1 of \cite{De}).} \emph{Let $(M_1, \omega_1)$ and
  $(M_2, \omega_2)$
  be $2d$-dimensional Hamiltonian toric manifolds with  moment maps
$\Phi_1: M_1 \to {\cal T}^*$ and $\Phi_2: M_2 \to {\cal T}^*$,
where $\cT$  is the Lie algebra of the
$d$-dimensional torus $\bT$ acting on $M_1$ and on $M_2$. If the images $\Phi_1(M_1)$ and
$\Phi_2(M_2)$ coincide, then there exists a $\bT$-equivariant  symplectomorphism
$\varphi: M_1 \to M_2$ such that $\Phi_2 \circ \varphi = \Phi_1$.}
\medskip

According to an earlier  result of Atiyah, Guillemin and Sternberg,
the images in question are convex
polytopes.
Delzant also obtained full classification of the moment polytopes associated with
Hamiltonian toric manifolds, which are now routinely called Delzant polytopes
\cite{Cannas}.

By now,  we have exhibited two effective Hamiltonian actions of the
torus $\bT_{n-1}=U(1)^{n-1}$ on the compact connected reduced phase space $(P, \hat \omega)$.
Referring to a fixed basis\footnote{Our base elements, $X_1, ..., X_{n-1}$, correspond to
a fixed product structure (\ref{torus}), and realize
$\bT_{n-1}$ as  $\cT_{n-1}$ factored by the lattice
$\mathrm{span}_\bZ\{ 2\pi X_1, ..., 2\pi X_{n-1}\}$,
i.e., the corresponding Hamiltonian flows are $2\pi$-periodic.}
of the Lie algebra $\cT_{n-1}$ of $\bT_{n-1}$,
the respective moment maps are $\hat \alpha: P \to \bR^{n-1}$ (\ref{hata}) and
$\hat \beta: P\to \bR^{n-1}$ (\ref{hatbeta}).
The Delzant polytopes are provided \emph{in both cases} by $\cP$ (\ref{poly}).

 Specialists of symplectic geometry can immediately recognize the Delzant
 polytope $\cP$ (\ref{poly}) as the one associated with a very
 standard Hamiltonian toric manifold:
 $\bC P(n-1)$ equipped with a multiple of the Fubini-Study form and the familiar
 `rotational action' of $\bT_{n-1}$. For the sake of keeping our paper self-contained,
 and also since we need to fix notations, we next explain how this Hamiltonian
 toric manifold comes about.

Let us start with the symplectic vector space $\bC^n \simeq \bR^{2n}$ endowed
with the Darboux form
\be
\Omega_{\bC^n} = \ri \sum_{k=1}^n d\bar u_k \wedge du_k,
\label{4.1}\ee
where $u_k$ ($k=1, ..., n$) are the components of  the vector $u$ that runs over $\bC^n$.
Then consider the Hamiltonian action $\psi$ of the group $U(1)$ on $\bC^n$ operating as
\be
\psi_{e^{\ri \gamma}}(u):= e^{\ri \gamma} u.
\label{4.6}\ee
This action is generated by the moment map $\chi: \bC^n \to \bR$,
\be
\chi(u)\equiv \sum_{k=1}^n \vert u_k \vert^2,
\label{4.7}\ee
since $d\chi= \Omega_{\bC^n}(V,\cdot)$ holds for the vector field
$V= \ri \sum_{k=1}^n (u_k \frac{\partial}{\partial u_k}-
\bar u_k \frac{\partial}{\partial \bar u_k})$
associated with the
infinitesimal action.
 For any fixed value $\chi_0>0$,
 usual  symplectic reduction yields the reduced phase space
\be
\chi^{-1}(\chi_0)/U(1) \equiv \bC P(n-1).
\label{4.9}\ee
For $\chi_0=1$,
the reduced symplectic form  is the standard Fubini-Study form $\omfs$ of $\bC P(n-1)$.
On the $\bC^{n-1}$ chart corresponding to those $u\in \chi^{-1}(1)$ for which
$u_n\neq 0$, the reduced symplectic form becomes
\be
 \omega_{\mathrm{FS}}(\bC^{n-1})= \ri \frac{\sum_{k=1}^{n-1} d\bar z_k \wedge d z_k}{ 1 + \vert z \vert^2}
- \ri \frac{\sum_{j,k=1}^{n-1} z_j \bar z_k d\bar z_j \wedge d z_k}{ (1 + \vert z \vert^2)^2}
= \ri {\bar \partial}{\partial} \log(\vert z\vert^2 + 1),
\label{4.20}\ee
where we use the `inhomogeneous coordinates' $z_j:=\frac{u_j}{u_n}$ and
$\vert z \vert^2 \equiv  \sum_{k=1}^{n-1} \vert z_k\vert^2$.
It is well-known that $\omega_{\mathrm{FS}}$ takes the form  (\ref{4.20})
in terms of  all the $n$ possible systems of inhomogeneous coordinates that together cover $\bC P(n-1)$.
For arbitrary $\chi_0>0$,  one has the following result.

\medskip
\no {\bf Lemma 4.}  \emph{The reduced symplectic manifold $\chi^{-1}(\chi_0)/U(1)$ obtained from
$(\bC^n,\Omega_{\bC^n})$ as described above is
the complex projective space $\bC P(n-1)$ equipped with the symplectic form
$\chi_0 \omega_{\mathrm{FS}}$.
}
\medskip

Now focus on the action
$R: \bT_{n-1} \times \bC^n \to \bC^n$
of the torus $\bT_{n-1}$ on $\bC^n$  furnished by
\be
R_\tau (u_1, ..., u_{n-1}, u_n) := (\tau_1 u_1, ..., \tau_{n-1}u_{n-1}, u_n),
\quad
\forall \tau\in \bT_{n-1},\,\, \forall u\in \bC^n.
\label{Ract}\ee
Defining
\be
J_k := \vert u_k \vert^2,
\quad
\forall k=1,..., n-1,
\ee
the corresponding moment map can be taken to be $J= (J_1, ..., J_{n-1}) : \bC^n \to \bR^{n-1}$.
Of course, the moment map of the torus action is unique only
up to a shift by an arbitrary constant, which we shall fix by convenience.

The  above  $\bT_{n-1}$-action  and  moment map survive the symplectic reduction
by the $U(1)$-action (\ref{4.6}) and descend to
the rotational $\bT_{n-1}$-action on $(\bC P(n-1), \chi_0 \omega_{\mathrm{FS}})$,
which thus becomes a Hamiltonian toric manifold.
This means that the rotational $\bT_{n-1}$-action, denoted as
$\cR: \bT_{n-1} \times \cp \to \cp$,
operates according to the rule
\be
\cR_\tau \circ \pi_{\chi_0} = \pi_{\chi_0} \circ R_\tau
\label{rotact}\ee
where $\pi_{\chi_0}: \chi^{-1}(\chi_0)\to \bC P(n-1)$ is the canonical projection.
We define its moment map $\cJ= (\cJ_1, ..., \cJ_{n-1}): \bC P(n-1) \to \bR^{n-1}$ by the formula
\be
\cJ_k \circ \pi_{\chi_0} = J_k + J_k^0,
\qquad
k=1,..., n-1,
\label{cJ}\ee
where the $J_k^0$ are constants.
It is obvious that
\be
0 \leq J_k
\qquad\hbox{and}\qquad
\sum_{k=1}^{n-1} J_k \leq \chi_0.
\ee
The point to note is that
\emph{if we choose  $\chi_0:=(\pi - \vert y \vert  n)$ and $J_k^0:= \vert y\vert$, then the
Delzant polytope $\cJ(\bC P(n-1))$ of the rotational
$\bT_{n-1}$-action coincides with the polytope $\cP$ (\ref{poly})}.
Therefore we    obtain the following main result of this section by combining
Delzant's theorem with the statements proved previously.

\medskip
\noindent
{\bf Theorem 3.}  \emph{Choose $y\in \bR $ for which $0 < \vert y \vert <  \frac{\pi}{n}$.
Consider the Hamiltonian toric manifold
$(\bC P(n-1), (\pi - \vert y \vert n) \omega_{\mathrm{FS}}, \cJ)$, where
 $\cJ$ defined by   $\cJ_k \circ \pi_{\chi_0} =
J_k + \vert y\vert$ is the moment map of the rotational
$\bT_{n-1}$-action, and consider also
the Hamiltonian toric manifolds $(P,\hat\om, \hat \alpha)$ and $(P, \hat \om, \hat \beta)$
that result from the quasi-Hamiltonian reduction according to Theorem 1 and Remark 3.
Then any two of these three Hamiltonian toric manifolds are $\bT_{n-1}$-equivariantly symplectomorphic.
More precisely, there exists a diffeomorphism $\phi_\alpha: P \to \bC P(n-1)$ such that
\be
\phi_\alpha^* ((\pi - \vert y \vert n) \omega_{\mathrm{FS}}) = \hat \omega,
\qquad \hat \alpha = \cJ \circ \phi_\alpha,
\ee
and also a diffeomorphism $\phi_\beta: P \to \bC P(n-1)$ such that
\be
\phi_\beta^* ((\pi - \vert y \vert n) \omega_{\mathrm{FS}}) = \hat \omega,
\qquad
\hat \beta = \cJ \circ \phi_\beta.
\label{eqvarb}\ee
The composed diffeomorphism $\phi:= \phi_\beta^{-1} \circ \phi_\alpha: P \to P$
converts $(P,\hat\om,\hat \alpha)$ into $(P,\hat \om, \hat \beta)$.}

\bigskip\no
{\bf Remark 5.} Theorem 3 says that  both completely integrable Hamiltonian systems
$(P, \hat \om, \hat \alpha)$ and $(P, \hat \om, \hat \beta)$ obtained from the quasi-Hamiltonian reduction
can be identified with the   system on
$(\bC P(n-1), (\pi - \vert y \vert n) \omega_{\mathrm{FS}})$ provided by the simple Hamiltonians
$\cJ_k$ that generate the rotational action of $\bT_{n-1}$.
As we shall see later,
the functions $\cJ_k\in C^\infty( \bC P(n-1))$ play the role of particle-positions
 in the
compactified  $\mathrm{III}_\mathrm{b}$ system \cite{RIMS95}.
It will turn  out  that if one converts $\hb$ into the particle-positions $\cJ$
of the $\mathrm{III}_\mathrm{b}$ system by the symplectomorphism $\phi_\beta$,
then  $\ha$ is converted by the same symplectomorphism  into
the action-variables of the system.
Roughly speaking,
the exchange
of the roles of $\ha$ and $\hb$ will then explain the Ruijsenaars self-duality
since the  other symplectomorphism $\phi_\alpha$ converts $\ha$ into the
particle-positions and $\hb$ into the action-variables.
From now on,  the symplectomorphisms  $\phi_\alpha$, $\phi_\beta$ that appear in Theorem 3
as well as their inverses and compositions will be referred to as
 \emph{Delzant symplectomorphisms}, or simply as Delzant maps.
 In the next section, we  explicitly construct the Delzant maps
 \be
 f_\alpha:= \phi_\alpha^{-1} \quad\hbox{and}\quad f_\beta:= \phi_\beta^{-1},
 \label{3.92}\ee
 which will be utilized in Section 5 where the statements of this remark
 will be elaborated.

\section{Construction of the  Delzant symplectomorphisms}
\setcounter{equation}{0}

The aim of this  section is to construct explicitly the Delzant
maps $f_\alpha, f_\beta: \cp \to P$  whose existence
has been established
by Theorem 3. We shall see that the  Ruijsenaars-Schneider Lax matrix
$L^{\mathrm{loc}}_y$  appears as a principal building
block of these symplectomorphisms. This remarkable fact will be further exploited
 in Section 5 where the emergence of the compactified
$\mathrm{III}_\mathrm{b}$ system as the fruit of the quasi-Hamiltonian reduction will be
 established and the Ruijsenaars self-duality
map of the $\mathrm{III}_\mathrm{b}$ system will be expressed in terms of the Delzant maps.

From the technical point of view,
below we first describe
a local version of the map $f_\beta$ defined in some  dense open subset of
$\cp$, and  then we construct its
global extension that will involve the global extension of the local
Lax matrix (appearing already in \cite{RIMS95}). Finally, we shall construct
$f_\alpha$ out of $f_\beta$ and certain involutions.

\subsection{Local version of the Delzant map $f_\beta$}

Let us denote by
$\bC P(n-1)_0$
the  dense open submanifold of $\cp=\chi^{-1}(\pi - n \vert y\vert )/U(1)$ where none of the $n$
homogeneous coordinates $u_k$  can vanish (cf.~Eq.~(\ref{4.9})).
In what follows we construct a symplectomorphism
\be
f_0: \cp_0 \to P_0^b,
\label{f0}\ee
where $P_0^b=\hb^{-1}(\cP^0)$ is the dense open submanifold of $P$ introduced in (\ref{3.79}).

On the subset of the constraint surface $\chi^{-1}(\pi - n \vert y\vert)$
covering $\cp_0$ we may impose the
gauge fixing condition $u_n >0$, and then $u_j$ ($j=1,...,n-1$)
parametrize $\bC P(n-1)_0$.
Adopting this condition, we
now introduce Darboux coordinates $\xi_j,\tau_j$ ($j=1,..., n-1$) on $\bC P(n-1)_0$
by setting
$u_j:= \tau_j \sqrt{\xi_j - \vert y \vert}$ for $j=1,..., n-1$,
where $\xi \in \cP^0$ (\ref{poly}), $\tau \in \bT_{n-1}$.
That is  we parametrize $\cp_0$  using the
diffeomorphism
\be
\cE: \cP^0  \times \bT_{n-1} \to \bC P(n-1)_0
\label{Edisp}\ee
 given by
\be
\cE(\xi,\tau) := \pi_{\chi_0}(\tau_1 \sqrt{\xi_1 - \vert y \vert},...,
\tau_{n-1} \sqrt{\xi_{n-1} - \vert y \vert}, \sqrt{\xi_n - \vert y\vert})
\quad\hbox{with}\quad
\xi_n\equiv \pi -\sum_{k=1}^{n-1} \xi_k
\label{C3}\ee
and $\pi_{\chi_0}$ defined in Subsection 3.4.
An easy calculation shows that  the
Fubini-Study symplectic structure on $\bC P(n-1)_0$
takes the Darboux form in the variables $\xi_j,\tau_j$.
Speaking more precisely, with the parametrization
$\tau_k := e^{\ri  \theta_k}$ $(k=1,..., n-1)$, there holds
the relation
\be (\pi - n \vert y\vert )\cE^*(\om_{\mathrm{FS}})= \ri \sum_{k=1}^{n-1} d \xi_k \wedge
d \tau_k \tau_k^{-1} = \sum_{k=1}^{n-1}  d  \theta_k \wedge d\xi_k.
\label{C4}\ee

As further pieces of preparation, recall
the isomorphism $\rho: \bT_{n-1} \to S\bT_n$  (\ref{rho}),
\be
\rho(\tau) \equiv \exp\Bigl( \ri \sum_{j=1}^{n-1}  \theta_j (E_{j,j}- E_{j+1, j+1})\Bigr)
\quad\hbox{for}\quad
\tau = (e^{\ri  \theta_1},..., e^{\ri  \theta_{n-1}}),
\label{C5}\ee
and consider the vector
\be
v_j(\xi,y):=\left[\frac{\sin y}{\sin ny}\right]^{\frac{1}{2}} W_j(\delta(\xi), y),
\qquad
\forall \xi\in \cP,\quad j=1,...,n,
\label{vW}\ee
with $W_j$ in (\ref{I.5}) where non-negative  square roots are taken.
Observe that if $\xi\in \cP^0$, then
all  $v_j(\xi,y)$ are strictly positive since their squares are the same as
the right-hand side of  (\ref{5.8}).
It is also important to notice that
these are $C^\infty$  functions on the open alcove
$\cP^0$, but their first
derivatives develop some singularities at the boundary of $\cP^0$.

It is readily checked that the following formulae yield a  unitary matrix
$g(v)\in U(n)$ for any vector $v\in \bR^n$ that has unit norm and component $v_n\neq -1$:
\bea
&&g(v)_{jn} := - g(v)_{nj}:= v_j,
\quad
\forall j=1,..., n-1,
\quad
g(v)_{nn} := v_n,\nonumber\\
&& g(v)_{jl} := \delta_{jl}- \frac{v_j v_l}{ 1 + v_n},
\quad
\forall j,l=1,..., n-1.
\label{pregy}\eea
Equations (\ref{I.5}), (\ref{5.8})--(\ref{norm}) imply that
the vector $v(\xi,y)$ in (\ref{vW}) has unit norm,
 and using this
we now introduce the unitary (actually real-orthogonal)
matrix $g_y(\xi)\in U(n)$ by setting
\be
g_y(\xi):= g(v(\xi,y)),
\qquad
\forall \xi\in \cP.
\label{gy}\ee

\medskip
\noindent
{\bf Theorem 4.} \emph{We can define a map $f_0: \cp_0 \to P_0^b$ by
the formula
\be
(f_0\circ \cE)(\xi,\tau) :=
p \left(g_y(\xi)^{-1} L^{\mathrm{loc}}_y(\delta(\xi),
\rho(\tau)^{-1})g_y(\xi),
g_y(\xi)^{-1} \delta(\xi)g_y(\xi) \right),
\label{C15}\ee
where $L^{\mathrm{loc}}_y$ is the Lax matrix given by (\ref{I.4}) and
$p: \mu^{-1}(\mu_0)\to P$ is the canonical projection.
This map is a symplectic diffeomorphism with respect to the restricted
symplectic forms,
\be
f_0^*(\hat\om ) =(\pi - n \vert y\vert ) \omfs,
\label{48}\ee
and it intertwines the restrictions of the corresponding toric moment
maps,
\be
f_0^*(\hat \beta ) =\cJ.
\label{49}\ee}

\medskip
The hardest part of the proof will be the verification of  Eq.~(\ref{48}), and
before dealing with this we present two lemmas.

\medskip
\no {\bf Lemma 5.} \emph{The $SU(n)$ matrix $L_{y}^{\mathrm{loc}}(\delta(\xi),\Theta)$
given by Eq.~(\ref{I.4}) verifies the relation
\be
L_{y}^{\mathrm{loc}}(\delta(\xi),\Theta) \dt(\xi) L_{y}^{\mathrm{loc}}
(\delta(\xi),\Theta)^{-1} =
\left[e^{2\ri y}\1_n+ (e^{2\ri(1-n) y}-e^{2\ri y}) v(\xi,y)v(\xi,y)^
\dagger\right]\dt(\xi)
\label{laxa}\ee
for all $\xi \in \cP^0$ and $\Theta\in S\bT_n$, with the vector
(\ref{vW}).}

\begin{proof}
We know from the proof of Theorem 2 (cf.~the discussion around
Eqs.~(\ref{mcon}) and (\ref{mg})) that
the unitary matrices $\dt(\xi)$ and
  $\left[e^{2\ri y}\1_n+ (e^{2\ri(1-n) y}-e^{2\ri y}) v(\xi,y)v(\xi,y)^
\dagger\right]\dt(\xi)$
  have the same spectra, and hence there exists a unitary matrix
$N(\xi,y)$ such that
\be
N(\xi,y)\dt(\xi)N(\xi,y)^{-1} =\left[e^{2\ri y}\1_n+
(e^{2\ri(1-n) y}-e^{2\ri y}) v(\xi,y)v(\xi,y)^\dagger\right]\dt(\xi).
\label{lax1}\ee
By conjugating the last relation by $N(\xi,y)^{-1}$  we obtain
\be
\dt(\xi)  =\left[e^{2\ri y}\1_n+ (e^{2\ri(1-n) y}-e^{2\ri y})
N(\xi,y)^{-1}v(\xi,y)
\left(N(\xi,y)^{-1}v(\xi,y)\right)^\dagger
\right]N(\xi,y)^{-1}\dt(\xi)N(\xi,y)
\label{lax2}\ee
and by inverting the term in square brackets we arrive at
\be
N(\xi,y)^{-1}\dt(\xi)N(\xi,y)=\left[e^{-2\ri y}\1_n+ (e^{2\ri(n-1) y}-
e^{-2\ri y})
N(\xi,y)^{-1}v(\xi,y)\big(N(\xi,y)^{-1}v(\xi,y)\big)^\dagger\right]
\dt(\xi).
\label{lax3}\ee
Let us rewrite Eq.~(\ref{lax1}) for $-y$ as
\be
N(\xi,-y)\dt(\xi)N(\xi,-y)^{-1} =\left[e^{-2\ri y}\1_n+ (e^{2\ri(n-1)
y}-e^{-2\ri y})
  v(\xi,-y)v(\xi,-y)^\dagger\right]\dt(\xi).
  \label{lax4}\ee
We have learned in proving Theorem 2 (cf.~Eq.~(\ref{dete})) that the
equality of the spectra of the
matrices $\dt(\xi)$ and $\left[e^{2\ri y}\1_n+ (e^{2\ri(1-n) y}-
e^{2\ri y}) ww^\dagger\right]\dt(\xi)$
fixes the absolute values of the components of the vector $w$ to be
given by the right-hand side of (\ref{vW}).
Comparing (\ref{lax4}) with (\ref{lax3}), we therefore see that the
absolute values of the components
of the vector  $N(\xi,y)^{-1}v(\xi,y)$ are the same as the (strictly
positive) components of the
vector $v(\xi,-y)$. Note that the matrix $N(\xi,y)$ verifying
Eq.~(\ref{lax1}) is not unique because it can be
multiplied
from the right by any diagonal element of $U(n)$ while keeping
(\ref{lax1}) valid. However, this ambiguity
can be completely fixed by requiring that the vector
$N(\xi,y)^{-1}v(\xi,y)$ has all components
real and strictly positive.
We denote the unique matrix $N(\xi,y)$ satisfying this requirement by $
\tN(\xi,y)$.
Thus we have
\be
  v(\xi,y)=\tN(\xi,y)v(\xi,-y)\quad\hbox{and hence}\quad
\tN(\xi,-y)=\tN(\xi,y)^{-1}.
\label{v12}\ee

By considering it for $N(\xi,y):= \tN(\xi,y)$, let us rewrite
(\ref{lax1}) as
\be
\tN(\xi,y)\dt(\xi)-e^{2\ri y}\dt(\xi)\tN(\xi,y) =
(e^{2\ri(1-n) y}-e^{2\ri y}) v(\xi,y)\big(\tN(\xi,y)^{-1}v(\xi,y)\big)^
\dagger
\tN(\xi,y)^{-1} \dt(\xi)\tN(\xi,y).
\label{lax5}\ee
With the help of (\ref{v12}), we can further rewrite the last relation
as
\be
\tN(\xi,y)\dt(\xi)-e^{2\ri y}\dt(\xi)\tN(\xi,y) =
(e^{2\ri(1-n) y}-e^{2\ri y}) v(\xi,y)  v(\xi,-y)^\dagger \tN(\xi,-y)
\dt(\xi)\tN(\xi,-y)^{-1}.
\label{lax6}\ee
Expressing $\tN(\xi,-y) \dt(\xi)\tN(\xi,-y)^{-1}$ from
Eq.~(\ref{lax4}) and using subsequently
Eq.~(\ref{norm}), we derive
\be
\tN(\xi,y)\dt(\xi)-e^{2\ri y}\dt(\xi)\tN(\xi,y) = (1-e^{2n\ri
y})v(\xi,y)v(\xi,-y)^\dagger\dt(\xi).
\label{lax7}\ee
By solving this for the components of $\tN(\xi,y)$  we obtain the
equality
\be
\tN(\xi,y)= e^{\ri (n-1) y} L_{y}^{\mathrm{loc}}(\delta(\xi),\1_n).
\ee
This implies the desired relation (\ref{laxa}).
The above argument also shows that $L_{y}^{\mathrm{loc}}(\delta(\xi),
\Theta)$ is unitary,
and the fact that its determinant equals $1$ is easily checked by the
determinant formula of Cauchy matrices.
\end{proof}

\medskip
\no {\bf Lemma 6.} \emph{Every element $(A,B)\in \mu^{-1}(\mu_0)$ such
that
  $p(A,B)$ belongs to $P_0^b$ (\ref{3.79}) has the form
\be
(A,B)=\Psi_{(g_y(\xi)\eta)^{-1}}
\left( L^{\mathrm{loc}}_y(\delta(\xi),\rho(\tau)^{-1}),
  \delta(\xi)\right)
\label{AB}\ee
with $\xi=\beta(A,B)$, $\tau\in \bT_{n-1}$ and $\eta\in U(n)$ for which
$\eta^{-1}\mu_0 \eta=\mu_0$,
using the notation (\ref{2conj}).
By this formula,
the pair $(\xi,\tau)\in \cP^0\times \bT_{n-1}$  uniquely parametrizes
the projection $p(A,B)\in P_0^b$.}

\begin{proof}
Conjugating the relation (\ref{laxa})  by $g_y(\xi)^{-1}$, and putting
$\Theta:= \rho(\tau)^{-1}$ (\ref{C5}),
we conclude by
using Eq.~(\ref{cosi}) that the pair (\ref{AB})
belongs to the constraint surface $\mu^{-1}(\mu_0)$.
It follows by tracing the definitions that, after the projection $p$, $
\tau$ parametrizes
a $\bT_{n-1}$ orbit in $P_0^b$ under the $\hat \Psi^b$-action.
As was noted in Remark 4, this action is transitive and free on $P_0^b$.
Hence the above solution of the constraint (\ref{con1})
projects to the most general element of the reduced phase space $P$
for which the
value of the function $\hb$ equals $\xi$,
and specifying the pair $(\xi,\tau)$ is equivalent to specifying the
projection $p(A,B)\in P_0^b$.
\end{proof}

\medskip
\no {\it Proof of Theorem 4:}
It follows directly from Lemma 6 that the formula
\be
(F_0\circ \cE)(\xi,\tau) :=
  \left(g_y(\xi)^{-1} L^{\mathrm{loc}}_y(\delta(\xi),
\rho(\tau)^{-1})g_y(\xi),
g_y(\xi)^{-1} \delta(\xi)g_y(\xi) \right),
\label{F0}\ee
defines a smooth map
\be
F_0: \cp_0 \to p^{-1}(P_0^b),
\ee
which is injective and its image intersects every gauge orbit in
$p^{-1}(P_0^b)$ precisely
in one point.
Thus $f_0 = p\circ F_0: \cp_0 \to P_0^b$  is an injective and surjective smooth map.
On account of (\ref{48}) (proved in what follows), the corresponding
Jacobian determinant
cannot vanish  and hence $f_0$  is a diffeomorphism.

Since the validity of (\ref{49}) is obvious, it remains to show
   that $f_0$  satisfies (\ref{48}).
   Because of Eqs.~(\ref{genred}) and (\ref{C4}), this amounts to
proving that
  the  restriction of the quasi-Hamiltonian $2$-form $\omega$ on $p^{-1}
(P_0^b)$ pulled back by the map $F_0\circ \cE$ on $\cP^0  \times
\bT_{n-1}$ is the Darboux
  $2$-form:
  \be (F_0\circ \cE)^*\om\vert_{p^{-1}(P_0^b)}=\ri \sum_{k=1}^{n-1} d
\xi_k \wedge
d \tau_k \tau_k^{-1}.\ee
  In fact, we here verify this by  direct computation, by inserting the
formula (\ref{F0}) into the formula (\ref{2.2}):
  \be
(F_0\circ \cE)^*\om\vert_{p^{-1}(P_0^b)} =\jp \langle A_{\xi,
\tau}^{-1} dA_{\xi,\tau} \stackrel{\wedge}{,} dB_\xi B_\xi^{-1}\rangle
+\jp\langle  dA_{\xi,\tau} A_{\xi,\tau}^{-1} \stackrel{\wedge}{,} B_
\xi^{-1} dB_\xi \rangle
\label{2.2bis}\ee
where $A_{\xi,\tau}:=g_y(\xi)^{-1} L^{\mathrm{loc}}_y(\delta(\xi),
\rho(\tau)^{-1})g_y(\xi)$ and $B_\xi:=g_y(\xi)^{-1}
\delta(\xi)g_y(\xi) $.
Note that we have omitted in (\ref{2.2bis}) the third term of the form
$\om$ displayed in  (\ref{2.2}). This term does not
contribute since, due to  the moment map constraint  (\ref{con1}), we
have $\mu_0B_\xi A_{\xi,\tau}=A_{\xi,\tau}B_\xi$ and
hence
\be
\langle (A_{\xi,\tau} B_\xi )^{-1} d (A_{\xi,\tau}B_\xi )
\stackrel{\wedge}{,} (B_\xi A_{\xi,\tau})^{-1} d (B_\xi A_{\xi,\tau})
\rangle=
  \langle (B_\xi A_{\xi,\tau})^{-1} d (B_\xi A_{\xi,\tau})
\stackrel{\wedge}{,} (B_\xi A_{\xi,\tau})^{-1} d (B_\xi A_{\xi,\tau}),
\ee
which vanishes as the scalar product is symmetric and the wedge product
 is anti-symmetric.

  In the following calculation, we set for simplicity $g\equiv g_y(\xi)
$, $L\equiv L^{\mathrm{loc}}_y(\delta(\xi),\rho(\tau)^{-1})$, $\rho
\equiv \rho(\tau)$ and $\delta\equiv\delta(\xi)$. Thus we obtain from
(\ref{2.2bis})
   \be (F_0\circ \cE)^*\om\vert_{p^{-1}(P_0^b)}=\jp\langle L^{-1}dL
+dgg^{-1}-L^{-1}dgg^{-1}L\stw d\delta\delta^{-1}+\delta
dgg^{-1}\delta^{-1}-dgg^{-1}\rangle- (L\leftrightarrow \delta),
\label{11:10}\ee
   where $(L\leftrightarrow \delta)$ means the first term on the
r.h.s. of (\ref{11:10})  with the role of $L$ and $\delta$ interchanged.
   By using the invariance of the scalar product  $\langle. ,.\rangle$
and the fact that $\langle \phi\stw \psi\rangle =-\langle \psi\stw \phi
\rangle$ for any $su(n)$-valued differential forms $\phi$ and $\psi$,
we can rewrite Eq.~(\ref{11:10}) as
    \be (F_0\circ \cE)^*\om\vert_{p^{-1}(P_0^b)}=\jp\langle \delta^{-
\jp}(L^{-1}dL-L^{-1}dgg^{-1}L)\delta^{\jp}+ \delta^{\jp}
(dLL^{-1}+Ldgg^{-1}L^{-1})\delta^{-\jp}\stw \kappa+\kappa^t\rangle.
\label{17:14}\ee
Here we have introduced the $su(n)$-valued differential form $\kappa$ by
\be \kappa:=\jp d\delta\delta^{-1} +\delta^{\jp}dgg^{-1}\delta^{-\jp},\ee
$\kappa^t$ denotes the transposed matrix, and by
 using (\ref{fund}) we have
$\delta^{\jp} \equiv \exp\left(- \ri \sum_{k=1}^{n-1} \xi_k \lambda_k\right)$.

We can write the matrix $L$ as
\be L\equiv L_1(\xi)
\rho(\tau)^{-1}.\label{13:04}\ee
Thus in Eq.~(\ref{17:14})  the dependence of the
form   $(F_0\circ \cE)^*\om\vert_{p^{-1}(P_0^b)}$ on the variable
$\tau$ is hidden in the (diagonal) matrix $\rho(\tau)$.
It will be convenient to employ also the decomposition
\be
(F_0\circ \cE)^*\om\vert_{p^{-1}(P_0^b)}\equiv V+\hat V,
\label{VhatV}\ee
where $V$ depends on $\rho$ differentially, i.e., $V$  collects the terms that contain $d\rho$.
The part $V$ is easily singled out  from
(\ref{17:14}) as
\be    V=-\jp\langle d\rho\rho^{-1} \stw d\delta\delta^{-1}+\delta
dgg^{-1}\delta^{-1}-dgg^{-1}+
   L_1^{-1}(d\delta\delta^{-1}+  dgg^{-1} -
\delta^{-1}dgg^{-1}\delta) L_1\rangle.
\ee
For $g$ is real orthogonal,  $dgg^{-1}$  is anti-symmetric.
Because
the  trace of the product of a symmetric matrix with an  anti-symmetric
one vanishes, we obtain
\be \langle d\rho\rho^{-1}\stw dgg^{-1}\rangle =0,\ee
and then $V$ can be rewritten as
\be    V =-\jp\langle d\rho\rho^{-1} \stw d\delta\delta^{-1}
+L_1^{-1}dgg^{-1}L_1-L_1^{-1} dL_1+
( \delta L_1)^{-1} d(\delta L_1)-(\delta L_1)^{-1}dgg^{-1}(\delta L_1)
\rangle.
\label{13:39}\ee
Now note that the constraint (\ref{con1}) implies
\be \delta L_1=\zeta^{-1} L_1\delta, \quad \zeta:=g\mu_0g^{-1}.
\label{con2}\ee
Inserting this in the last two terms of (\ref{13:39}) gives directly the Darboux form:
\be    V=-\langle d\rho\rho^{-1} \stw d\delta\delta^{-1}\rangle -\jp
\langle  d\rho\rho^{-1} -\delta d\rho\rho^{-1} \delta^{-1}\stw
L_1^{-1}d \zeta \zeta^{-1}L_1-L_1^{-1}dL_1\rangle=-\langle d\rho
\rho^{-1}\stw d\delta\delta^{-1}\rangle.
\label{Vequal}\ee
The restricted form $\om\vert_{p^{-1}(P_0^b)}$ is  closed
because $p^{-1}(P_0^b)$ is a subset of a level set of the moment
map, and thus its pull-back (\ref{VhatV}) is closed as well.
Moreover,  the Darboux differential form $V$ is also closed.
Then we
observe that the  part $\hat V$ in (\ref{VhatV})  cannot depend on $\rho$ because
otherwise it would not be closed.
By taking this into account,
(\ref{17:14}) gives
\be \hat V=\jp\langle \delta^{-\jp}(L_1^{-1}dL_1-
L_1^{-1}dgg^{-1}L_1)\delta^{\jp}+ \delta^{\jp}
(dL_1L_1^{-1}+L_1dgg^{-1}L_1^{-1})\delta^{-\jp}\stw \kappa+\kappa^t
\rangle.\label{9:34}\ee
Now we again use the constraint (\ref{con2}) to derive
  \be  L_1^{-1}dL_1-L_1^{-1}dgg^{-1}L_1=\delta \left(L_1^{-1}dL_1
+L_1^{-1}(\delta^{-1}d\delta-\delta^{-1}dgg^{-1}\delta)L_1  \right)\delta^{-1}-
d\delta\delta^{-1}.\label{18:27}\ee
Inserting the expression (\ref{18:27}) into (\ref{9:34}) yields
immediately
\be \hat V=\jp\langle L_1^{-1}dL_1 +L_1^{-1}(\delta^{-1}d\delta-
\delta^{-1}dgg^{-1}\delta) L_1   +dL_1L_1^{-1}+L_1dgg^{-1}L_1^{-1}\stw
\delta^{-\jp}(\kappa+\kappa^t)\delta^{\jp}\rangle.\label{18:42}\ee
Coming back to the formula (\ref{I.4}), we notice that the unitary matrix
$L_1$ can be cast as
\be L_1=\delta^{-\jp}\cM\delta^{\jp}\ee
where $\cM$ is a real orthogonal matrix. With this representation of
$L_1$, the expression $\hat V$ can be rewritten as
\be \hat V=\jp\langle d\cM\cM^{-1}+\cM^{-1}d\cM +\cM\kappa \cM^{-1}
+\cM^{-1}\kappa^t\cM\stw\kappa +\kappa^t\rangle.\ee
By using  again that the  trace of the product of a symmetric matrix with an
anti-symmetric one vanishes,
and using also  the invariance of the scalar product and that
$\langle \phi\stw \psi\rangle =\langle \phi^t\stw \psi^t\rangle$ for any $su(n)$-valued forms,
the last equation implies that $\hat V=0$.
Having calculated $V$ and $\hat V$ in (\ref{VhatV}),
we finally obtain the desired equality:
  \be (F_0\circ \cE)^*\om\vert_{p^{-1}(P_0^b)}=-\langle d\rho
\rho^{-1}\stw d\delta\delta^{-1}\rangle=\ri \sum_{k=1}^{n-1} d \xi_k
\wedge
d \tau_k \tau_k^{-1}.\ee
\rightline{$\square$}

\bigskip\no
{\bf Remark 6.}
We know from the theory of the quasi-Hamiltonian reduction that the
reduced spectral Hamiltonians $\hat\alpha_i$  Poisson commute,
and Theorem 4 permits to identify the $\hat \alpha_i$ on $P_0^b$ with the spectral functions
of the Ruijsenaars-Schneider Lax matrix.
The proof of Theorem 4 shows  that this commutativity
property of the $\hat \alpha_i$
can be viewed as
a consequence of the Darboux form of the reduced symplectic structure.
The fact that the spectral invariants of the Ruijsenaars-Schneider Lax
matrix Poisson commute with respect to the Darboux structure
was also proved  previously by means of
different methods (see \cite{RS,SR-CRM,rmat} and references therein).

It follows easily from Theorem 4 that  the local Delzant map $f_0$
converts $\hat \beta$ into
  particle-position variables and converts $\hat \alpha$ into
  action-variables of the {\it local}
$\mathrm{III}_\mathrm{b}$ system.
Consequently,  the {\it full} reduced phase space $P$
must carry  a
 {\it completion} of
the
local $\mathrm{III}_\mathrm{b}$ system.
Eventually this completion will be identified with the one
introduced by Ruijsenaars, but before explaining this
further effort is needed in order to work out
certain details of our picture that will enable us to give  precise
comparison with the results of
\cite{RIMS95}  regarding also the self-duality
of the completed $\mathrm{III}_\mathrm{b}$ system.
In particular, we need to prove that $f_0$ extends to a global Delzant map.

\subsection{Global extension of the Lax matrix}

The local Lax matrix $L^{\mathrm{loc}}_y(\delta(\xi),\rho(\tau)^{-1})$,
viewed as a function on $\cp_0$,
is the crucial ingredient of the local Delzant map of Theorem 4.
The global Delzant map that we shall construct later
will involve an extension
of (a conjugate of) this Lax matrix to a smooth function on $\cp$.
We here present this extension, which appears also in \cite{RIMS95}.
In order to save space, from now on we assume that
\be
y>0.
\ee

We continue to realize  $\cp$ as the factor space
$\cp= S_{\chi_0}^{2n-1}/U(1)$
with
\be
S_{\chi_0}^{2n-1} = \{ (u_1,..., u_n)\in \bC^n \,\vert\,
\sum_{k=1}^n \vert u_k \vert^2 =\chi_0\},
\qquad
\chi_0 = \pi - n \vert y \vert.
\label{S0}\ee
In the subsequent arguments we identify the $U(1)$-invariant functions defined
on $S_{\chi_0}^{2n-1}$  by
 \be
 r_i:= \vert u_i\vert
 \quad\hbox{and}\quad
 \xi_i := \vert u_i\vert^2 +\vert y\vert,
 \quad i=1,...,n,
 \label{functions}\ee
 as functions on $\cp$.
Regarded in this way, $\xi_i$
belongs to $C^\infty(\cp)$,
while $r_i$ is not even differentiable at its zero locus.
(For $i=1,...,n-1$, the function $\xi_i$ is  just another name for
the moment map component $\cJ_i$.)

Now we  give a simple technical lemma, whose
proof contains the essential observation that will lead to the global Lax matrix.
Its statement will be utilized also in Subsection 4.3.

\medskip
\noindent
\textbf{Lemma 7.} \emph{By combining  equations (\ref{I.5}), (\ref{par1}) and
(\ref{functions}),  with $0<y< \frac{\pi}{n}$,
consider the expressions  $W_k(\delta(\xi), \pm y)$ as functions on $\cp$.
Then    $W_k(\delta(\xi),y)$ can be written as
\be
W_k(\delta(\xi),y) = r_k w_k^y(\xi),
\label{D1.1}\ee
where  $w_k^y(\xi)$ represents a positive
 $C^\infty$ function on $\cp$ for each $k=1,...,n$.
Similarly,
\be
W_k(\delta(\xi),-y) = r_{k-1} w_k^{-y}(\xi),
\qquad
r_0:= r_n,
\label{D1.2}\ee
where the function $w_k^{-y}$ has the same properties as those mentioned for $w_k^{y}$.}

\begin{proof}
First restricting to $\cp_0$ where $r_k\neq 0$,
we directly  spell out  $W_k(\delta(\xi), y)$ in the form (\ref{D1.1}) with
\be
w_k^y(\xi)= \left[\frac{\sin(r_k^2)}{r_k^2 \sin(\xi_k)} \right]^{\frac{1}{2}} R_k^y(\xi),
\label{wR}\ee
where, introducing   the shorthand
$\xi_{i,l} := \sum_{m=i}^{l} \xi_m$ for all $1\leq i \leq l \leq n$, we have
\be
R_k^y(\xi)=
\left(\prod_{1\leq j\leq k-1} \left[
\frac{\sin( \xi_{j,k-1} + y)}{ \sin(\xi_{j,k-1})}
\right]^{\frac{1}{2}}\right)
\left(\prod_{k+2\leq  j\leq n} \left[
\frac{\sin( \xi_{k,j-1} - y)}{ \sin(\xi_{k,j-1})}
\right]^{\frac{1}{2}}\right).
\label{Rj}\ee
By using (\ref{functions}),
it is easily  checked that
 all arguments of the sinus-functions involved in (\ref{Rj}) lie strictly
in the interval $(0,\pi)$  even when running over the full  $\cp$,
which immediately implies that $R_k^y (\xi)$
 represents a  positive $C^\infty$-function
on $\cp$.
Since the function $\sin x/x$ remains smooth and positive at $x=0$,
and $y\leq \xi_k  \leq \pi- (n-1) y$,
we then
see from (\ref{wR}) that $w_k^y(\xi)$ also represents a
positive $C^\infty$ function on $\cp$.
The claim about $W_k(\delta(\xi), - y)$  can be verified in an analogous manner.
\end{proof}

Next, using $L_{y}^{\mathrm{loc}}$ given in (\ref{I.4}),
we introduce the functions $\Lambda^y_{k,l}(\xi)$ ($1\leq k,l\leq n$) by the equations
\be
 r_k r_{l-1} \Lambda^y_{k,l}(\xi) \equiv
L_{y}^{\mathrm{loc}}(\delta(\xi), \1_n)_{k,l}
\quad\hbox{for}\quad
l\neq k+1,\quad (k,l)\neq (n,1),
\label{431}\ee
\be
\Lambda_{k,k+1}^y(\xi) \equiv L^{\mathrm{loc}}_{y}(\delta(\xi),\1_n)_{k,k+1},
\qquad
\Lambda_{n,1}^y(\xi)\equiv L^{\mathrm{loc}}_{y}(\delta(\xi), \1_n)_{n,1}.
\label{432}\ee
These equations directly define $\Lambda^y_{k,l}$ as functions on $\cp_0$, where all $r_i$ are
non-zero. Then, by using Lemma 7 and a similar analysis for the denominators
in the formula (\ref{I.4}), we find that
$\Lambda^y_{k,l}(\xi)$ extends to a $C^\infty$ function on $\cp$ for each $1\leq k,l\leq n$.
The extended function, which we denote by the same letter, vanishes nowhere on $\cp$.
The last statement follows by easy inspection, and will be utilized later.

\medskip
\noindent
\textbf{Lemma 8.} \emph{By using the above functions
$\Lambda^y_{k,l}$  and (\ref{functions}), and setting $u_0:= u_n$,
we can define  $C^\infty$
functions $\cK^y_{k,l}$ on $S^{2n-1}_{\chi_0}$ by the formulae
\be
\cK^y_{k,l}(u) := \bar u_k u_{l-1}\Lambda^y_{k,l}(\xi)
\quad\hbox{for}\quad
l\neq k+1,\quad (k,l)\neq (n,1),
\label{443}\ee
\be
\cK^y_{k,k+1}(u) := \Lambda_{k,k+1}^y(\xi),
\qquad
\cK^y(u)_{n,1} := \Lambda_{n,1}^y(\xi).
\label{442}\ee
The functions $\cK^y_{k,l}$ are $U(1)$-invariant and thus yield
$C^\infty$ functions on $\cp$ that together form  an $SU(n)$-valued
$C^\infty$ function on $\cp$, also denoted as $\cK^y$.
The restriction of this  matrix function to $\cp_0$ satisfies
the following identity:
\be
(\cK^y\circ \cE)(\xi,\tau) = \Delta(\tau)^{-1}
L^{\mathrm{loc}}_y(\delta(\xi), \rho(\tau)^{-1})\Delta(\tau),
\label{428}\ee
where $\cE$ denotes the parametrization introduced in (\ref{C3}) and
\be
\Delta(\tau):= \operatorname{diag}(\tau_1, ..., \tau_{n-1},1).
\label{Delta}\ee}
\begin{proof}
It follows from what we established before that
the formulas (\ref{443}) and (\ref{442}) yield
$C^\infty$ functions on $S^{2n-1}_{\chi_0}$, which are obviously invariant
under the $U(1)$-action $u\mapsto e^{\ri \gamma} u$.
By using (\ref{C5}) and (\ref{Delta}),
one readily checks that
the right-hand-side of (\ref{428}) is equal to the matrix
\be
\operatorname{diag}(\tau_1^{-1},..., \tau_{n-1}^{-1}, 1) L^{\mathrm{loc}}_y(\delta(\xi), \1_n)
 \operatorname{diag}(1, \tau_1,..., \tau_{n-1}).
\ee
On account of (\ref{431}) and (\ref{432}),
this is further equal to the matrix $\cK^y(u)$ at $u$ given by
\be
u_i= \tau_i \sqrt{\xi_i - \vert y\vert }
\quad
(i=1,...,n-1),
\quad
u_n = \sqrt{\xi_n -\vert y\vert }.
\ee
By the definition of the map $\cE$ (\ref{C3}), this proves the equality
(\ref{428}).   Finally, note
that  $\cK^y$ is $SU(n)$ valued
since  its restriction to $\cp_0$ is $SU(n)$ valued.
\end{proof}

\medskip
\noindent
\textbf{Remark 7.} The $C^\infty$ function $\cK^y:\cp \to SU(n)$ specified
by Lemma 8 will be referred to as the \emph{global Lax matrix}.
Since by (\ref{428}) it reduces to a conjugate of the local Lax matrix
$L^{\mathrm{loc}}_y$
on $\cp_0$,   $\cK^y$ can serve as the Lax matrix
of an integrable system defined on $\cp$.
Apart from slight differences of conventions,
our $\cK^y$ actually coincides with the  Lax matrix
of the compactified $\mathrm{III}_\mathrm{b}$
system
constructed  in \cite{RIMS95} (pages 311-312 loc.~cit.) relying on
arguments similar to the above.
It might be worth noting  $\cK^y$ is not only $C^\infty$ but real-analytic on $\cp$,
as  follows by inspection of the above proof and established also in \cite{RIMS95}.

\subsection{Construction of the global Delzant map $f_\beta$}

Let us remember that $\mu^{-1}(\mu_0)$ is the total space of a principal bundle
with projection
\be
 \phi_\beta \circ p: \mu^{-1}(\mu_0) \to P\to \cp.
\label{4.48}\ee
We established the characteristic properties (\ref{eqvarb}) of $\phi_\beta$, but
not yet its explicit form.
Now we wish to give a construction of the inverse map $f_\beta=\phi_\beta^{-1}$.
Our plan to achieve this is as follows.
We first cover $\cp$ by $n$ `coordinate charts' $\cp_j$, where
$\cp_j \subset \cp$ ($j=1,...,n$)
is by definition the set of those $U(1)$ orbits
in $S_{\chi_0}^{2n-1}$ (\ref{S0}) for which $u_j\neq 0$.
\emph{By an explicit formula}, we then introduce a map
\be
F_j: \cp_j \to \mu^{-1}(\mu_0),
\label{4.49}\ee
which turns out to define a local section of the principal bundle over $\cp_j$.
These maps have the property that the projected maps
\be
 p \circ F_j: \cp_j \to P
\ee
coincide on the overlaps of their domains, and engender the desired global
symplectomorphism
\be
f_\beta: \cp \to P.
\ee
We shall also see that $f_\beta$ extends the map
$f_0:\cp_0 \to P$   described in Theorem 4.
This will be shown by using that $\cp_0 = \cap_{j=1}^n \cp_j$ and
$f_0$  has the form $f_0 = p\circ F_0$ with the local section
$F_0: \cp_0 \to \mu^{-1}(\mu_0)$ given  in equation (\ref{F0}).

To begin,
we introduce coordinates on $\cp_j$ for each $j$
by considering the $n$-tuples
\be
u^j := (u^j_1,..., u^j_n)
\ee
subject to the conditions
\be
\sum_{k\neq j}^n  \vert u_k^j \vert^2 < \chi_0,
\quad
u_j^j:= \sqrt{ \chi_0 - \sum_{k\neq j}^n  \vert u_k^j \vert^2},
\quad
\chi_0 =\pi - n \vert y \vert.
\ee
Since $u^j_j$ is a function of the other components of $u^j$, we may think of $u^j$ as
a variable running over the open ball
$\mathfrak{B}_{\chi_0} \subset \bC^{n-1}$ defined by
\be
\mathfrak{B}_{\chi_0}:= \Bigl\{ (z_1,...,z_{n-1})\in \bC^{n-1}\,\vert\,
\sum_{k=1}^{n-1} \vert z_k \vert ^2  < \chi_0 \Bigr\}.
\ee
Accordingly, we let
\be
(\cp_j, \mathfrak{B}_{\chi_0}^j)
\label{chart}\ee
denote the dense open subset $\cp_j$ of $\cp$ equipped with the coordinates $u^j_k$.
For notational convenience,  we also keep the component $u_j^j$, although
it is a function of the true coordinates $u_k^j$ ($k\neq j$) on $\cp_j$.
In this notation the formula for the change of coordinates between the charts in
especially simple.
For example, the $n$ alternative coordinates $u^j$ ($j=1,...,n)$
of the same point $\cE(\xi,\tau)\in \cp_0$ (\ref{C3}) can be written briefly as
\be
u^j_k = r_k \bar \tau_j \tau_k
\quad\hbox{with}\quad r_k= \sqrt{\xi_k -  \vert y\vert },\quad k=1,...,n,  \quad \tau_n:=1.
\label{altern}\ee
Since it is a reduction of $\Omega_{\bC^n}$ (\ref{4.1}),  the
symplectic form  $\chi_0 \omfs$ is represented by the Darboux form
 $\ri \sum_{k\neq j} d \bar u_k^j \wedge du_k^j$
on the chart (\ref{chart})

Consider now the component  $v_j$ of the vector (\ref{vW}).
Notice from Lemma 7  that (since $y>0$)  $v_j(\xi,y)$ yields
a $C^\infty$ function on $\cp_j$.
Then, for each $j=1,...,n$, define the $U(n)$ valued function
$g_y^j(\xi)$ as follows. First of all, set $g_y^n:= g_y$ in (\ref{gy}).
For $1\leq j <n$, let $T^j$ denote the $n$ by $n$ transposition matrix given explicitly by
\be
T^j:= \1_n - E_{j,j} - E_{n,n} + E_{j,n} + E_{n,j}.
\ee
Then using the formula (\ref{pregy}) and the vector $v(\xi,y)$ in (\ref{vW}) define
\be
g_y^j(\xi):= T^j g(T^j v(\xi,y)).
\ee
It is clear that $g_y^j(\xi)$ is actually a real-orthogonal
matrix for all $j$.
Moreover, we point out that
\be
g_y^j(\xi) = g_y^n(\xi) \eta_y^j(\xi)
\quad\hbox{with \quad $\eta^j_y(\xi)\in U(n)$ \quad for which
\quad $\eta_y^j(\xi) \mu_0 \eta_y^j(\xi)^{-1} = \mu_0$},
\label{etay}\ee
which holds simply because the matrices $g_y^j(\xi)$  have the same last column for all $j$.

As examples that illustrate well the general case, for $n=3$ we display the matrices
\be
g_y^1 = \begin{bmatrix}
-v_3 & -  v_2 & v_1 \\
- \frac{v_2 v_3}{d_1} & (1 - \frac{v_2^2}{d_1}) & v_2 \\
(1 - \frac{v_3^2}{d_1})  & - \frac{v_3 v_2}{d_1} &  v_3
\end{bmatrix}
\quad\hbox{and}\quad
g_y^2 = \begin{bmatrix}
(1 - \frac{v_1^2}{d_2})  & -\frac{v_1 v_3}{d_2} & v_1 \\
-v_1 & -  v_3 & v_2 \\
- \frac{v_3 v_1}{d_2} & (1 - \frac{v_3^2}{d_2})   &  v_3
\end{bmatrix},
\ee
where $d_1:= 1 + v_1$, $d_2:= 1 + v_2$ and $v:= v(\xi,y)$ in (\ref{vW}).
The point is that,
in general, $g_y^j$ contains the denominator $d_j = 1 + v_j$, which yields a $C^\infty$ function on $\cp_j$.

The rationale behind the definition of $g_y^j$ is
that using $\Delta_n:= \Delta$ in (\ref{Delta})
and introducing
\be
\Delta_j(\tau):= T^j \Delta_n(\tau) T^j=\operatorname{diag}(\tau_1,...,1,..., \tau_{n-1}, \tau_j),
\qquad \forall j=1,...,n-1,
\label{Deltaj}\ee
where the entry $1$ appears in the $jj$ position,
one can verify the following lemma.

\medskip
\noindent
\textbf{Lemma 9.} \emph{The
$U(n)$-valued $C^\infty$ function $\cG_y^j$ defined on $\cp_0$ by the formula
\be
(\cG_y^j\circ\cE)(\xi,\tau):= \Delta(\tau)^{-1} g_y^j(\xi) \Delta_j(\tau),
\qquad
j=1,...,n,
\ee
extends to a $C^\infty$ function on $\cp_j$.
The extended function is denoted by the same letter,  $\cG_y^j: \cp_j\to U(n)$.}

\medskip\noindent
\begin{proof}
This is a simple inspection of the matrix elements of $\cG_y^j$ based on
the properties of $w_j^y(\xi)$ in (\ref{D1.1}) and the formula (\ref{altern}).
Indeed, if $u^j$ is  representative of $\cE(\xi,\tau)$ according to (\ref{altern}),
then for $j\neq n$ one finds for example that
\be
\cG_y^j(u^j)_{kn} \sim \bar u^j_k,
\quad
\cG_y^j(u^j)_{nj} \sim  1,
\quad
\cG_y^j(u^j)_{jj} \sim u^j_n,
\quad
\cG_y^j(u^j)_{kl} \sim  \delta_{kl} + (1- \delta_{kl}) \bar u^j_k u^j_l
\,\,\hbox{for}\,\,
k,l\notin\{j,n\},
\label{proprel}\ee
where the symbol $\sim$ means proportionality by a function of $\xi$ that extends to a
$C^\infty$, nowhere zero function on $\cp_j$.
The relations (\ref{proprel}), and similar relations that hold for all  matrix elements of
$\cG_y^j(u^j)$,
including $j=n$, imply that $\cG^j_y$
extends from $\cp_0$ to a $C^\infty$ function on $\cp_j$.
\end{proof}

\medskip

For any fixed $j=1,...,n$ and $y>0$, define the
$SU(n) \times SU(n)$ valued $C^\infty$
map $F_j$ on the chart $(\cp_j, \mathfrak{B}_{\chi_0}^j)$
by the following formula:
\be
F_j(u^j):= \left(\cG_y^j(u^j)^{-1} \cK^y(u^j) \cG_y^j(u^j),
\cG_y^j(u^j)^{-1} \delta(\xi) \cG_y^j(u^j)\right),
\label{Fj}\ee
where $\xi_i = \vert u^j_i\vert^2 + y$ for every $i=1,...,n$ and $\cK^y$ is the global
Lax matrix given in Lemma 8.

\medskip\noindent
{\bf Theorem 5.} \emph{The maps $F_j$ (\ref{Fj}) enjoy the following properties:
\begin{enumerate}
\item{$F_j(u^j)$ belongs to $\mu^{-1}(\mu_0)$ and $p\circ F_j: \cp_j \to P$ is a
smooth map.
}
\item{$p\circ F_j$ coincides with $p \circ F_k$ on  $\cp_j \cap \cp_k$
and it coincides with $f_0$ of Theorem 4  on $\cp_0=\cap_{j=1}^n \cp_j$.}
\item{One can define a smooth map $f_\beta: \cp \to P$ by requiring
that $f_\beta$ coincides with
$p \circ F_j$ on $\cp_j$.
The so-obtained map satisfies
\be
f_\beta^* (\hat \omega)= \chi_0 \omega_{\mathrm{FS}}
\quad\hbox{and}\quad f_\beta^*(\hat \beta) = \cJ.
\label{Delzprop}\ee}
\item{The map $f_\beta$ is surjective and injective.}
\end{enumerate}
Consequently, $f_\beta$ is a symplectomorphism
that extends the local Delzant map $f_0$ of Theorem 4 and its inverse
$\phi_\beta:= f_\beta^{-1}$
is a Delzant symplectomorphism satisfying  equation (\ref{eqvarb}).}

\medskip
\noindent
\begin{proof}
It follows directly from the definitions that
\be
(F_j \circ \cE)(\xi,\tau) = \Psi_{\eta^j_y(\xi,\tau)^{-1}} \left( (F_0\circ \cE)(\xi,\tau)\right),
\quad \forall (\xi,\tau)\in \cP^0 \times \bT_{n-1},
\label{FjF0}\ee
where we use (\ref{F0}) and the definition $\eta_y^j(\xi,\tau):= \eta_y^j(\xi) \Delta_j(\tau)$
with (\ref{etay}).
Since $\eta_y^j(\xi,\tau)$ belongs to the little group of $\mu_0$
in $U(n)$, this entails that $(F_j \circ \cE)(\xi,\tau)\in \mu^{-1}(\mu_0)$.
Then we obtain property 1
since $\mu^{-1}(\mu_0) \subset D$ is closed,
the values $\cE(\xi,\tau)$ cover $\cp_0$ which is dense in $\cp_j$,
and
$F_j:\cp_j \to D$ is a $C^\infty$ map as shown by its formula (\ref{Fj}).

Property 2 holds since $\cp_0 \subset \cp_j\cap \cp_k$ is dense,  and $p\circ F_j$ coincides with
$f_0 = p \circ F_0$
on $\cp_0$ because of (\ref{FjF0}).

Property 3 is immediate from the preceding properties and the fact that $f_0$ is a local
Delzant map satisfying (\ref{48}) and (\ref{49}).

To establish the surjectivity of $f_\beta$,
notice that  equivariance
with respect to the torus actions (\ref{rotact}) and (\ref{hatpsib}),
\be
 f_\beta\circ \cR_\tau =  \hat\Psi^b_\tau \circ f_\beta,
\qquad
\forall \tau\in \bT_{n-1},
\label{fbequivar}\ee
follows from (\ref{Delzprop}).
Then $f_\beta^*(\hat \beta) = \cJ$ and
(\ref{fbequivar}) imply that the image of $f_\beta$ contains each $\bT_{n-1}$-orbit
in $P$.
This entails  the surjectivity.

Our final task it to demonstrate the injectivity of $f_\beta$.
To do this, we remark  that if $f_\beta$ takes the same values on two
elements of $\cp$, then those elements must belong to the same chart
$\cp_j$ at least for one $j$. Indeed, this is a consequence of
the second equality in (\ref{Delzprop}) and the definition of $\cp_j$.
Then, assume that $f_\beta(u^j)= f_\beta(z^j)$ for two elements
$u^j, z^j \in \mathfrak{B}_{\chi_0}^j$.
By the definition of $f_\beta$,
this is equivalent to
the existence of an element $\eta$ from the little group of $\mu_0$ in $U(n)$ such that
\bea
&&\left(\eta^{-1} \cG_y^j(u^j)^{-1}
\cK^y(u^j) \cG_y^j(u^j)\eta, \eta^{-1} \cG_y^j(u^j)^{-1} \delta(\xi) \cG_y^j(u^j)\eta \right) =
\nonumber\\
&&{}\qquad\qquad =
\left( \cG_y^j(z^j)^{-1}
\cK^y(z^j) \cG_y^j(z^j), \cG_y^j(z^j)^{-1} \delta(\xi) \cG_y^j(z^j)\right),
\label{assum}\eea
where $\xi$ is given by
$\xi_i = \vert u_i^j\vert^2 + \vert y\vert  =  \vert z_i^j\vert^2 +\vert y\vert$.
We see from the second component of (\ref{assum}) that
\be
T:= \cG_y^j(u^j) \eta  \cG_y^j(z^j)^{-1}
\ee
must belong to the torus $\bT_n \subset U(n)$.
Then the first component of  (\ref{assum}) and  the fact that
$\cK_{k,k+1}^y$ (\ref{442}) depends
only on $\xi$ and never vanishes for any $k$ imply that
$T = \lambda \1_n$ for some $\lambda \in U(1)$.
Upon re-substitution into the first component of (\ref{assum}),
this gives the equality
\be
\cK^y(u^j) = \cK^y(z^j).
\label{469}\ee
By using  that the components of $\Lambda^y(\xi)$ in (\ref{443}) are non-zero,
we infer from the inspection of  $\cK_{k,l}^y(u^j) = \cK^y_{k,l}(z^j)$  for the fixed
index $l=j+1$ ($l:=1$ if $j=n$)
that $\bar u_k^j u_j^j= \bar z_k^j z_j^j$ holds for each $k$.
Since $u_j^j= z_j^j$ as this component depends only on $\xi$, and $u_j^j\neq 0$,
we conclude that $z^j = u^j$, whereby the injectivity of $f_\beta$ follows.

As  an alternative to the above self-contained reasoning, we can also give
a shorter proof of the injectivity of $f_\beta$ by invoking
that  for any Hamiltonian toric manifold and the pre-image of
any moment map value
there exists a certain subtorus that acts \emph{freely} on that pre-image  \cite{Cannas}.
The subtorus in question (which is the whole torus for the interior of the
Delzant polytope) depends only on the moment map value, and by using this
the injectivity of $f_\beta$ follows easily from
$f_\beta^*(\hat \beta) = \cJ$  and (\ref{fbequivar}).
\end{proof}

\medskip
\noindent
{\bf Remark 8.}
It is readily seen from the above
that the maps $F_j:\cp_j\to \mu^{-1}(\mu_0)$ defined by (\ref{Fj})
are indeed local sections of the principal bundle in (\ref{4.48})
whose base  is $\cp$ and total space is the constraint surface $\mu^{-1}(\mu_0)$.
We constructed the \emph{global} Delzant symplectomorphism $f_\beta: \cp\to P$
by patching together the projected maps $p\circ F_j$.

\subsection{The global Delzant map $f_\alpha$ and involution properties}

We here present the construction of $f_\alpha$ in terms of $f_\beta$, and establish
the involution properties (\ref{478}) and (\ref{Simonid}) for the maps (\ref{fSR})
whose significance
will become clear in Section 5.

We need some preparations.
First, let us define the map $\nu$ on the unreduced double by
\be
\nu(A,B):= (\bar B, \bar A),
\label{B1}\ee
where `bar' means complex conjugation.
This is an involution of $D$ that enjoys the properties
\be
\nu^* (\omega)= -\omega,
\qquad
\nu^*(\mu) = {\bar \mu}^{-1},
\qquad
 \nu \circ \Psi_g  = \Psi_{\bar g} \circ \nu
\quad (\forall g\in G).
\label{B2}\ee
By using these properties and the fact that $(\bar \mu_0)^{-1} = \mu_0$, we see that
$\nu$ maps $\mu^{-1}(\mu_0)$ to itself and it
induces an \emph{anti-symplectic involution}, $\hat \nu$,  of the reduced phase space $P$.

Second, let us define the anti-symplectic involution $\Gamma$ of the symplectic vector space
$\bC^n$ by
\be
\Gamma(u_1, ..., u_{n-1}, u_n):= (\bar u_{n-1},..., \bar u_1, \bar u_n),
\label{B3}\ee
which acts as  `reflection composed with complex conjugation' on the first $(n-1)$
coordinates.
Straightforwardly, $\Gamma$ induces an \emph{anti-symplectic involution}, $\hat \Gamma$,
of $(\bC P(n-1), \chi_0 \omfs)$.
We also need the anti-symplectic involution
$\hat C:\cp \to \cp$
that descends from componentwise complex conjugation on $\bC^n$, i.e.~for which
 \be
\hat C \circ \pi_{\chi_0}(u_1,..., u_{n-1}, u_n) =
\pi_{\chi_0}(\bar u_1,..., \bar u_{n-1}, \bar u_n)
\quad
\hbox{with}\quad u \in S^{2n-1}_{\chi_0},\label{wise}
\ee
as well the symplectic involution
$\hat \sigma$ furnished by\footnote{
The value $n=2$ is special since in this case $\hat C=\hat \Gamma$ and
$\hat \sigma$ (as well as $\sigma$ in (\ref{B9})) becomes the identity map.}
\be
\hat \sigma := \hat C \circ \hat \Gamma = \hat \Gamma \circ \hat C.
\label{compinv}\ee

Third, note that the spectral functions $\Xi_k$ defined in Section 2 (Eq.~(\ref{bigXi}))
verify the
identity
\be
\Xi_k(\bar A) = \Xi_{n-k}( A)\quad \forall k=1,..., n-1,\quad
\Xi_n(\bar A)= \Xi_n(A),
\quad
\forall A\in G.
\label{B8}\ee
This can be checked by direct calculation starting from (\ref{par1}),
 and also follows from well-known
group theoretic facts via the formula (\ref{fund}).

Fourth, let $\sigma$ denote the involutive map on $\bC^{n-1}$ given by
\be
\sigma(x)_k := x_{n-k},
\quad
\forall x\in \bC^{n-1},
\quad
\forall k=1,..., n-1.
\label{B9}\ee
As a result of (\ref{B8}), the $\bR^{n-1}$-valued spectral Hamiltonians, and their respective
reductions, are subject to the relations
\be
\alpha  = \sigma \circ \beta \circ \nu,
\qquad
\hat \alpha = \sigma \circ \hat \beta \circ \hat \nu.
\label{B10}\ee

Fifth, the moment map $\cJ: \bC P(n-1) \to \bR^{n-1}$ of the rotational $\bT_{n-1}$-action
used in Theorem 3 obeys
\be
\cJ \circ \hat \Gamma = \cJ \circ \hat \sigma = \sigma \circ \cJ.
\label{B12}\ee

Now we can construct $f_\alpha$ in terms of $f_\beta$.

\medskip
\noindent
{\bf Theorem 6.} \emph{If $f_\beta$ is a Delzant map in the sense
of Theorem 3 (cf.~also Eq.~(\ref{3.92})), then
\be
f_\alpha:= \hat \nu \circ f_\beta \circ \hat \Gamma
\label{B4}\ee
is also a Delzant map in the sense of Theorem 3.
Equation (\ref{B4})  implies the involution property
\be
(\hat \Gamma \circ f_\alpha^{-1} \circ f_\beta)^2= {\mathrm{id}}_{\bC P(n-1)}.
\label{478}\ee}
\begin{proof}
It is clear that $f_\alpha$ as defined by (\ref{B4}) is a symplectomorphism
provided that $f_\beta$ is a symplectomorphism.
By assuming that $\hat \beta \circ f_\beta = \cJ$, (\ref{B4})  and the previous relations entail
\be
\hat \alpha \circ f_\alpha = \hat \alpha \circ \hat \nu \circ f_\beta\circ \hat\Gamma =
\sigma \circ\hat \beta \circ f_\beta \circ \hat \Gamma = \sigma \circ \cJ \circ \hat \Gamma = \cJ,
\label{B13}\ee
which is the required  property of the Delzant map $f_\alpha$.
Eq.~(\ref{478}) follows directly from (\ref{B4}).
\end{proof}

It is worth noting that we did not use the explicit formula of $f_\beta$
to establish Theorem 6.

\medskip\noindent
\textbf{Lemma 10.} \emph{It follows from the local formula (\ref{C15}) of $f_\beta$
that the restriction of the map $f_\alpha$  (\ref{B4})
  to $\cp_0$ operates according to
\be
(f_\alpha\circ \cE)(\xi,\tau) =
\left( p \circ \Psi_{g_{-y}(\xi)^{-1}}\right) \left( \delta(\xi), L^{\mathrm{loc}}_{-y}(\delta(\xi),\rho(\tau))
 \right),
\label{faloc}\ee
using the  same notations as in
(\ref{C15}) and $\Psi$ in (\ref{2conj}).
Moreover,  the following identity holds:
\be
f_\beta^{-1} \circ f_\alpha = \hat \Gamma \circ (f_\beta^{-1} \circ f_\alpha) \circ \hat C.
\label{inter}\ee}
\begin{proof}
For typographic  reasons,
in this proof we use both alternative notations
\be
M^*\equiv \bar M
\ee
to denote the complex conjugate of any matrix $M$; $M^t$ denotes transpose
and $M^\dagger$ adjoint.
We begin the proof of formula (\ref{faloc}) by  remarking that the map $\cE$ (\ref{C3})
 satisfies
\be
(\hat \Gamma \circ \cE)(\xi, \tau)= \cE( \sigma(\xi), \sigma(\bar \tau)),
\qquad
\forall (\xi,\tau)\in \cP^0 \times \bT_{n-1},
\label{B15}\ee
with $\hat \Gamma$ and $\sigma$ defined earlier.
Then the combination of equations (\ref{C15}), (\ref{B4}) and (\ref{B15}) gives
\be
(f_\alpha\circ \cE)(\xi,\tau) =
p \circ \Psi_{{g_y(\sigma(\xi))}^{-1}} \left( \delta(\sigma(\xi))^*,
 L^{\mathrm{loc}}_y\left(\delta(\sigma(\xi)), \rho(\sigma(\bar \tau))^{-1}\right)^*\right).
\label{B17}\ee
Here,  we have taken into account that $g_y(\sigma(\xi))^* = g_y(\sigma(\xi))$
holds since $g_y$ (4.7) is real.
Let $\eta_0$ be the $n\times n$ matrix whose non-zero entries are
$(\eta_0)_{j, n+1-j}=1$ for all $j=1,..., n$; $\eta_0=\eta_0^{-1} = \eta_0^t$.
It is not difficult, although somewhat long,   to check that
\be
\delta(\sigma(\xi))^* = \eta_0 \delta(\xi) \eta_0
\label{B18}\ee
and
\be
 L^{\mathrm{loc}}_y\left(\delta(\sigma(\xi)), \rho(\sigma(\bar \tau))^{-1}\right)^*
= \eta_0 L^{\mathrm{loc}}_{-y}\left(\delta(\xi), \rho(\tau) \right) \eta_0.
 \label{B19}\ee
In the course of deriving these relations we utilized that
\be
W_j(\delta(\sigma(\xi)), y)= W_{n+1-j}(\xi, -y),
\quad
\forall j=1,...,n,
\label{B20}\ee
and
\be
\rho(\sigma(\bar \tau))^{-1}=\rho(\sigma(\tau)) = \eta_0 \rho(\tau)^{-1} \eta_0.
\label{B21}\ee
By using (\ref{B18}) and (\ref{B19}), we can rewrite (\ref{B17}) as
\be
(f_\alpha \circ \cE)(\xi,\tau) =
p \circ \Psi_{(\eta_0 g_y(\sigma(\xi)))^{-1}}  \left(\delta(\xi),
L^{\mathrm{loc}}_{-y}(\delta(\xi), \rho(\tau) )\right).
\label{B22}\ee
It follows from (\ref{B20}) that $\eta_0 g_y(\sigma(\xi))$ is a unitary matrix whose last column
is given by the vector $v(\xi, -y)$  defined in (\ref{vW}).
This permits  to conclude that
$\eta_0 g_y(\sigma(\xi)) =  g_{-y}(\xi) g$
 with some ($\xi$ and $y$-dependent) $g\in U(n)$
 for which $g \mu_0 g^{-1}=\mu_0$.
 Taking into account that $\Psi_{g}$ is a gauge transformation,
 equation (\ref{B22}) implies
 the desired formula (\ref{faloc}).

Now are are going  to prove the identity (\ref{inter}).
It is enough to verify this identity on the dense open submanifold of $\cp_0$ whose image
under the map $f_\beta^{-1} \circ f_\alpha$ is also contained in $\cp_0$.
For any $\cE(\xi,\tau)$ from this submanifold, we
define $(\xi', \tau')\in \cP^0\times \bT_{n-1}$ by the equation
\be
(f_\beta^{-1} \circ f_\alpha)(\cE(\xi, \tau))= \cE(\xi', \tau'),
\label{B27}\ee
which can be rewritten equivalently as
\be
 f_\alpha (\cE(\xi, \tau))= f_\beta(\cE(\xi', \tau')).
\label{B28}\ee
Then the claim (\ref{inter}) can be reformulated as the statement that
the relation (\ref{B28}) is equivalent to the relation
\be
(\hat \Gamma \circ (f_\beta^{-1} \circ f_\alpha)\circ \hat C)
(\cE (\xi, \tau))= \cE (\xi', \tau'),
\label{B29}\ee
which (by taking into account (\ref{B15})) is in turn  equivalent to
\be
 f_\alpha(\cE(\xi, \bar \tau))=  f_\beta(\cE (\sigma(\xi'), \sigma(\tau')^{-1})).
\label{B30}\ee
Consequently, we have to show that (\ref{B28}) is equivalent to (\ref{B30}).

We now introduce the notation
$(A_1, B_1) \sim (A_2,B_2)$
for elements of the double $D$ for which there exists $g\in U(n)$ such that
$gA_1 g^{-1}= A_2 $ and $g B_1 g^{-1}=B_2$.
We notice that two pairs $(A_1,B_1)$ and $(A_2,B_2)$ in
$\mu^{-1}(\mu_0)$ represent the same element of $P$ if and only if
$(A_1,B_1) \sim (A_2,B_2)$.
Therefore, by using the local formulae (\ref{C15}) of $f_\beta$ and (\ref{faloc}) of  $f_\alpha$,
we can reformulate the equivalence of (\ref{B28}) and (\ref{B30}) as the equivalence
between the relation
\be
\left( \delta( \xi), L^{\mathrm{loc}}_{-y}(\delta( \xi), \rho(\tau)) \right) \sim
\left(L^{\mathrm{loc}}_{y}(\delta( \xi'), \rho(\tau')^{-1}), \delta(\xi')\right)
\label{B31}\ee
and the relation
\be
\left( \delta( \xi), L^{\mathrm{loc}}_{-y}(\delta( \xi), \rho(\bar \tau)) \right) \sim
\left(L^{\mathrm{loc}}_{y}(\delta( \sigma(\xi')), \rho(\sigma(\tau'))), \delta(\sigma(\xi'))\right).
\label{B32}\ee
By applying (\ref{B18}) and (\ref{B19}),
we observe that (\ref{B32}) is equivalent to
\be
\left( \delta( \xi), L^{\mathrm{loc}}_{-y}(\delta( \xi), \rho(\bar \tau)) \right) \sim
\left(L^{\mathrm{loc}}_{-y}(\delta( \xi'), \rho(\tau'))^*, \delta(\xi')^*\right).
\label{B34}\ee

To finish the proof, we need the identities
\be
L^{\mathrm{loc}}_{-y}(\delta(\xi'),\rho(\tau'))^* = \rho(\tau')
\left[ L^{\mathrm{loc}}_{y}(\delta(\xi'),\rho(\tau')^{-1})\right]^t  \rho(\tau')^{-1},
\label{B36}\ee
\be
L^{\mathrm{loc}}_{-y}(\delta(\xi),\rho(\tau))^* = \delta(\xi)
 L^{\mathrm{loc}}_{-y}(\delta(\xi),\rho(\tau)^{-1})  \delta(\xi)^{-1},
\label{B38}\ee
which can be readily verified.

Now suppose that (\ref{B31}) holds (for some arbitrarily fixed $(\xi,\tau)$).
This assumption is equivalent to the existence of a unitary matrix $g$ for which
\be
g \delta(\xi) g^{-1} = L^{\mathrm{loc}}_{y}(\delta( \xi'), \rho(\tau')^{-1})
\quad\hbox{and}\quad
g^{-1} \delta(\xi') g =L^{\mathrm{loc}}_{-y}(\delta( \xi), \rho(\tau)).
\label{B35}\ee
Then, by means of (\ref{B36}),
the validity of the first equality in (\ref{B35})  implies that
\be
L^{\mathrm{loc}}_{-y}(\delta(\xi'),\rho(\tau'))^* = \rho(\tau') (g \delta(\xi) g^{-1})^t \rho(\tau')^{-1}
= [\rho(\tau') \bar g \delta(\xi)] \delta(\xi) [\rho(\tau') \bar g \delta(\xi)]^{-1},
\label{B37}\ee
which can be recognized as the `first component' of the relation (\ref{B34}).
By using (\ref{B38}), the second equality in (\ref{B35}) becomes
\be
 \delta(\xi')^* =\bar g L^{\mathrm{loc}}_{-y}(\delta( \xi), \rho(\tau))^* \bar g^{-1}
 =\bar g \delta(\xi) L^{\mathrm{loc}}_{-y}(\delta( \xi), \rho(\tau)^{-1}) (\bar g \delta(\xi))^{-1},
 \label{B39}\ee
 which (since $\delta$ and $\rho$ take values in $\bT_n$) entails that
 \be
 \delta(\xi')^*
 =[\rho(\tau') \bar g \delta(\xi)] L^{\mathrm{loc}}_{-y}(\delta( \xi), \rho(\tau)^{-1})
 [\rho(\tau')\bar g \delta(\xi)]^{-1}.
\label{B40}\ee
Thus we have derived (\ref{B37}) and (\ref{B40}) from (\ref{B35}), which tells us
that (\ref{B31}) implies (\ref{B32}). The converse implication can be demonstrated  by following the
above equations in reverse order, whereby the proof is complete.
\end{proof}

In Section 5, we shall identify the maps
\be
\fS:= f_\alpha^{-1}\circ f_\beta
\quad\hbox{and}\quad
\fR:= \hat C \circ \fS
\label{fSR}\ee
as the symplectic and respectively the anti-symplectic version of Ruijsenaars' self-duality map
of the compactified $\mathrm{III}_\mathrm{b}$  system.
Then the  properties (\ref{478}) and (\ref{inter}) will reproduce certain
relations established in \cite{RIMS95}.
The same is true regarding the following identities  that can be derived easily from the above:
\be
\fS^2 = \hat \sigma
\quad\hbox{and} \quad
\fR^2 ={\mathrm{id}}_{\cp}.
\label{Simonid}\ee
As for their derivation,  the first identity in (\ref{Simonid}) is obtained
by recasting  (\ref{478}) as
\be
{\mathrm{id}}_{\cp} = \hat \Gamma \circ \fS \circ \hat\Gamma \circ \fS =
\hat \Gamma \circ \hat C \circ \fS^2 = \hat \sigma \circ \fS^2,
\label{4119}\ee
where we applied (\ref{inter}) to establish the second equality.
Similarly, we can write
\be
\fR= \hat C \circ \fS  = \fS \circ \hat \Gamma =
\hat\Gamma \circ (\hat \Gamma \circ \fS) \circ \hat \Gamma,\label{novy1}
\ee
whereby the second identity in (\ref{Simonid}) follows from (\ref{4119}).

Finally, let us record also the following useful identities:
\be
\delta \circ \cJ \circ \hat C = \delta\circ \cJ,
\quad
\delta \circ \cJ \circ \hat \Gamma =
\delta\circ \cJ \circ \hat \sigma = \eta_0 (\delta \circ \cJ)^\dagger \eta_0,
\label{deltaid}\ee
\be
\cK^y \circ \hat C= \bar \cK^y,
\quad
\cK^y \circ \hat \Gamma= \eta_0 (\cK^y)^t \eta_0,
\quad
\cK^y \circ \hat \sigma= \eta_0 (\cK^y)^\dagger \eta_0,
\label{Kid}\ee
with $\eta_0$ defined above (\ref{B18}). The identities in (\ref{deltaid})
are equivalent to (\ref{B12}) and (\ref{B18}), and those
in (\ref{Kid}) can be
verified directly by using the definition of the global Lax matrix $\cK^y$.

\section{Self-duality of the compactified $\mathrm{III}_\mathrm{b}$  system}
 \setcounter{equation}{0}

The construction of the compactified $\mathrm{III}_\mathrm{b}$  system and
the discovery  of its self-duality properties are due to Ruijsenaars \cite{RIMS95}.
Below we first recall  the definition of this system
from \cite{RIMS95}, interlaced with some explanatory comments in terms of our
present work.
Then we establish that our reduction yields the
 compactified  $\mathrm{III}_\mathrm{b}$ system, which is the message of Theorem 7.
The subsequent Theorem 8 reproduces the symplectic version of Ruijsenaars' self-duality
map as an automatic consequence of our construction.
Finally, we briefly explain how the anti-symplectic self-duality involution of Ruijsenaars
fits into our framework.

Theorems 7 and 8 represent the first main result of this paper
(the second will be encapsulated in Theorem 9 of Section 6).
We can be relatively brief here, since these theorems are just easy corollaries
of our preceding technical results. In particular, they follow from our detailed
description of the Delzant symplectomorphisms $f_\beta$ and $f_\alpha$ that
incorporate the Lax matrix (\ref{I.4}) and its global extension (\ref{428}).
Although we gained inspiration from the seminal paper \cite{RIMS95} with which
we must compare our results,  we wish to emphasize that our approach is self-contained.

The space of particle-positions  $\cD_y\subset S\bT_n$ of the local
$\mathrm{III}_\mathrm{b}$ system can be identified with the open polytope
$\cP^0$ by means of the map $\delta$ (cf.~(\ref{fund})).
Although the true particle-positions are the components
of $\delta(\xi)$, we may (and often do) regard the equivalent $\xi\in \cP^0$ as the local
particle-position variable of the system.
The canonical conjugates  of the coordinates $\xi_k$ are the $\theta_k$ that parametrize
$\Theta = \rho(e^{\ri \theta})^{-1}$ by (\ref{C5}).
By inserting these parametrizations into (\ref{I.3}) one indeed obtains
\be
\Omega^{\mathrm{loc}} =
- \frac{1}{2} \tr\!\left(\delta( \xi)^{-1} d \delta(\xi) \wedge
 \rho(\tau)^{-1} d \rho(\tau)\right)
= \ri \sum_{k=1}^{n-1} d \xi_k \wedge \tau_{k}^{-1} d \tau_k =
\sum_{k=1}^{n-1}  d \theta_k \wedge d\xi_k.
\label{52}\ee
Taking $\xi$ and $\tau= e^{\ri \theta}$ as the basic variables,
from now on we identify the local phase space  as
\be
M_y^{\mathrm{loc}}\equiv \cP^0 \times \bT_{n-1} = \{ (\xi, \tau)\}.
\label{51}\ee
The same variables  were used in \cite{RIMS95} to describe
the relative motion of the particles  governed by the Hamiltonian (\ref{I.1}).
The local $\mathrm{III}_\mathrm{b}$  system of \cite{RIMS95}  is thus encapsulated by the triple
\be
(M_y^{\mathrm{loc}}, \Omega^{\mathrm{loc}}, \cL_{y}^{\mathrm{loc}}),
\ee
 where the value of
the Lax matrix $\cL_{y}^{\mathrm{loc}}$ at  $(\xi,\tau) \in M_y^{\mathrm{loc}}$ is defined to be
\be
\cL_{y}^{\mathrm{loc}}(\xi,\tau):=L_{y}^{\mathrm{loc}}(\delta(\xi), \rho(\tau)^{-1})
\label{53}\ee
 with the expression (\ref{I.4}).
The commuting Hamiltonians of the system are provided by the functions
$h\circ \cL_{y}^{\mathrm{loc}}$
for all $h\in C^\infty(G_\mathrm{reg})^G$.
The composed function $h\circ \cL_{y}^{\mathrm{loc}}$ belongs to $C^\infty(M_y^{\mathrm{loc}})$
since
 $\cL_{y}^{\mathrm{loc}}: M_y^{\mathrm{loc}} \to G_\mathrm{reg}$, as was shown in \cite{RIMS95}
 and follows also from our Theorem 2 combined with Lemma 6.

Expressed in our notations, the compactified
 $\mathrm{III}_\mathrm{b}$ system was defined in \cite{RIMS95}  as the
 triple
\be
(\cp, \chi_0 \omfs, \cK^y),
\qquad
\chi_0 = \pi - n \vert y\vert,
\label{54}\ee
where $\cK^y$ is the global Lax matrix described in Subsection 4.2.
The commuting Hamiltonians of this system were identified in \cite{RIMS95} as
the $C^\infty$ functions\footnote{Actually in \cite{RIMS95} (Eq.~(5.50) loc.~cit.)
real-analytic Hamiltonians were studied, which is  a negligible difference.}
of the form
$h \circ \cK^y$ for all $h\in C^\infty(G_\mathrm{reg})^G$.
In particular, it was established in \cite{RIMS95} (and is obvious in our setting)
that the functions
\be
\Xi_k \circ \cK^y,
\qquad
k=1,..., n-1,
\label{55}\ee
represent action-variables for the system (\ref{54}), where  $\Xi_k\in C^\infty(G_\mathrm{reg})^G$
was defined in (\ref{bigXi}).
The crucial fact \cite{RIMS95} is that by the identification (\ref{51}) the
map $\cE$ (\ref{Edisp}) embeds the local $\mathrm{III}_\mathrm{b}$  system into $\cp$,
converting $M_y^{\mathrm{loc}}$ into the dense open submanifold $\cp_0$.
As seen upon comparison
of (\ref{C4})  and (\ref{52}), this embedding is symplectic.
Because of Lemma 8, the commuting Hamiltonians of the local system
extend to those of the compactified one. The compactified system has complete flows, simply
since $\cp$ is compact and the Hamiltonians of interest are smooth.

\medskip
\noindent
\textbf{Theorem 7.} \emph{For any $h\in C^\infty(G_\mathrm{reg})^G$, let $\hat h_1$
and $\hat h_2$ denote the reduced
Hamiltonians defined by means of equation (\ref{genred}) with (\ref{f1f2}).
Consider the Delzant symplectomorphisms $f_\beta$ and $f_\alpha$ that
 map $(\cp, \chi_0 \omfs)$  to the reduced phase space
 $(P,\hat\om)$  as given by
 Theorems 5 and 6.  Then, using  the preceding notations, the following relations hold
 for $f_\beta$:
\be
\hat h_1 \circ f_\beta =  h \circ \cK^y,
\qquad
\hat h_2 \circ f_\beta = h \circ \delta \circ \cJ.
\label{57}\ee
In particular, with $h=\Xi_k$ for any $k=1,...,n-1$,
\be
\hat \alpha_k \circ f_\beta = \Xi_k \circ \cK^y,
\qquad
\hat \beta_k \circ f_\beta = \cJ_k.
\label{58}\ee
Regarding $f_\alpha$, there hold the analogous relations,
\be
\hat h_1 \circ f_\alpha =  h \circ \delta \circ \cJ,
\qquad
\hat h_2 \circ f_\alpha = h \circ (\cK^y)^\dagger,
\label{59}\ee
and in particular
\be
\hat \alpha_k \circ f_\alpha = \cJ_k,
\qquad
\hat \beta_k \circ f_\alpha = \Xi_{n-k}\circ \cK^y.
\label{510}\ee}
\begin{proof}
After recalling the underlying definitions,
all these relations follow in a direct and  straightforward way from the
formulae for the maps $f_\alpha$ and $f_\beta$.
Actually it is enough to use the local formulae (\ref{C15}) for  $f_\beta$  and
(\ref{faloc}) for $f_\alpha$, since
any smooth function is determined by its restrictions to
a dense open submanifold. What was  non-trivial to establish is that the local
formulae just mentioned
 represent the restrictions of  \emph{global} Delzant maps.
\end{proof}

Let us discuss the meaning of Theorem 7.
First note that the function $\delta\circ \cJ$ (or equivalently just $\cJ$)
on $\cp$ represents the global analogue of the particle-positions, since
under the embedding $\cE: M_y^{\mathrm{loc}}\to \cp$ we have
\be
(\cJ_k\circ \cE)(\xi,\tau)= \xi_k.
\label{511}\ee
According to (\ref{57}),  $f_\beta$ converts  the reductions of the
invariant functions of the form $h_1$, where $h_1(A,B)= h(A)$ before reduction,
into the respective  functions of the global Lax matrix $\cK^y$, and converts
the reductions of the invariant functions $h_2$, where $h_2(A,B)= h(B)$ before reduction, into the
respective functions
of the particle-position matrix $\delta\circ \cJ$.
In particular, $f_\beta$ converts the  reduced spectral Hamiltonian $\hat \alpha_k$
into the action-variable $\Xi_k\circ \cK^y$ of the
compactified $\mathrm{III}_\mathrm{b}$ system, and at the same time it converts the reduced
spectral Hamiltonian
$\hat\beta_k$ into the global particle-position
variable $\cJ_k$, respectively for each $k=1,...,n-1$.
The formula (\ref{59}) for the map $f_\alpha$ works similarly, but in addition to the exchange
of the subscripts
$1$ and $2$ on the function $\hat h$, which goes back to the exchange of the two factors of the double,
it applies the adjoint of the unitary Lax matrix
$\cK^y$ instead of $\cK^y$.
This implies that, for each $k=1,...,n-1$, $f_\alpha$  converts  $\hat \alpha_k$  into the particle-position
variable $\cJ_k$, and at the same time converts
$\hat \beta_k$ into the `flipped action-variable'  $\Xi_{n-k}\circ \cK^y$.
Such a flip, which arises from the involutions inevitably involved in the relation (\ref{B4})
between $f_\alpha$ and $f_\beta$, is necessary in order to ensure the symplectic property
 of the map.

The phase space $\cp$ of the compactified $\mathrm{III}_\mathrm{b}$ system
is equipped with the two Abelian Poisson algebras formed by the respective functions of the form
\be
h \circ \delta \circ \cJ
\quad\hbox{and}\quad
h\circ \cK^y
\quad\hbox{with}\quad
h\in C^\infty(G_\mathrm{reg})^G.
\ee
These Abelian algebras
are generated respectively by the global particle-position variables
$\cJ_k$ and action-variables $\Xi_k \circ \cK^y$.
We now describe the behaviour of these algebras under the Delzant symplectomorphism $\fS$,
\be
\fS \equiv f_\alpha^{-1} \circ f_\beta: \cp \to \cp
\label{513}\ee
as  introduced in (\ref{fSR}).
In the following theorem, we shall use  the `flip-involution' $h\mapsto h^\sharp$ of
$C^\infty(G_\mathrm{reg})^G$ defined by
\be
h^\sharp(g) := h(g^\dagger),
\qquad
\forall g\in G_\mathrm{reg}.
\label{514}\ee
With this notation, the property (\ref{B8}) of $\Xi_k \in C^\infty(G_\mathrm{reg})^G$ implies
\be
\Xi_k^\sharp = \Xi_{n-k},
\qquad
k=1,...,n-1.
\label{515}\ee

\medskip
\noindent
\textbf{Theorem 8.} \emph{The Delzant symplectomorphism $\fS$ given by (\ref{513})
satisfies the identities
\be
(h \circ (\delta\circ \cJ))\circ \fS = h \circ \cK^y
\quad\hbox{and}\quad
(h\circ \cK^y)\circ \fS = h^\sharp \circ (\delta \circ \cJ)
\label{516}\ee
for every $h\in C^\infty(G_\mathrm{reg})^G$.
In particular, with $h= \Xi_k$ ($k=1,...,n-1$)  this yields
\be
\cJ_k \circ \fS = \Xi_k \circ \cK^y
\quad\hbox{and}\quad
(\Xi_k \circ \cK^y)\circ \fS = \cJ_{n-k}.
\label{517}\ee
In this way, $\fS$ converts the particle-position variables into the action-variables
and converts the action-variables into the flipped particle-position variables.}
\begin{proof} Both equalities in (\ref{516}) follow by trivial one line calculations from our
preceding results.
In fact, by using the definition of $\fS$ and Theorem 7 we can write
\be
(h \circ \delta\circ \cJ) \circ \fS=
(\hat h_1 \circ f_\alpha) \circ (f_\alpha^{-1} \circ f_\beta) =\hat h_1 \circ f_\beta= h\circ \cK^y,
\label{518}\ee
and similarly
\be
(h\circ \cK^y)\circ \fS = (h^\sharp \circ (\cK^y)^\dagger) \circ (f_\alpha^{-1} \circ f_\beta)
= (h^\sharp \circ (\cK^y)^\dagger \circ f_\alpha^{-1})\circ f_\beta = h^\sharp_2 \circ f_\beta
= h^\sharp \circ (\delta \circ \cJ).
\label{519}\ee
The equalities in (\ref{517}) are special cases of those in (\ref{516}), using also  (\ref{515}).
\end{proof}

The message of Theorem 8 is that  $\fS$
\emph{is the symplectic version of Ruijsenaars'
self-duality map\footnote{To give a precise comparison, our map
$\fS$ corresponds to $f= k\circ \phi $ in Eq.~(4.125) of \cite{RIMS95}.
The anti-symplectic involution
$\mathfrak{R}= \hat C\circ \fS$ defined here in Eq.~(\ref{fSR}) corresponds to $\phi$
in \cite{RIMS95}.
The involutions $\hat C, \hat \Gamma, \hat \sigma$ that we introduced in Subsection 4.4
correspond respectively
to $k, \hat k, p$ in Subsection 4.4 of \cite{RIMS95}, which also contains
the equivalents of the relations (\ref{478}), (\ref{inter}) and (\ref{Simonid})
enjoyed by $\fS$ and $\fR$.
The identification of $k$ and $p$ with time reversal and
parity is further discussed in  \cite{SR-CRM} (page 69).}
of the
compactified $\mathrm{III}_\mathrm{b}$ system}.
The symplectic property of this map is an automatic  consequence of
our construction, while it was  proved in \cite{RIMS95}  in a
very different manner.

It is worth noting that, as follows immediately from Theorem 8,
the symplectomorphism $\fS^2$ satisfies the relations
\be
(h \circ (\delta\circ \cJ))\circ \fS^2 = h^\sharp \circ (\delta\circ \cJ),
\qquad
(h\circ \cK^y)\circ \fS^2 = h^\sharp \circ \cK^y,
\label{520}\ee
\be
\cJ_k \circ \fS^2 = \cJ_{n-k},
\qquad
(\Xi_k \circ \cK^y)\circ \fS^2 = \Xi_{n-k} \circ \cK^y.
\label{521}\ee
These are consistent with the fact that $\fS^2$ \emph{is equal to the symplectic
involution} $\hat \sigma$  of
$\cp$  according to (\ref{Simonid}).

Ruijsenaars  focused mainly on the \emph{anti-symplectic} version
of his self-duality map that has somewhat simpler (but equivalent) properties
as the symplectic map $\fS$.
As we now explain, this map is provided in our setting by  $\mathfrak{R}\equiv \hat C \circ \fS$,
defined already in (\ref{fSR}).
Our map $\mathfrak{R}$ is obviously anti-symplectic, $\mathfrak{R}^*\omfs= - \omfs$,
and it is involutive  according to (\ref{Simonid}).
For any $h\in C^\infty(G_\mathrm{reg})^G$, we readily derive that
\be
(h \circ (\delta\circ \cJ))\circ \fR = h \circ \cK^y,
\qquad
(h\circ \cK^y)\circ \fR = h \circ (\delta \circ \cJ),
\label{524}\ee
which in particular implies that
\be
\cJ_k \circ \fR = \Xi_k \circ \cK^y\quad\hbox{ and}\quad
(\Xi_k \circ \cK^y)\circ \fR = \cJ_k \quad\hbox{for all}\quad  k=1,...,n-1.
\label{523}\ee
These are precisely the characteristic properties  of the anti-symplectic
self-duality involution of Ruijsenaars \cite{RIMS95}.
The derivation of the first relation in (\ref{524}) goes simply as
\be
h \circ \delta \circ \cJ \circ \hat C \circ \fS =
h \circ \delta \circ \cJ \circ \fS = h \circ  \cK^y,
\label{526}\ee
by combining (\ref{516}) with  $\cJ \circ \hat C= \cJ$.
The second relation in (\ref{524}) can be derived similarly
using (\ref{516}) and the  equality $\cK^y \circ \hat C = \bar \cK^y$ (\ref{Kid}).

The relative advantage of $\fR$ over $\fS$ is that it simply interchanges
the respective action and particle-position variables, with the price that $\fR$
is anti-symplectic while $\fS$ is symplectic.
The relationship between $\fS$ and $\fR$ is similar to that
between the symplectic map $(q,p)\mapsto (p, -q)$ and  the anti-symplectic
map $(q,p) \mapsto (p,q)$ on the canonical phase space $\bR^{2n}$.
We note that the seminal paper \cite{RIMS95} contains further
interesting properties
of the maps $\fS$, $\mathfrak{R}$ and  `discrete symmetries'
(including but not exhausted by the time reversal $\hat C$  and the parity  $\hat \sigma$),
which  we could also reproduce.
This, perhaps together with the investigation of open questions mentioned in \cite{RIMS95},
will be reported in a survey that we plan to write about
the reduction approach to Ruijsenaars dualities in  a few years.

\section{Duality and the mapping class group}
\setcounter{equation}{0}

As was already mentioned,
quasi-Hamiltonian geometry   provides universal finite-dimensional
  `master' objects  by reductions of which one may obtain the symplectic moduli spaces of
  non-Abelian  flat connections  on Riemann surfaces with
prescribed  boundary conditions \cite{AMM}.
Originally those moduli spaces were derived  by infinite-dimensional
reductions of infinite-dimensional symplectic manifolds $\cA(\Sigma)$ consisting of
all (suitably smooth) non-Abelian connections on
the Riemann surface $\Sigma$.  If $\Sigma$ is a  torus with an open disc removed
(called `one-holed torus'), then the relevant  finite-dimensional  master
object is precisely the internally fused quasi-Hamiltonian double of  the non-Abelian   group.

The treatment of the compactified $\mathrm{III}_\mathrm{b}$ system
by quasi-Hamiltonian reduction
 now permits us  to  establish the validity of a remarkable interpretation
of this  system that goes back to Gorsky and Nekrasov \cite{GN}.
Namely, it is immediate from \cite{AMM} and our results that the compactified Ruijsenaars-Schneider system
is the natural integrable system on the moduli  space of flat $SU(n)$ connections on the one-holed torus,
where the holonomy around the hole is constrained to the conjugacy class of the
moment map value $\mu_0$ that we used to define our reduction.
In this picture, the pair $(A,B)$ in the double
represents the holonomies of the flat connection along the standard cycles on the torus.
This interpretation of the $\mathrm{III}_\mathrm{b}$ system
 was anticipated in \cite{GN} based on some formal, non-rigorous arguments.
Utilizing the quasi-Hamiltonian framework, we here further develop
this interpretation and  clarify the connection (conjectured in \cite{JHEP})
between the mapping class group of the one-holed torus and the Ruijsenaars duality.

In the next two
subsections 6.1 and 6.2 we briefly summarize necessary background material  \cite{AMM,Gol},
and present new results in the last two subsections 6.3 and 6.4.
In the background material  $G$ can be taken to be any connected and simply connected compact
 Lie group, with invariant scalar product $\langle\ , \rangle $ on its Lie algebra $\G$,
 but when dealing with our example
 it will be understood without further notice that $G\equiv SU(n)$.

\subsection{Flat connections and the quasi-Hamiltonian double}

 Let $\cA(\Sigma)$ be the  space of smooth  $\G$-valued connection forms on the one-holed torus  $\Sigma$.
 The space $\cA(\Sigma)$
carries a natural symplectic form $\Omega$ with respect to which the gauge group  $G(\Sigma)$
of  smooth maps from $\Sigma$ to $G$ acts
in the Hamiltonian way.
The symplectic form reads
\be
\Omega(a,b)=\int_\Sigma \langle
a\stackrel{\wedge}{,} b\rangle ,\ee
where  $a$ and $b$ are $\G$-valued  $1$-forms on $\Sigma$ interpreted as tangent vectors to $\cA(\Sigma)$.
Consider then
the space $\cA_{\flat}(\Sigma)$ of connections  $\phi$ verifying
the flatness condition $d \phi+[\phi,\phi]=0$ and   the subgroup $G_{\res}(\Sigma)$ of
$G(\Sigma)$ mapping a chosen point $p_0$
on the boundary of the hole  to the unit element of the group $G$. The quotient
 $\cA_{\flat}(\Sigma)/G_{\res}(\Sigma)$ can be identified with the finite-dimensional manifold $G\times G$ because
 every flat connection  is completely determined by its ($G_{\res}(\Sigma)$-independent)
holonomies along
two  generators of the fundamental group $\pi_1(\Sigma,p_0)$.  Furthermore, as shown
in \cite{AMM}, the symplectic  form $\Omega$
induces the quasi-Hamiltonian form (\ref{2.2}) on the double $G\times G$ such that every quasi-Hamiltonian reduction
of the double $(G\times G,\omega)$  gives the same outcome as a corresponding  symplectic
reduction of $(\cA(\Sigma),\Omega)$.
In particular, the quasi-Hamiltonian reduction based on the moment map
value $\mu_0$  (\ref{1.2}) that we have performed in
Section 3 gives the symplectic manifold $P$ that can  be also  interpreted as  the moduli space
$\cA_{\flat}(\Sigma,C)/G(\Sigma)$
consisting of flat connections with  holonomies around the boundary of the hole belonging
to the conjugacy class $C$ of $\mu_0$ modulo $G(\Sigma)$ gauge transformations.

Since the concepts of quasi-Hamiltonian geometry and
  quasi-Hamiltonian reduction  make no reference to flat connections on Riemann surfaces,
  the following question appears.
Despite technical advantages of the quasi-Hamiltonian vantage point,
 are there some relevant structural insights concerning the symplectic geometry of the
  moduli spaces that are still  more transparent in the infinite-dimensional reduction approach?
  The answer to this question
  is affirmative and this section is devoted to an important  illustration of this fact.
  Indeed, we shall show that the symplectic
  version $\fS$ of the Ruijsenaars duality map introduced in Section 4 is nothing but the natural
  symplectomorphism induced by a particular
  element of the mapping class group   of the one-holed torus.

  Recall that the group $ {\rm Diff}^+(\Sigma)$ of orientation-preserving diffeomorphisms acts symplectically on
  the space of all connections
 $(\cA(\Sigma),\Omega)$ by simply associating to a connection 1-form $\phi$ its pull-back by the diffeomorphism.
 This infinite-dimensional  group
 of symplectomorphisms  has an immense `reduction kernel' consisting of all diffeomorphisms  which
 can be connected  to the  identity.
  This means that effectively only the corresponding discrete
 quotient group (known as the mapping class group) acts  on the   moduli space $\cA_{\flat}(\Sigma,C)/G(\Sigma)=P$.
 We shall explain how this symplectic action  on the reduced phase
 space $P$ can be understood  entirely in the quasi-Hamiltonian context as well.
 In fact, it turns out that an appropriate central extension of the mapping class group acts by automorphisms on
 the quasi-Hamiltonian double, and upon  quasi-Hamiltonian reduction
 the center factors out yielding the symplectic action of
 the orientation-preserving mapping classes on the reduced phase space.

\subsection{Mapping class group of the one-holed torus}

The diffeomorphisms considered below include also those which reverse the orientation of the
Riemann surface. We shall need this generality later on in order  to explain
the geometrical origin of both the symplectic version
$\fS$ and the anti-symplectic version $\fR$ of the duality map.

We recall (see e.g.~\cite{Gol})  that the fundamental group  $\pi_1:=\pi_1(\Sigma,p_0)$ of the one-holed torus
is a free group with two generators $X$ and $Y$
 corresponding  to the pathes passing through the point $p_0$ and forming the standard homology basis of the torus.
 It admits a redundant presentation in terms of three generators
 $X,Y,K$:
 \be
 \pi_1 =\left\{X,Y,K\,\vert\, XYX^{-1}Y^{-1}=K\right\},
 \ee
 where $K$ corresponds to the generator of the boundary fundamental group $\pi_1(\partial\Sigma,p_0)$.
Every
element  from the group of diffeomorphisms ${\rm Diff}_{p_0}(\Sigma)$ preserving the point $p_0$ induces an
automorphism of
the fundamental group $\pi_1$ and two diffeomorphisms
from  ${\rm Diff}_{p_0}(\Sigma)$ which can be connected
by a path
in  ${\rm Diff}(\Sigma)$ induce automorphisms
of $\pi_1$ differing by an inner automorphism.
Let ${\rm Diff}_{p_0,0}(\Sigma)$ be the normal subgroup of ${\rm Diff}_{p_0}(\Sigma)$
consisting of the elements connected to the identity by a path in ${\rm Diff}(\Sigma)$.
The mapping class group can be defined as
\be
 {\rm MCG}_{p_0}(\Sigma)\equiv {\rm Diff}_{p_0}(\Sigma)/{\rm Diff}_{p_0,0}(\Sigma),
 \ee
 and there exists a natural homomorphism
\be
{\cal N}: {\rm MCG}_{p_0}(\Sigma)\to {\rm Out}(\pi_1)\equiv {\rm Aut}(\pi_1)/{\rm Inn}(\pi_1).
\label{627}\ee
Nielsen \cite{Nie} has proved  that ${\cal N}$ is actually an  isomorphism identifying
${\rm MCG}_{p_0}(\Sigma)$ with   ${\rm Out}(\pi_1)$. Furthermore one has the standard Hurewicz homomorphism   from
$\pi_1$  into  the  homology group  $H_1(\Sigma;\bZ)=\pi_1/[\pi_1,\pi_1]\cong \bZ^2$. Since
$H_1(\Sigma;\bZ)$ is Abelian, its outer automorphisms obviously form the discrete linear group $\gl$.
Hence one obtains  a homomorphism
\be
\cH:{\rm Out}(\pi_1) \to \gl,
\label{628}\ee
which again  turns out to be an isomorphism \cite{Nie1}.

In addition to the isomorphisms ${\cal N}$ and ${\cal H}$, there exists  another   isomorphism (see e.g.~\cite{Gol})
\be
{\cal I}:\gl\to{\rm Aut}(\pi_1,K)/\langle \iota_K \rangle,
\label{629}\ee
which will be especially useful for us.
Here ${\rm Aut}(\pi_1,K)$ denotes the subgroup of automorphisms of $\pi_1$ which take $K$ either to $K$ or to $K^{-1}$
and $\langle \iota_K \rangle$ is the normal subgroup of ${\rm Aut}(\pi_1,K)$ formed by conjugations by powers of $K$.
The existence of the
isomorphism $\cI$ is based on the fact that  the centralizer of $K$ in $\pi_1$ is the infinite cyclic  group
$\langle K\rangle$ and on the result of Nielsen \cite{Nie1} stating that   any automorphism of $\pi_1$ takes $K$
to a conjugate of either $K$ itself or to a conjugate of its inverse $K^{-1}$.
Notice that as the one-holed torus has just one boundary component, which is  therefore preserved by every element of
${\rm Diff}_{p_0}(\Sigma)$,  in  the image of ${\cal N}$ (\ref{627}) there may occur only such
elements of Out$(\pi_1)$   that preserve  the union
of the conjugacy classes of $K$ and of $K^{-1}$ in $\pi_1$.
The orientation-preserving mapping classes
are mapped by ${\cal N}$
into outer automorphisms respecting   the conjugacy class of $K$ alone.
(For the results recalled so far,  the reader may also consult the reviews \cite{Ivanov,Lyndon}.)

By composing the isomorphisms $\cN,\cH$ and $\cI$ we identify  the mapping class group ${\rm MCG}_{p_0}(\Sigma)$
with the group  ${\rm Aut}(\pi_1,K)/ \langle \iota_K \rangle$.
Four particular elements $F, S$, $T$ and $\tilde T$ of  ${\rm Aut}(\pi_1,K)$ will feature subsequently.
We define them as follows:
\bea
&&F:  X\to Y, \quad Y\to X;\qquad
 S:X\to Y^{-1},\quad Y\to YXY^{-1};
 \nonumber\\
 && T:X\to XY, \quad Y\to Y; \qquad \tilde T: X \to X, \quad Y\to YX^{-1}.
 \label{class}\eea
These elements (redundantly) generate ${\rm Aut}(\pi_1,K)$ \cite{Gol}.
Since ${\cal N}$ (\ref{627}) is an isomorphism, there exist diffeomorphisms inducing the
actions of their equivalence classes on $\pi_1$.
The diffeomorphisms (and their mapping classes) that yield $[T]$ and $[\tilde T]$
are commonly referred to as Dehn twists around the cycles $Y$ and $X$, respectively.

By an easy calculation we find
\be
F^2=(FS)^2=(FST)^2={\rm Id},  \quad S^2=(ST)^3, \quad S^4: X\to  K^{-1}XK, \quad Y\to K^{-1}YK,
\ee
and therefore the equivalence classes $[F],[S]$ and $[T]$ in ${\rm Aut}(\pi_1,K)/\langle \iota_K \rangle$ verify
\be
[F]^2=([F][S])^2=([F][S][T])^2={\rm Id},  \quad [S]^2=([S][T])^3, \quad [S]^4={\rm Id}.
\label{gen}\ee
It is well-known (\cite{Cox}, Eq.~(7.21)) that the  relations in  (\ref{gen})
are defining relations of the group $GL(2,\bZ)$
and the last two of them define the subgroup $SL(2,\bZ)$.
Correspondingly, we have the identifications
\be
SL(2,\bZ) \simeq \Aut^+(\pi_1,K)/\langle \iota_K \rangle \simeq  {\rm MCG}^+_{p_0}(\Sigma),
\label{634}\ee
where  $\Aut^+(\pi_1,K)$ is generated by $S$ and $T$  and
${\rm MCG}^+_{p_0}(\Sigma)$ contains the orientation-preserving mapping classes.
It is helpful to display the integer matrices associated to the classes $[F],[S]$
and $[T]$ by the isomorphism $\cI$ (\ref{629}).
Following \cite{Gol}, we obtain
\be [F]\to   \begin{bmatrix}
0 &1  \\
1 & 0
\end{bmatrix},\quad  [S]\to   \begin{bmatrix}
0 &1  \\
-1 & 0
\end{bmatrix},\quad  [T]\to   \begin{bmatrix}
1 &0  \\
1 & 1
\end{bmatrix}.\label{homol}\ee
We did not need all four automorphisms $F,S,T$ and $\tilde T$ to characterize the mapping class group, since
the relation  $\tilde T=(TST)^{-1}$ is valid.
Instead of $F,S$ and $T$ we can use  as  generators $F,T$ and $\tilde  T$, whereby
 $[S]$ can be expressed as the composition
 of three successive Dehn twists:
\be
[S]= ([T][\tilde T][T])^{-1}.
\label{three}
\ee

\subsection{The duality map $\fS$ as a mapping class symplectomorphism}

 By definition, an {\it automorphism} of a quasi-Hamiltonian space
$(M,G,\om,\mu)$ is  a  diffeomorphism of $M$
that preserves $\om$ and $\mu$ and commutes  with the $G$-action.
We next exhibit a natural homomorphism from the group  $\Aut^+(\pi_1,K)$ generated by $S$ and $T$  (\ref{class})
into the
group Aut$(D(G))$ of automorphisms of the double.
For this purpose, we set
\be
S(A,B):=(B^{-1},BAB^{-1}),
\quad T(A,B):=(AB,B),
\quad \tilde T(A,B):=(A,BA^{-1}).
\label{mpg}\ee
One readily verifies  that the maps $S$, $T$  and $\tilde T$ defined by (\ref{mpg}) preserve
the basic structures (\ref{2conj}),
(\ref{2.2}) and (\ref{1.1}) of the double, giving  automorphisms of $D(G)$.
Note that we slightly abuse    the notation by using the same symbols $S$, $T$, $\tilde T$ for the elements of
$\Aut^+(\pi_1,K)$ and of Aut$(D(G))$,
and later we shall use them even for corresponding elements of  factor groups of ${\rm Aut}^+(\pi_1,K)$ and
of Aut$(D(G))$. We believe however  that the reader prefers to figure out from the context which group we have
in mind rather than to remember which of  several different putative notations corresponds to which group.

The similarity of the formulae (\ref{class}) and (\ref{mpg})  implies that in
(\ref{mpg}) we have constructed a homomorphism
of  $\Aut^+(\pi_1,K)$ into Aut$(D(G))$.
This similarity is not accidental, of course. Indeed,
the assignment of a pair of $G$-elements
$A$ and $B$ to a flat connection
is not unique but it is fixed by the choice of the homology cycles along which one computes the
holonomies of the flat connection. It is the
standard choice of  homology basis which we have adopted previously.
The transition from the standard homology basis to another one  following the matrices
(\ref{homol}) induces   the automorphisms
(\ref{mpg})  of the double.

As we know, the fourth power $S^4$ in $\Aut^+(\pi_1,K)$ is  the conjugation by the element $K^{-1}$,
which belongs to the center of $\Aut^+(\pi_1,K)$.
On the other hand,
calculating $S^4$ (\ref{mpg}) in Aut$(D(G))$ gives the
  so-called twist automorphism $Q$ introduced in \cite{AMM}:
\be
Q(A,B)=\Psi_{\mu(A,B)^{-1}}(A,B).
\label{twist}\ee
It is easy to check directly that the
 twist  automorphism  is in the center of the group Aut$(D(G))$
 (this is by the way true for any quasi-Hamiltonian manifold).
The Abelian group $\langle Q\rangle$ generated by $Q$ is thus a distinguished normal subgroup of Aut$(D(G))$,
and Aut$(D(G))$ can be viewed as a central extension of the group Aut$(D(G))/\langle Q\rangle$.
On account of (\ref{634}),  we may conclude that
the orientation-preserving  subgroup of the mapping class group
 acts on the double $D(G)$  `projectively'.

Turning to quasi-Hamiltonian reduction, note that
the automorphisms   $S$, $T$ and $\tilde T$ defined in (\ref{mpg})  respect the constraint surface
$\mu^{-1}(\mu_0)$. Moreover, $\mu_0$ belongs to the isotropy group $G_0$  and therefore
 the automorphism $Q$ descends
to the trivial identity map on the reduced phase space  $P=\mu^{-1}(\mu_0)/G_0$.
Thus $S$, $T$ and $\tilde T$ descended on $P$ generate a true action
of the orientation-preserving  subgroup of the  mapping class group,
${\rm MCG}^+_{p_0}(\Sigma)$ given by (\ref{634}).
By construction, this action operates via symplectomorphisms of $(P, \hat\om)$.

We are now going to demonstrate
that in the $\cp$ parametrization of the reduced phase space $P$ of our interest
the mapping class generator $S$ yields just the Ruijsenaars self-duality symplectomorphism $\fS$ of $\cp$.

\medskip
\noindent \textbf{Theorem 9.} \emph{With the choice $\mu_0$ in (\ref{1.2}), let us denote by
$S_P:P\to P$ the mapping class symplectomorphism that
descends from the automorphism $S$  (\ref{mpg}). Then
\be  \fS=f_\beta^{-1}\circ S_P\circ f_\beta,
\label{638}\ee
where   $f_\beta: \cp\to P$ is the Delzant symplectomorphism constructed in Section 4 and
$\fS:\cp\to\cp$  is the Ruijsenaars symplectomorphism defined in (\ref{fSR}).}

\begin{proof}
   We have obviously
\be (S_P\circ p)(A,B)=p(B^{-1},BAB^{-1}), \quad \forall (A,B)\in\mu^{-1}(\mu_0),\ee
where $p$ is the projection from $\mu^{-1}(\mu_0)$ to $P= \mu^{-1}(\mu_0)/G_0$.
Recall  the involution $\hat \nu$ that verifies $(\hat\nu\circ p)(A,B)=p(\bar B,\bar A)$ and descends from
the map $\nu(A,B)=(\bar B,\bar A)$
defined on the double in (\ref{B1}).  We can thus write
\be
(\hat\nu \circ S_P\circ p)(A,B)= p(\bar B \bar A \bar B^{-1} , \bar B^{-1}),
\quad \forall (A,B) \in \mu^{-1}(\mu_0).
\label{7-2}\ee
Let us show that
\be
\hat \nu\circ S_P\circ f_\beta=f_\beta\circ \hat C,
\label{Lip}\ee
where $\hat C$ is the complex conjugation on $\cp$ introduced in
(\ref{wise}). By continuity, it is sufficient to verify (\ref{Lip}) on $\cp_0$, where
(thanks to the formula (\ref{C15}))
 it can be rewritten as the equality
\be
p\circ \Psi_{g_y(\xi)^{-1}} \left( \delta(\xi)^{-1} L_y^{\mathrm{loc}}(\delta(\xi), \rho(\tau)^{-1})^* \delta(\xi),
\delta(\xi)\right) =
p\circ \Psi_{g_y(\xi)^{-1}} \left(  L_y^{\mathrm{loc}}(\delta(\xi), \rho(\tau)),
\delta(\xi)\right),
\ee
for all $(\xi,\tau) \in \cP^0 \times \bT_{n-1}$. But this holds simply on account of equation (\ref{B38})
applied to $-y$ instead of $y$.

Now we use the involutivity of $\hat\nu$, then  Theorem 6, which states that
$f_\alpha=\hat\nu\circ f_\beta\circ \hat\Gamma$,
then also equation (\ref{4119}), rewritten
as $\fS^2=\hat\Gamma\circ\hat C$, and the definition  of the map $\fS:=f_\alpha^{-1}\circ f_\beta$ (\ref{fSR})
to  conclude  from (\ref{Lip})
\be
f_\beta^{-1} \circ S_P\circ f_\beta= f^{-1}_\beta\circ \hat\nu\circ f_\beta\circ \hat C=
f_\beta^{-1} \circ f_\alpha \circ \hat \Gamma \circ \hat C=
f_\beta^{-1}\circ f_\alpha\circ \fS^2=\fS^{-1}\circ \fS^2=\fS.\ee
\end{proof}

As is well-known, and is evident from (\ref{three}),
the Dehn twists $T$ and $\tilde T$ can be used as alternative  generators of the (orientation-preserving)
mapping class
 group instead of $T$ and $S$.
 This directly leads to the  decomposition
 \be
 S_P= (T_P\circ \tilde T_P\circ T_P)^{-1}.
 \label{644}\ee
  Applying ideas of Goldman \cite{Gold} (see also \cite{Meus}), we now show
  that the Dehn twist symplectomorphisms
   $T_P$ and $\tilde T_P$ themselves are specializations  of simple Hamiltonian flows
  on $P$. More precisely, we can realize already the automorphisms $T$ and $\tilde T$ of $D$ given in (\ref{mpg}) by
  means of quasi-Hamiltonian flows as stated by the following lemma.

 \medskip
 \noindent
 {\bf Lemma 11.}  \emph{Employing the notations of Subsection 2.2, define the functions $h\in C^\infty(D_b)^G$ and
 $\tilde h \in C^\infty(D_a)^G$  by
   \be
     h=\tr( \sum_{k=1}^{n-1} \beta_k\lambda_k)^2, \qquad
     \tilde h=\tr( \sum_{k=1}^{n-1} \alpha_k\lambda_k)^2,
   \ee
   and let $\phi_{h,s}$ and $\phi_{\tilde h, s}$ be the corresponding quasi-Hamiltonian flows.
   Then, respectively on $D_b$ and on $D_a$,  there hold the equalities
   $T(A,B) = \phi_{h,1}(A,B)$ and $\tilde T(A,B) =\phi_{\tilde h,1}(A,B)$.}
   \medskip

\begin{proof}
Using the definitions of Subsection 2.2,
introduce  arbitrary real powers of any $C\in G_\mathrm{reg}$ by setting
\be
C^s:=  g(C)^{-1}\exp\left(-2 \ri s\sum_{k=1}^{n-1} \Xi_k(C) \lambda_k\right)g(C),
\qquad
\forall s\in \bR.
\ee
Then it follows from equations (\ref{flow1}), (\ref{flow2}) and (\ref{A6}) that
\be
\phi_{h,s}(A,B)= (AB^s, B),
\qquad
\phi_{\tilde h,s}(A,B)= (A, B A^{-s}),
\ee
and comparison with (\ref{mpg})  entails the claim.
 \end{proof}

 The functions $h$ and $\tilde h$ descend to $P$, and when transferred to the model
$\cp$ they become functions of the global particle-positions, $\hat \beta_k \circ f_\beta$, and action-variables,
$\hat \alpha_k \circ f_\beta$.
The decomposition of the Ruijsenaars duality map $\fS$ implied by (\ref{638}) and (\ref{644})
  represents a new result.
  This is a simple  by-product of the reduction approach, which would have been difficult
to notice
in the direct approach \cite{RIMS95} to the compactified $\mathrm{III}_\mathrm{b}$ system.

\subsection{The anti-symplectomorphism $\fR$ as a $\gl$ generator}
In Section 6.3, we have  implemented the generators $S$ and $T$ of  the
orientation-preserving $SL(2,\bZ)$ part of the full mapping class group $\gl$  as
 automorphisms of the double that descend upon reduction to the symplectomorphisms $S_P$ and $T_P$
 of our reduced phase space $P$.
We now observe that, on the one hand,
the third generator $[F]$ of $\gl$  implemented, according to (\ref{class}),
as the map $F_D(A,B):=(B,A)$
is not an automorphism of the double and it does not survive the quasi-Hamiltonian reduction.
On the other hand,
if we consider instead of $F_D$  the related
 map $\nu:D(G)\to D(G)$ defined by (\ref{B1}) as  $\nu(A,B)=(\bar B,\bar A)$, then
$\nu$ maps $\mu^{-1}(\mu_0)$ to itself and induces the anti-symplectic involution $\hat\nu$
of the reduced phase space. Moreover, it is readily checked that
under the assignment
\be
[F] \mapsto \hat \nu,
\quad
[S]\mapsto S_P,
\quad
[T]\mapsto T_P
\ee
$\hat\nu$, $S_P$ and $T_P$  fulfil the generating relations
(\ref{gen}) of $\gl$, and thus they induce   an action of $\gl$ on $P$.
 Note also that $\gl$ can be written as  the semi-direct product
\be
GL(2,\bZ) = \bZ_2 \ltimes SL(2,\bZ),
\ee
where the $\bZ_2$ subgroup is generated by $[F]$.
Correspondingly, there are two kinds of elements of
$\gl$: $(+,\rho)$ and $(-,\rho)$ where $\rho\in SL(2,\bZ)$ and $\pm$ is the sign of the determinant of the
$GL(2,\bZ)$ matrix.
The elements $(+,\rho)$ acting on $P$ are
symplectomorpisms and $(-,\rho)$ are anti-symplectomorphisms. This follows from
the fact that $\hat\nu\equiv(-,e)$ reverses
the sign of the symplectic form on $P$ (since $\nu$ reverses the sign of $\om$ on $D(G)$)
while $S_P$ and $T_P$ preserve it.

Parametrizing our reduced phase space $P$ as $\cp$
by means of the Delzant symplectomorphism $f_\beta$, as before,  the generator
$\hat\nu$ of $\gl$ becomes $f_\beta^{-1}\circ \hat\nu\circ f_\beta$ and it is directly related to the
involutive Ruijsenaars  anti-symplectomorphism $\fR:\cp\to\cp$.
Indeed, we
find the following identities by combining Theorem 6, which states that
$f_\alpha=\hat\nu\circ f_\beta\circ \hat\Gamma$,
the definition  of the map $\fS:=f_\alpha^{-1}\circ f_\beta$ (\ref{fSR})
and Eq.~(\ref{novy1})  saying that $\fS\circ\hat\Gamma=\fR$:
\be
\fR=\fS\circ\hat\Gamma= f_\alpha^{-1}\circ f_\beta\circ \hat\Gamma =
f_\alpha^{-1}\circ \hat\nu \circ f_\alpha=
\fS\circ f_\beta^{-1}\circ \hat\nu\circ f_\beta\circ \fS^{-1}.
\ee
Hence, the Ruijsenaars map $\fR$ can be viewed as an alternative anti-symplectic generator of the $\gl$
action on $\cp$.

We have exhibited an (anti)-symplectic action of the full group $\gl$ on our
reduced phase space. This action
does not descend from the (projective) action of the full mapping class group on the double since
we have replaced the generator $F_D$
by the  new generator $\nu$.
It is thus natural to ask the  following question: Does the $\gl$ generator $\hat \nu$
has its origin in some natural
(anti)-symplectomorphism of the space of connections $(\cA(\Sigma),\Omega)$? It turns out that
the answer to this question is positive. Indeed,
take any orientation-reversing diffeomorphism of $\Sigma$ which is in the class $F$ of the
mapping class group and compose it with
the complex conjugation acting on the $su(n)$-valued connection 1-forms (without touching their argument).
  Since the complex conjugation is an automorphism
 of the group $SU(n)$ and of its Lie algebra, this composed map is an
 anti-symplectomorphism of $(\cA(\Sigma),\Omega)$ which
 descends to the involutive anti-symplectomorphism $\hat\nu$.

 \section{Discussion}
\setcounter{equation}{0}

In this paper we have demonstrated that an appropriate quasi-Hamiltonian reduction
\cite{AMM} of the internally  fused double
$D=SU(n) \times SU(n)$   yields a
reduced phase space $P$
that turns into a Hamiltonian toric manifold (i.e.~a compact completely integrable system) in two
different but equivariantly symplectomorphic  ways.
The underlying two toric moment maps on $P$,
with respective components $\hat \alpha_k$ and $\hat \beta_k$,
 arise from the reductions of
the two sets of spectral Hamiltonians on $D$ generated by the two components of the pair $(A,B) \in D$.
On the other hand, the phase space $\cp$ also carries two distinguished toric structures, with
moment maps $\cJ_k$ and $\Xi_k \circ L^y$ that encode, respectively,
 the particle-positions and the action-variables
of the compactified $\mathrm{III}_\mathrm{b}$ system as discovered in \cite{RIMS95}.
\emph{We have explicitly constructed two `Delzant symplectomorphisms'
$f_\alpha$ and $f_\beta$ from  $\cp$ to  $P$ that relate these toric moment maps according to
Eqs.~(\ref{58}) and (\ref{510}), and have identified the composed map
$\fS= f_\alpha^{-1} \circ f_\beta$ as the symplectic self-duality map \cite{RIMS95}  of
the compactified  $\mathrm{III}_\mathrm{b}$ system, which satisfies  Eq.~(\ref{517})}.
In our setting  the symplectic property of the pertinent self-duality map is obvious, while in the
original approach  of  \cite{RIMS95} it required  a special proof.
 We have also recovered the anti-symplectic
version $\fR= \hat C \circ \fS$ of Ruijsenaars' self-duality map,
which satisfies Eq.~(\ref{523}).

In addition, we have rigorously established the
interpretation of the compactified  $\IIIb$ system in terms of flat $SU(n)$ connections
on the one-holed torus suggested by Gorsky and Nekrasov \cite{GN} and,
 by proving the formula  $\fS= f_\beta^{-1} \circ S_P \circ f_\beta$ (\ref{638}),
\emph{we have demonstrated that the Ruijsenaars self-duality map $\fS$ represents
the natural action of the mapping class generator $S\in SL(2,\bZ)$ on $\cp\simeq P$}.
As for the map $\fR$,  we have shown that it arises from a  $GL(2,\bZ)$
extension of the $SL(2,\bZ)$ mapping class group action on our reduced phase space.

The interpretation of the Ruijsenaars self-duality  as the reduction remnant
of the $SL(2,\bZ)$ mapping class
generator $S$  is a long-expected result that we
finally succeeded to prove thanks to the
quasi-Hamiltonian technique. For the sake of objectivity, we should
mention that  Gorsky and his collaborators
were very close to establish this  interpretation; formula (4.31)
of their paper \cite{JHEP} coincides essentially with our formula
(\ref{mpg}). However, they remarked that their definition (4.31) of $S$
violates the $SL(2,\bZ)$ relations and  they could recover
a true $SL(2,\bZ)$ action only for the rational Calogero limiting case
of the reduced system \cite{JHEP}.  Since they have not furnished more quantitative
details we cannot
extract from their paper the precise  cause of the trouble, but we
believe that it may be related to the fact that our formula (\ref{mpg})
also defines only  the action of a  suitable central extension
of $SL(2,\bZ)$  on the double and  not a true
   $SL(2,\bZ)$ action. Our point is, however,  that upon the
   quasi-Hamiltonian reduction this projective action descends
to a true $SL(2,\bZ)$ action on the reduced phase space.

Besides the coupling constant, $y$, a second parameter, $\Lambda$, can be
introduced
into the $\mathrm{III}_\mathrm{b}$ system by replacing the symplectic form
(\ref{I.3})
by $\Omega^{\mathrm{loc}}_{\Lambda}:= \Lambda \Omega^{\mathrm{loc}}$.
The local Darboux variables  $p_j$ and $x_j$ then parametrize
$\delta$ and $\Theta$  in (\ref{I.2}) as $\delta_j=e^{2 \ri x_j/\Lambda}$
and $\Theta_j = e^{-\ri p_j}$,
whereby $x_j$ becomes $x_j/\Lambda$ also in the Hamiltonian (\ref{I.1}).
The parameter $\Lambda$ can be encoded in the reduction approach by
choosing the invariant scalar
product on $su(n)$ to be $-\frac{\Lambda}{2}\tr$, which  scales the 2-form
$\omega$ (\ref{2.2})
as well as the reduced symplectic form and the corresponding toric
moment polytope.
Being a mere scale parameter of the symplectic structure, $\Lambda$
essentially plays
no role at the classical level, and we omitted it to simplify the notations.
However, this parameter is important at the quantum level (see \cite{vDV}).

After the present paper,   just one from the list of the known Ruijsenaars
dualities remains to be derived in the reduction approach.
It is the self-duality of the hyperbolic Ruijsenaars-Schneider system
for which  a suitable `double' to be reduced is still to be discovered.
It may appear tempting to search for distinct real forms of the complex holomorphic
constructions of \cite{FR,Ob}, but this scenario  does not seem to work and the problem is wide open.
An intriguing reformulation of the problem is to enquire whether the known self-duality map of the
hyperbolic system \cite{SR-CMP} can be factorized similarly to the
representation $\fS = f_\alpha^{-1} \circ f_\beta$
that we obtained here in the case of
the compactified $\mathrm{III}_\mathrm{b}$ system.

Another interesting problem for the future is to study the Ruijsenaars duality in relation to root
systems different from $A_n$.
For progress in this direction, we refer to the paper \cite{Pusztai}.

\newpage
\noindent{\bf Acknowledgements.}
We wish to thank A. Alekseev for relevant remarks and for bringing reference \cite{MW}
to our attention,
and are also grateful to  J.~Huebschmann for useful information on moduli spaces.
We are indebted to V. Fock, who in relation to an earlier draft posed us a question about
the role of the mapping class group,
which motivated our research described in Subsections 6.3 and 6.4.
We thank the referee for suggesting the simple proof of the free action of $\bar G_0$,
which prompted us to present here a short and rather general proof of Theorem 1.
This work was supported in part
by the Hungarian
Scientific Research Fund (OTKA, K77400).

\end{document}